\documentclass[twoside,11pt]{article}

\usepackage[abbrvbib]{jmlr2e}

\hypersetup{hidelinks}

\usepackage{amsmath,amssymb,mathrsfs,dsfont,mathtools,amsfonts}%
\usepackage{graphicx}
\usepackage[table]{xcolor}
\usepackage{booktabs}
\usepackage{multirow}
\usepackage{tabularx}
\usepackage{array}

\usepackage{bm}
\usepackage{natbib}
\usepackage{xcolor}
\usepackage{colortbl}
\usepackage{algorithm}
\usepackage{algpseudocode}

\usepackage{comment}
\usepackage{booktabs,subcaption}
\usepackage[font=footnotesize,labelfont=bf]{caption}
\usepackage{dsfont}
\usepackage[all]{nowidow}
\usepackage{tikz}

\newcommand{\barwidth}[1]{\rule{#1cm}{6pt}}

\newcolumntype{L}[1]{>{\raggedright\arraybackslash}p{#1}} % Fixed-width column
\newcolumntype{M}{>{\raggedright\arraybackslash}X} % Auto-stretch column
\newcolumntype{Y}{>{\centering\arraybackslash}X} % Auto-stretch column
\newcolumntype{Z}[1]{>{\centering\arraybackslash}p{#1}} % Fixed-width column

\algblock{ParFor}{EndParFor}
\algnewcommand\algorithmicparfor{\textbf{for}}
\algnewcommand\algorithmicpardo{\textbf{do Parallel}}
\algnewcommand\algorithmicendparfor{\textbf{end\ Parallel for}}
\algrenewtext{ParFor}[1]{\algorithmicparfor\ #1\ \algorithmicpardo}
\algrenewtext{EndParFor}{\algorithmicendparfor}
\def\NoNumber#1{{\def\alglinenumber##1{}\State #1}\addtocounter{ALG@line}{-1}}

\newcommand \dd  { \,\textup d}   % infintesimal
\def\T{{ \mathrm{\scriptscriptstyle T} }}

\newcommand{\given}{\,|\,}

\newcommand{\Exp}{\mbox{E}}
\newcommand{\E}{\mbox{E}}

\usepackage{lastpage}
\jmlrheading{27}{2026}{1-\pageref{LastPage}}{1/26; Revised 7/26}{7/26}{26-0307}{Luca Presicce and Sudipto Banerjee}

\ShortHeadings{Bayesian Transfer Learning for GeoAI Systems}{Presicce and Banerjee}
\firstpageno{1}

\begin{document}

\title{Bayesian Transfer Learning for Artificially Intelligent Geospatial Systems: A Predictive Stacking Approach}

\author{\name Luca Presicce \email l.presicce@campus.unimib.it \\
       \addr Department of Economics, Management and Statistics\\
       University of Milano-Bicocca\\
       Milan, Italy
       \AND
       \name Sudipto Banerjee \email sudipto@ucla.edu \\
       \addr Department of Biostatistics\\
       University of California, Los Angeles\\
       Los Angeles, CA 90025, USA}

\editor{Christoph Lampert}

\maketitle

\begin{abstract}%   <- trailing '%' for backward compatibility of .sty file
Building artificially intelligent geospatial systems requires rapid delivery of spatial data analysis on massive scales with minimal human intervention. Depending on their intended use, learning about underlying spatial processes can also involve model assessment and uncertainty quantification. We devise transfer learning frameworks for deployment in artificially intelligent systems, where a massive data set is split into smaller data sets that stream into the analytical framework to propagate learning and assimilate learning for the entire data set. Specifically, we develop Bayesian predictive stacking for multivariate spatial data and demonstrate rapid automated probabilistic learning from massive spatial data sets.
We illustrate the effectiveness of our approach through extensive simulation experiments and through the analysis of a massive dataset on vegetation index that are indistinguishable from traditional (and more expensive) statistical approaches.
\end{abstract}

\begin{keywords}
  Bayesian predictive stacking, climate science, Gaussian process, geospatial systems, matrix-variate spatial models, transfer learning
\end{keywords}

%%%%%%%%%%%%%%%%%%%%%%%%%%%%%%%%%%%%%%%%%%%%
\section{Introduction}\label{sec:intro}

Geospatial artificial intelligence (\textsc{GeoAI}) is a rapidly evolving discipline at the interface of machine learning and spatial data science that attempts to harness the analytical capabilities of Artificial Intelligence (\textsc{ai}) to analyze massive amounts of geographic data for data-driven scientific discoveries in the environmental and physical sciences. The area, while still fledgling, presents opportunities to devise essential data analytic tools that should comprise an artificially intelligent ``geospatial'' system. 
This manuscript focuses primarily on statistical learning tools for artificially intelligent \textsc{GeoAI} systems. 

What should be the role of probabilistic learning in \textsc{GeoAI}? Formal inference for spatial random fields enjoys a venerable intellectual presence spanning several decades of theoretical developments within classical and Bayesian paradigms \citep[see, e.g.,][]{cressie_statistics_1993, stein_interpolation_1999, gelfand_handbook_2010, cressie_statistics_2011, gelfandBanerjee2017annurev, banerjee_hierarchical_2025}. Statisticians have built richly structured hierarchical models in pursuit of ``full'' probabilistic inference for richly structured spatial data. The term ``full'' loosely refers to the estimation of \emph{all} unknowns in the model, including all parameters (irrespective of how well they are informed by the data), random effects, and predictive random variables for probabilistic interpolation at arbitrary locations (and time points in spatial-temporal data). 

Spatial data analysis employs Gaussian processes (GPs) to model dependence and achieves superior predictive inference. While GPs offer flexibility and are a conspicuous choice in spatial modeling, they generally do not offer computationally exploitable structures for covariance matrices. The computational requirements for full inference become onerous and impractical for large datasets. This presents a conundrum: should we strive to completely retain the probabilistic rigors of theoretical inference, which may be computationally burdensome or even impracticable given the massive amounts of data, or should we concede uncertainty quantification and adopt algorithm-driven data analysis that would easily scale massive data sets? Statistical learning on unprecedented scales requires some concessions from conventional decision-theoretic paradigms, as has been reflected, for example, in a recent comprehensive case study by \citet{zhou_measuring_2022}, who synthesize statistical models with machine learning algorithms to measure temporal trends and spatial distribution of housing vitality by exploiting information from multiple sources \citep[also see the discussion by][for conventional Bayesian modeling perspectives and its challenges]{banerjee_discussion_2022}. Other efforts in combining machine learning methods with formal spatial statistics include deep learning for spatial data \citep[][with the latter offering a comprehensive review of spatial machine learning]{zammit-mangion_deep_2022, wikle_statistical_2023} and, in a related domain, deep GP emulation in computer experiments \citep{sauer_active_2023}, and spatial random forests \citep{georganos_geographical_2021,talebi_truly_2022,saha_random_2023}.  

Even a cursory review reveals various methods for massive spatial datasets, which are too vast to be summarized here \citep[see, e.g.,][]{banerjee_high-dimensional_2017, heaton_case_2017}. 
Examples include reduced-rank or subsets of regression approaches \citep[see, e.g.,][]{quinonero-candela_unifying_2005, cressie_fixed_2008, banerjee_gaussian_2008, wikle_low-rank_2010}, multi-resolution approaches \citep{nychka_multiresolution_2015, katzfuss_multi-resolution_2017}, and graph-based models \citep{vecchia_estimation_1988, datta_hierarchical_2016, katzfuss_general_2021, peruzziBanerjeeFinley2022, dey_graphical_2022, sauer_vecchia-approximated_2023}. Full inference typically requires Markov chain Monte Carlo (\textsc{mcmc}) \citep{finley_efficient_2019}, variational approximations \citep{ren_variational_2011, wu_variational_2022, cao_variational_2023} or Gaussian Markov random field approximations \citep[][]{rue_approximate_2009, lindgren_explicit_2011} and integrated nested Laplace approximations (\textsc{inla}) for computing the marginal posterior distributions of the process.

The aforementioned methods focus on the richness of statistical inference but can be encumbered by the challenges presented by spatial data analysis. To be specific, we seek inference for the spatial process, which is an uncountable set of random variables, based only on a partial realization of the process manifested in the form of measurements at a finite set of spatial locations \citep[see, e.g., Chapter~3 in][]{banerjee_hierarchical_2025}. In particular, a careful choice of prior distributions and subsequent tuning of the iterative estimation algorithms are needed because spatial covariance kernel parameters are typically not statistically identifiable or consistently estimable from the observed data \citep{zhang_inconsistent_2004, tang_identifiability_2021}. Exploratory data analysis (\textsc{eda}) methods \citep[such as the variogram analysis presented in Appendix~\ref{sec:eda} of this article; also see, e.g., Chapter~2 in][for details on spatial \textsc{eda}]{banerjee_hierarchical_2025} provide some insights into the values of spatial covariance kernel parameters generating the data. Geostatistical models can then be fixed over a finite set of values for these parameters, and the inference can be assimilated or ``stacked'' over these models \citep{zhangEtAl2025jasa, pan_bayesian_2025}.

Our approach relies on two basic tenets: (i) model-based statistical inference for underlying spatial processes (including multivariate processes) in a robust and largely automated manner with minimal human input; and (ii) achieving such inference for truly massive amounts of data without resorting to iterative algorithms that may require significant human intervention to diagnose convergence (such as in \textsc{mcmc}).    

We retain the benefits of posterior learning of the underlying stochastic process. In particular, we obtain predictive learning over the uncountable collection of points in the spatial domain while propagating uncertainty about the model's parameters. However, some compromises regarding the richness of models are inevitable from the viewpoint of \textsc{GeoAI}. At the same time, statistical distribution theory has much to offer, and we harness analytical closed-form distribution theory within a family of matrix-variate distributions to deliver inference. This is only possible by ``fixing'' certain spatial correlation kernel parameters that are weakly identified and hinder automated inference. Hence, we infer using closed-form matrix-variate distributions corresponding to a set of fixed values of hyperparameters and subsequently assimilate posterior learning using Bayesian predictive stacking or \textsc{bps} \citep{wolpert_stacked_1992, yao_using_2018, zhangEtAl2025jasa, pan_bayesian_2025}. Stacking \citep{breiman_stacked_1996} is primarily related to predictions by minimizing the generalization error rate of one or more models.
Bayesian predictive stacking (\textsc{bps}) is a particular type of Bayesian model averaging and, more broadly, Bayesian predictive synthesis \citep[see, e.g.,][for different perspectives]{clarke2003jmlr, clyde2013bayesian, le2017bayes, mcalinn_dynamic_2019, mcalinn_multivariate_2020, yaoEtAl2022, tallman_bayesian_2023, cabel_bayesian_2025}.

A \textsc{GeoAI} system will access massive amounts of data. We devise Bayesian transfer learning on a streaming set of spatial datasets. Data flows into the \textsc{GeoAI} system in the form of $K$ subsets, each with $\approx n/K$ locations. We analyze and propagate inference from one subset to the next and assimilate inferences from these subsets. Delegating Bayesian calculations to a group of independent datasets is intuitive and has been explored from diverse perspectives. Examples include Consensus Monte Carlo (\textsc{cmc}) \citep{scott_bayes_2016} and its kernel-based adaptations \citep{rendell_global_2021}, Bayesian meta-analysis in clinical applications \citep[see, e.g., Chapter~4 in][]{parmigiani_modeling_2002}, and in spatially-temporally structured data \citep{bell_meta-analysis_2005, kang_meta_2011}; and more recently, ``meta'' approaches using diverse distributed computing architectures \citep{srivastava_wasp_2015, deisenroth_distributed_2015, minsker_robust_2017, srivastava_scalable_2018, guhaniyogi_meta-kriging_2018, guhaniyogi_multivariate_2019, guhaniyogi_divide-and-conquer_2019, guhaniyogi_distributed_2022, guhaniyogi_distributed_2023}. Such approaches have typically relied on iterative algorithms to estimate weakly identifiable parameters. Instead, we propose transfer learning using a double stacking approach. We first obtain stacked posterior distributions within each subset and then assimilate inference across subsets by a second stacking algorithm. In Figure~\ref{fig:double_stacking}, a flow chart is provided comprising three steps: (i) partition the data into smaller subsets; (ii) analyze each subset by stacking conjugate posteriors in closed form; and (iii) stack across datasets to obtain inference for all data.

\usetikzlibrary{arrows.meta, positioning, calc, decorations.pathmorphing, backgrounds, fit, petri}
\begin{figure}[t]
    \centering
    \scalebox{0.9}{
    \begin{tikzpicture}[node distance=2cm, auto]
% Nodes
\node at (-3.5, 0) [shape=circle,draw] (D) {$\mathscr{D}$};
\node at (-1, 2) [shape=circle,draw] (D1) {$\mathscr{D}_1$};
\node at (-1, 0) (Dkk) {$\vdots$};
\node at (-1,-2) [shape=circle,draw] (Dk) {$\mathscr{D}_K$};

\node at (1.5,3) [shape=rectangle,draw]  (M11){$\mathscr{M}_1$};
\node at (1.5,1) [shape=rectangle,draw]  (Mj1) {$\mathscr{M}_J$};
\node [below=0.25cm of M11] (M1) {$\vdots$};
\node [right=2cm of Dkk] (Mj) {$\vdots$};
\node at (1.5,-1) [shape=rectangle,draw]  (M1k){$\mathscr{M}_1$};
\node at (1.5,-3) [shape=rectangle,draw]  (Mjk) {$\mathscr{M}_J$};
\node [below=0.25cm of M1k] (Mk) {$\vdots$};

\node (Y1) [shape=rectangle,draw,right=1cm of M1] {\tiny$\begin{aligned}
    \hat{p}(Y_{\mathcal{U}}\;; \mathscr{D}_{1}) = \sum_{j=1}^{J}\hat{z}_{1,j}\; p(Y_{\mathcal{U}}\mid \mathscr{D}_{1},\mathscr{M}_j)\\
    \hat{p}(\Theta\;; \mathscr{D}_{1}) = \sum_{j=1}^{J}\hat{z}_{1,j}\; p(\Theta\mid \mathscr{D}_{1},\mathscr{M}_j)
\end{aligned}$};
\node [right=3cm of Mj] (Ykk) {$\vdots$};
\node (Yk) [shape=rectangle,draw,right=1cm of Mk] {\tiny$\begin{aligned}
    \hat{p}(Y_{\mathcal{U}}\;; \mathscr{D}_{K}) = \sum_{j=1}^{J}\hat{z}_{K,j}\; p(Y_{\mathcal{U}}\mid \mathscr{D}_{K},\mathscr{M}_j)\\
    \hat{p}(\Theta\;; \mathscr{D}_{K}) = \sum_{j=1}^{J}\hat{z}_{K,j}\; p(\Theta\mid \mathscr{D}_{K},\mathscr{M}_j)
\end{aligned}$};

\node (Yg) [shape=rectangle,draw,right=10cm of Dkk] {\tiny$\begin{aligned}
    \hat{p}(Y_{\mathcal{U}}\;; \mathscr{D}) = \sum_{k=1}^{K}\hat{w}_{k}\; \hat{p}(Y_{\mathcal{U}}\;; \mathscr{D}_{k})\\
    \hat{p}(\Theta\;; \mathscr{D}) = \sum_{k=1}^{K}\hat{w}_{k}\; \hat{p}(\Theta\;; \mathscr{D}_{k})
\end{aligned}$};

% Data partition arrows
\draw[->] (D) -- (D1);
\draw[->] (D) -- (Dkk);
\draw[->] (D) -- (Dk);

% Local inference arrows
\draw[->] (D1) -- (M11);
\draw[->] (D1) -- (M1);
\draw[->] (D1) -- (Mj1);
\draw[->] (Dkk) -- (Mj);
\draw[->] (Dk) -- (M1k);
\draw[->] (Dk) -- (Mk);
\draw[->] (Dk) -- (Mjk);

% Local inference results arrows
\draw[->] (M11) -- (Y1);
\draw[->] (M1) -- (Y1);
\draw[->] (Mj1) -- (Y1);
\draw[->] (Mj) -- (Ykk);
\draw[->] (M1k) -- (Yk);
\draw[->] (Mk) -- (Yk);
\draw[->] (Mjk) -- (Yk);

% Global inference arrows
\draw[->] (Y1) -- (Yg);
\draw[->] (Ykk) -- (Yg);
\draw[->] (Yk) -- (Yg);

% Section dividers
\draw[dashed, blue] ($(D1.west) + (2,1.5)$) -- ($(Dk.west) + (2,-2.5)$);
\draw[dashed, blue] ($(Y1.west) + (5,1.5)$) -- ($(Yk.west) + (5,-2.5)$);

% Text labels
\node at (-2,-4) [node distance=1.5cm, blue] {DATA PARTITION};
\node at (4,-4) [node distance=1.5cm, blue] {LOCAL INFERENCE};
\node at (11,-4)[node distance=1.5cm, blue] {GLOBAL INFERENCE};

\node at (2.15,2.75) (ss1) {};
\node at (4.15,3.9) [red] (s1) {STACKING};
\draw[->,red,thick,decorate,decoration=snake] (ss1) to node [] {} (s1.west);
% \draw[->,red,very thick,dotted] (ss1) to node [] {} (s1.west);
\node at (8.5,0.75) (ss2) {};
\node at (10.5,1.9) [red] (s2) {STACKING};
\draw[->,red,thick,decorate,decoration=snake] (ss2) to node [] {} (s2.west);
% \draw[->,red,very thick,dotted] (ss2) to node [] {} (s2.west);

\end{tikzpicture}
    }
    \caption{Double Bayesian predictive stacking approach representation}
    \label{fig:double_stacking}
\end{figure}

This article evolves according to Figure~\ref{fig:double_stacking}. Section~\ref{sec:smk} outlines Bayesian transfer learning using \textsc{double bps} (\textsc{dbps}), Section~\ref{sec: comp_programs} provides details on computation, Section~\ref{sec:sim} presents simulation experiments to illustrate and evaluate \textsc{double bps} and its applicability to amortized inference.
Section~\ref{sec:dataappl} analyzes a vegetation index data set with observed locations in the millions on global scales. Finally, Section~\ref{sec:discussion} concludes with some discussion. References to an accompanying Supplement are provided throughout the article and after Section~\ref{sec:discussion}.

%%%%%%%%%%%%%%%%%%%%%%%%%%%%%%%%%%%%%%%%%%%%
\section{Bayesian Transfer Learning}\label{sec:smk}
Transfer learning (TL) broadly refers to the propagation of knowledge from one task to accomplish a different task \citep[][]{suder_bayesian_2023}. We transfer inference through a stream of subsets to analyze the entire dataset. 
Learning from spatial random fields is challenging because of the limited information available for the latent spatial process (uncountable collection of unobserved random variables) using a partially observed finite realization. We devise an automated approach for amortized learning \citep{zammit-mangion_neural_2024}.

%-------------------------------------------
\subsection{Divide-and-Conquer Multivariate Bayesian inference}\label{sec:DivConBays}
Let $Y_{n\times q}$ be an $n\times q$ random matrix that is endowed with a probability law from the matrix-normal distribution, $\operatorname{MN}(M,V,U)$, with a probability density function
\begin{equation}\label{eq:MatrixNormal}
    p(Y\mid M,V, U)= \frac{\exp\left[ -\frac{1}{2} \operatorname{tr}\left\{  U^{-1} (Y-M)^\T V^{-1} (Y-M) \right\} \right]}{(2\pi)^{\frac{nq}{2}} \mid  U\mid^{\frac{n}{2}} \mid V\mid^{\frac{q}{2}} },
\end{equation}
where $\operatorname{tr}(\,\cdot\,)$ is the trace operator on a square matrix, $M$ is the mean matrix, and $V$ and $U$ are the $n\times n$ row-covariance and $q\times q$ column covariance matrices, respectively. We consider the matrix-variate Bayesian linear regression model
\begin{equation}\label{eq:MatrixNormLik}
\begin{split}
Y &= X\beta + E_{Y}\;,\; E_{Y} \sim \operatorname{MN}(O, V, \Sigma)\;;\;
\beta = M_0m_0 + E_{\beta}\;,\; E_{\beta} \sim \operatorname{MN}(O, M_0, \Sigma)\;, 
\end{split}
\end{equation}
where $Y$ is $n\times q$, $X$ is $n\times p$ comprising explanatory variables, $\beta$ is $p\times q$ consisting of regression slopes, $E_{Y}$ and $E_{\beta}$ are zero-centered random matrices with row covariances $V$ and $M_0$, respectively, and a shared column covariance matrix $\Sigma$. We assign an inverse-Wishart distribution $\Sigma\sim\operatorname{IW}(\Psi_{0} ,\nu_{0})$ and denote the joint density of $\beta$ and $\Sigma$ by $\operatorname{MNIW}(M_0m_0, M_0, \Psi_{0}, \nu_{0})$, which provides a closed-form posterior for $\{\beta,\Sigma\}$ in the same family.

Let $\mathscr{D}=\{Y,X\}$ be the entire dataset, which is too large to even be accessed, let alone analyzed using \eqref{eq:MatrixNormal} within the \textsc{GeoAI} system. Therefore, we envisage $K$ disjoint and exhaustive subsets $\mathscr{D}=\{\mathscr{D}_{1},\dots,\mathscr{D}_{K}\}$ streaming into \textsc{GeoAI} as a sequence. Each $\mathscr{D}_{k} =  \{Y_{k},X_{k}\}$ consists of $n_{k}$ rows of $Y$ and $X$, where $n=\sum_{k=1}^K n_{k}$, $Y_{k}$ is $n_{k}\times q$ and $X_{k}$ is $n_{k}\times p$ for each $k=1,\dots,K$. We now fit \eqref{eq:MatrixNormLik} to each subset using $Y_{k} = X_{k}\beta + E_{k}$, $E_{k} \sim \operatorname{MN}(O, V_{k}, \Sigma)$, where $V_{k}$ is the $n_{k}\times n_{k}$ row-covariance matrix corresponding to the rows in $Y_{k}$; the specification for $\{\beta,\Sigma\}$ remains as in \eqref{eq:MatrixNormLik}. Starting with $\operatorname{MNIW}(\beta,\Sigma \given M_k m_k, M_k, \Psi_k, \nu_k)$ at $k=0$ (the prior), we use Bayesian updating $p(\beta,\Sigma\mid\mathscr{D}_{1:k+1}) \propto p(\beta,\Sigma\mid\mathscr{D}_{1:k}) \times p(Y_{k+1} \mid X_{k+1}\beta, V_{k+1}, \Sigma)$ to obtain $\beta,\Sigma \mid \mathscr{D}_{1:k+1}\sim \operatorname{MNIW}(M_{k+1}m_{k+1}, M_{k+1}, \Psi_{k+1}, \nu_{k+1})$ with $M_{k+1}^{-1} = M_{k}^{-1} + X_{k+1}^\T  V_{k+1}^{-1} X_{k+1}$, $ m_{k+1} = m_k + X_{k+1}^\T  V_{k+1}^{-1} Y_{k+1}$, $\nu_{k+1} = \nu_{k} + n_{k+1}$, and $\Psi_{k+1} = \Psi_{k} + Y_{k+1}^\T  V_{k+1}^{-1} Y_{k+1} + m_{k}^\T M_{k}m_{k} - m_{k+1}^\T M_{k+1} m_{k+1}$. Upon terminating at $k=K$, we exactly recover the posterior distribution $p(\beta,\Sigma \given \mathscr{D})$. We do not interact between subsets, and computational complexity is determined solely by the dimension of the subsets. 

Spatial random fields immediately present a challenge. The above method delivers inference without loss of information only if the $Y_k$'s are independent between blocks. If each row of $Y$ corresponds to a spatial location, and $V$ is an $n\times n$ spatial correlation matrix, then each $V_k$ is the spatial correlation matrix constructed from spatial locations $n_k$ in $\mathscr{D}_k$. The independence among blocks may yield reasonable inference if we can design the blocks such that spatial correlation between blocks does not affect inference \citep[see Section~4 of][for a ``seasons and episodes'' framework to adapt the Markovian Forward Filter Backward Sampling (FFBS) algorithm for spatial-temporal learning]{banerjeeEtAl2025jmlr}. However, designing such blocks requires significant human intervention. We seek to avoid this in transfer learning and, in particular, in amortized inference, which will require rapid estimation of datasets to train deep networks (see Section~\ref{sec:sim_abi}).

Instead, we assimilate statistical learning from each of these blocks using predictive stacking. We exploit the fact that $V$ is indexed by a small number of parameters in a spatial correlation kernel. Fixing these parameters fixes $V$, and hence $V_k$ for each $k$, producing closed-form posterior inference on $\beta$ and $\Sigma$ based on the entire dataset as described above. Stacking combines these analytically accessible distributions using an optimal set of weights that are computed using a convex optimization algorithm. These weights are then used to reconstruct the posterior and predictive distributions for the spatial random field without imposing block independence. Existing ``meta-kriging'' and related ``divide and conquer'' approaches \citep[e.g.,][and other references on ``divide and conquer'' methods provided in Section~\ref{sec:intro}]{guhaniyogi_meta-kriging_2018, scott_bayes_2016} analyze subsets of data using \textsc{mcmc}, which is expensive and not fully automated. A key distinction of the current manuscript is that we abandon all iterative estimation algorithms, let alone expensive \textsc{mcmc}, and focus on assimilating inference using closed form distributions.     

What remains to be resolved is the issue of fixing the parameters in $V$. These parameters govern the strength of association across the spatial random field and possibly the smoothness of the field. Unfortunately, these parameters are weakly identified by the data, and posterior learning struggles due to the slower convergence of iterative algorithms. In addition, exploratory spatial data science tools, such as variograms, used to glean information about these parameters will also be less helpful, as they may suggest different values for each of the variables indexed by columns, while we insist on retaining a single parameter to exploit conjugate distribution theory. Hence, we collect the closed-form posterior distributions obtained for a collection of fixed values of the spatial parameters and subsequently average these posterior distributions.

%-------------------------------------------
\subsection{Bayesian Stacking of Predictive Densities}\label{sec:BPS}
% \todo[inline]{Introduce BPS}

Bayesian predictive stacking (\textsc{bps}) assimilates models using a weighted distribution in the convex hull, $C=\left\{\sum_{j=1}^J w_j p(\,\cdot \mid \mathscr{D}, \mathscr{M}_j ): \sum_j w_j=1, w_j \geq 0 \right\}$, of individual posterior distributions by maximizing the score \citep{gneiting_strictly_2007, yao_using_2018} to fetch  
\begin{equation}\label{eq:OptProbBPSLogScore}
(w_1,\ldots,w_J)^{\T} = \arg\max _{ w \in  S_1^J} \frac{1}{n} \sum_{i=1}^n \log \sum_{j=1}^J w_j p\left(Y_i \mid  \mathscr{D}_{-i}, \mathscr{M}_j\right)\;,
\end{equation}
where $\mathscr{D}_{-i}$ are the data that exclude the $i$-th block (indexed by a row) of observations in $Y$, and $\mathscr{M}=\left(\mathscr{M}_1, \ldots, \mathscr{M}_J\right)$ are $J$ different models. For any given data set $\mathscr{D}$, these $J$ different models correspond to fixed spatial correlation kernel parameters in $V$. Solving \eqref{eq:OptProbBPSLogScore} minimizes the Kullback-Leibler divergence from the true predictive distribution by using convex optimization \citep{grant_disciplined_2005,cvx_research_cvx_2012}. 
Although the true predictive distribution is unknown, we target a leave-one-out (\textsc{loo}) estimate of the expected value of the score \citep[see. e.g.,][for details]{yao_using_2018} as the theoretical criterion for computing stacking weights. Since exact \textsc{loo} requires fitting the model $n$ times, as we exclude one row of data $\mathscr{D}$ at a time, we employ K-fold cross-validation as a more cost-effective and tractable method for generating predictions \citep{breiman_stacked_1996}.

%-------------------------------------------
\subsection{Accelerated Learning for Spatial Random Fields}\label{sec:AccelerMultivarMod}

Let $\mathcal{S}=\left\{ s_1, \ldots,  s_n\right\} \subset \mathcal{D}$ be a set of $n$ locations that produce observations on $q$ possibly correlated outcomes collected into a $q\times 1$ vector $y(s) = (y_1(s),\ldots,y_q(s))^{\T}$ for each $s\in \mathcal{S}$. We collect these measurements in the matrix $n\times q$ $Y=[y_j(s_i)^{\T}]$ for $i=1,\ldots,n$ and $j=1,\ldots,q$. Let $X = [x(s_i)^{\T}]$ be $n\times p$ with rows $x(s_i)^{\T}$ consisting of $p < n$ explanatory variables at location $s_i\in \mathcal{S}$; we assume that $X$ has rank $p$. We introduce latent spatial processes, $\omega_j(s)$, for each outcome $y_j(s)$ to capture spatial dependence and a $q\times q$ covariance matrix, $\Sigma$, to capture non-spatial dependence among the elements of $y(s)$ within $s$. This matrix is typically adjusted by a scale factor $\left(\alpha^{-1}-1\right)$ to accommodate additional variation on local scales. Thus, setting $\alpha=\sigma^2 /\left(\sigma^2+\tau^2\right)$ where $\sigma^2$ and $\tau^2$ denote variances for the spatial process and measurement error (``nugget'') implies $\alpha$ is the ratio of the spatial variance (partial sill) to the total variance (sill) gleaned from a variogram.  

We cast this into \eqref{eq:MatrixNormLik}, but we explicitly introduce a latent $q\times 1$ spatial process $\omega(s)$ as
\begin{equation}\label{eq:MatrixNormHier}
\begin{split}
Y &= X\beta + \Omega + E\;,\quad E \given \Sigma \sim \operatorname{MN}(O, (\alpha^{-1} - 1)\mathbb{I}_n, \Sigma)\;;\quad \Sigma \sim \operatorname{IW}(\Psi_{0},\nu_{0})\;;\\ 
\beta &= M_0m_0 + E_{\beta}\;,\quad E_{\beta} \given \Sigma \sim \operatorname{MN}(O, M_0, \Sigma)\;;\quad \Omega \given \Sigma \sim \operatorname{MN}(O, V, \Sigma)\;, 
\end{split}
\end{equation}
where $\Omega = [\omega(s_i)^{\T}]$ is $n\times q$ with rows $\omega(s_i)^{\T}$. To capture spatial dependence, $V$ is an $n\times n$ spatial correlation matrix with the $(i,j)$-th element equal to the value of a positive definite spatial correlation function $\rho(s_i,s_j;\phi)$ indexed by the parameter(s) $\phi$. We account for measurement errors in observations so that the spatial component of the variation in the elements of $Y$ is modeled using $\rho(s_i,s_j;\phi) + (\alpha^{-1} - 1)\mathds{1}_{s_i=s_j}$, where $\alpha \in [0,1]$ is the proportion of total variability attributed to the spatial process. 

Letting $\gamma^{\T}=\left[\beta^{\T}, \Omega^{\T}\right]$ be $q \times(p+n)$, we assume $\{\gamma, \Sigma\} \sim \operatorname{MNIW}\left(\mu_\gamma,  V_\gamma, \Psi_{0}, \nu_{0}\right)$, where
\begin{equation}\label{eq:MultivJoinPrior}
\operatorname{MNIW}\left(\gamma,\Sigma\mid \mu_{\gamma},  V_{\gamma}, \Psi_{0}, \nu_{0}\right) = \operatorname{IW}(\Sigma\mid\Psi_{0},\nu_{0}) \times \operatorname{MN}_{p+n,q}(\gamma\mid\mu_{\gamma},  V_{\gamma},\Sigma),
\end{equation}
with $\mu_\gamma^{\T}=\left[m_{0}^{\T}M_{0},  0_{q \times n}\right]$ and $ V_\gamma=$ blockdiag $\left\{ M_{0}, \rho_\phi(\mathcal{S}, \mathcal{S})\right\}$. The \textsc{mniw} prior is conjugate with respect to the matrix-normal likelihood. Thus, for any fixed $\{\alpha,\phi\}$ and hyperparameters in the prior density, we obtain an \textsc{mniw} posterior density for $\{\gamma, \Sigma\}$,
\begin{equation}\label{eq:MultivLatPost}
p(\gamma, \Sigma \mid \mathscr{D})  = \operatorname{MNIW}\left(\gamma, \Sigma \mid \mu_{\gamma}^{\star},  V^{\star}_{\gamma}, \Psi^{\star}, \nu^{\star}\right),
\end{equation}
where $V^{\star}_{\gamma}=\begin{bsmallmatrix}
    \frac{\alpha}{1-\alpha} X^{\T} X + M_{0}^{-1} & \frac{\alpha}{1-\alpha} X^{\T} \\ \frac{\alpha}{1-\alpha} X & \rho_\phi^{-1}(\mathcal{S}, \mathcal{S})+\frac{\alpha}{1-\alpha} \mathbb{I}_n \end{bsmallmatrix} ^{-1}$ and $\mu_\gamma^{\star}= V^{\star}_{\gamma}\begin{bsmallmatrix} \frac{\alpha}{1-\alpha} X^{\T} Y+ m_{0} \\ \frac{\alpha}{1-\alpha} Y \end{bsmallmatrix}$, $\Psi^{\star}=\Psi_{0}+\frac{\alpha}{1-\alpha} Y^{\T} Y+m_{0}^{\T}M_{0} m_{0}-\mu_\gamma^{* \T}  V^{\star -1}_{\gamma} \mu_\gamma^{\star}$ and $\nu^{\star}=\nu_{0}+n$.

The framework in \eqref{eq:MatrixNormHier} is equivalent to \eqref{eq:MatrixNormLik} with $V = R_{\phi} + (\alpha^{-1} - 1)\mathbb{I}_{n}$ and $R_{\phi} = [\rho(s_i,s_j;\phi)]$. We recover posterior samples of $\Omega$ by using each posterior draw of $\{\beta,\Sigma\}$ to draw a corresponding $\Omega$ from $p(\Omega \given \mathscr{D}, \beta, \Sigma, \mathscr{M}_j)$. This seamlessly renders itself to the Bayesian transfer learning framework described in Section~\ref{sec:DivConBays}, provided that $V$ or $\{\alpha,\phi\}$ is fixed. For \textsc{GeoAI}, we seek to minimize human intervention. Rather than fixing them at one particular value, perhaps gleaned from a spatial variogram that requires human inspection, we use a set of $J$ candidate values $\{\alpha_j,\phi_j\}$ specifying model $\mathscr{M}_j$ for $j=1,\ldots,J$. We now obtain analytical closed forms for $p(\beta,\Sigma \given \mathscr{D}, \mathscr{M}_j)$ for each $j$, as described in Section~\ref{sec:DivConBays}, and use Bayesian predictive stacking to evaluate the stacked posterior distribution.    

Turning to prediction, let $\mathcal{U}=\{ u_1,\dots, u_{n^{\prime}}\}$ be a finite set of locations where we seek to predict or impute the value of $Y$ based on an observed ${n^{\prime}}\times p$ design matrix $X_\mathcal{U}$ associated with the locations in $\mathcal{U}$. The joint posterior predictive for $Y_\mathcal{U}$ and the unobserved latent process $\Omega_{\mathcal{U}}=[\omega(u_i)^{\T}]$ for $i=1,\dots,n^{\prime}$ can be recast by integrating out $\{\gamma,\Sigma\}$ from the conditional posterior predictive distribution to yield
\begin{equation}\label{eq:MultivLatPred}
\begin{split}
    p(Y_{\mathcal{U}}, \Omega_{\mathcal{U}} \mid \mathscr{D}) &= \int \mathrm{MN}_{{n^{\prime}}, q}\big(Y_{\mathcal{U}} \mid X_{\mathcal{U}} \beta+\Omega_{\mathcal{U}},\left(\alpha^{-1}-1\right) \mathbb{I}_{{n^{\prime}}}, \Sigma\big) \\ &\times \operatorname{MN}_{{n^{\prime}}, q}\big(\Omega_{\mathcal{U}} \mid M_{\mathcal{U}} \Omega,  V_{\Omega_{\mathcal{U}}}, \Sigma\big) \times \operatorname{MNIW}\left(\gamma, \Sigma \mid \mu_{\gamma}^{\star} ,  V^{\star}_{\gamma}, \Psi^{\star}, \nu^{\star}\right) d \gamma d \Sigma,
\end{split}
\end{equation}
where $M_{\mathcal{U}}=\rho_\phi(\mathcal{U}, \mathcal{S}) \rho_\phi^{-1} (\mathcal{S}, \mathcal{S})$ and $ V_{\Omega_{\mathcal{U}}}=\rho_\phi(\mathcal{U}, \mathcal{U})-\rho_\phi(\mathcal{U}, \mathcal{S}) \rho_\phi^{-1}(\mathcal{S}, \mathcal{S}) \rho_\phi(\mathcal{S}, \mathcal{U})$. This is a matrix-variate %Student's 
$\operatorname{t}$ distribution $\operatorname{T}_{2n^{\prime},q}(\nu^{\star}, \mu^{\star}, V^{\star}, \Psi^{\star})$, with degrees of freedom $\nu^{\star}$, location matrix $\mu^{\star}=M\mu_{\gamma}^{\star}$, row-scale matrix $V^{\star}$, and column-scale matrix $\Psi^{\star}$, where $M = \begin{bsmallmatrix} 0 & \; M_{\mathcal{U}} \\ X_\mathcal{U} &  \; M_{\mathcal{U}} \end{bsmallmatrix}$ and $V^{\star} = MV^{\star}_{\gamma} M^\T + V_E$ with $V_E = \begin{bsmallmatrix} V_{\Omega_{\mathcal{U}}} & V_{\Omega_{\mathcal{U}}} \\ V_{\Omega_{\mathcal{U}}} & \;\;\; V_{\Omega_{\mathcal{U}}} + (\alpha^{-1}-1)\mathbb{I}_{n^{\prime}} \end{bsmallmatrix}$. See Section~\ref{sec:matrixT} for details. While the conditional posterior predictive distributions take the following forms
$p(\Omega_{\mathcal{U}} \mid \mathscr{D}, \gamma, \Sigma) = \operatorname{MN}_{{n^{\prime}},q}\left(\Omega_{\mathcal{U}} \mid M_{\mathcal{U}} \Omega,  V_{\Omega_{\mathcal{U}}}, \Sigma\right)$ and $
p(Y_\mathcal{U} \mid \mathscr{D}, \Omega_{\mathcal{U}}, \gamma, \Sigma) = \operatorname{MN}_{{n^{\prime}},q}(Y_\mathcal{U} \mid X_{\mathcal{U}} \beta+\Omega_{\mathcal{U}},\left(\alpha^{-1}-1\right) \mathbb{I}_{{n^{\prime}}}, \Sigma)$.
Hence, we can proceed with posterior predictive inference by sampling from the closed-form joint predictive distribution or sampling from the conditional distributions. We draw one instance of $\Omega_{\mathcal{U}} \sim p(\Omega_{\mathcal{U}} \mid \mathscr{D}, \gamma, \Sigma)$ for each posterior draw of $\{\gamma, \Sigma\}$ and then draw a matrix $Y_{\mathcal{U}}$ from $p(Y_\mathcal{U} \mid \mathscr{D}, \Omega_{\mathcal{U}}, \gamma, \Sigma)$ for each drawn $\left\{\Omega_{\mathcal{U}}, \gamma, \Sigma\right\}$. The resulting samples are exactly drawn from the posterior predictive distribution $p(Y_{\mathcal{U}}\given \mathscr{D})$.

% ALGO %%%%%%%%%%%%%%%%%%%%%%%%%%%%%%%%%%%%%%%%%
\begin{algorithm}[!t]
\caption{Computing stacking weights within subsets using \textsc{bps}}\label{alg:first_weights}
%\vspace{0.25em}
\textbf{Input:} $Y$ ($n\times q$ matrix of outcomes), $X$ ($n\times p$ design matrix), $\mathcal{S}$ (coordinates of $n$ locations); $\{m_{0}, M_{0},\Psi_{0}, \nu_{0}\}$: Prior parameters; $G_{\alpha}\times G_{\phi}$: Grids of $\{\alpha, \phi\}$; $n$ (no. of locations), $q$ (no. of outcomes), $p$ (no. of predictors); $K$ (no. of subsets), $J$ (no. of models), $L$ (no. of folds).\\
\textbf{Output:} $\hat{z} = \{\hat{z}_{k}=\{\hat{z}_{k,j}\}: k=1,\dots,K, j=1,\dots,J \}$: Stacking weights within subsets; $\{pd_{k,j,i}:k=1,\dots,K, j=1,\dots,J, i=1,\dots,n\}$:  point-wise predictive density of $Y$; $G_{all}$: Grid of dimension $J$, spanned by $G_{\alpha}, G_{\phi}$
\begin{algorithmic}[1]
\State Partition $Y,X,\mathcal{S}$ into $\mathscr{D}_k=\{Y_{k},X_{k},\mathcal{S}_{k}\}$, $k=1,\dots,K$
\State Store $n_{k}$, as cardinality of $\mathcal{S}_{k}$; Compute $G_{all}$ by expanding $G_{\alpha},G_{\phi}$
\ParFor{$k=1,\dots,K$}
\For{$j=1,\dots,J$}
\State Extract $\{\alpha_{j},\phi_{j}\}$ from $j$-th row of $G_{all}$
\State Form $L$ folds: $\mathscr{D}_{k,[l]}=\{Y_{k,[l]},X_{k,[l]},\mathcal{S}_{k,[l]}\}$ and $\mathscr{D}_{k,[-l]}=\{Y_{k,[-l]},X_{k,[-l]},\mathcal{S}_{k,[-l]}\}$
\State Store $n_{k,[-l]}$, as cardinality of $\mathcal{S}_{k,[-l]}$
\For{$l=1,\dots,L$}
\State Compute $R_{\phi_j}([-l])=\rho_{\phi_{j}}(\mathcal{S}_{k,[-l]},\mathcal{S}_{k,[-l]})$,  $R^{-1}_{\phi_j}([-l])$ and $M_{0}$ for $M_{0}^{-1}$
\State Construct $V^{\star -1}_{\gamma,[-l]}=${$\begin{bsmallmatrix}
    \frac{\alpha_{j}}{1-\alpha_{j}}X^{\T}_{k,[-l]}X_{k,[-l]}+M_{0}^{-1} & \frac{\alpha_{j}}{1-\alpha_{j}}X^{\T}_{k,[-l]} \\
    \frac{\alpha_{j}}{1-\alpha_{j}}X_{k,[-l]} & R^{-1}_{\phi_j}([-l]) + \frac{\alpha_{j}}{1-\alpha_{j}}\mathbb{I}_{n_{k,[-l]}}
\end{bsmallmatrix}$}
\State Solve for $\mu^{\star}_{\gamma,[-l]}$: $V^{\star -1}_{\gamma,[-l]}\mu^{\star}_{\gamma,[-l]} = \begin{bsmallmatrix}
    \frac{\alpha_{j}}{1-\alpha_{j}}X^{\T}_{k,[-l]}Y_{k,[-l]}+m_{0}\\
    \frac{\alpha_{j}}{1-\alpha_{j}}Y_{k,[-l]}
\end{bsmallmatrix}$
\State Calculate $\Psi^{\star}_{[-l]}=\Psi_{0} + (\alpha_{j}^{-1}-1)^{-1}Y^{\T}_{k,[-l]}Y_{k,[-l]} + m_{0}^{\T}M_{0}m_{0} - \mu^{\T\star}_{\gamma,[-l]}V^{\star -1}_{[-l]}\mu^{\star}_{\gamma,[-l]}$
\State Calculate $\nu^{\star}_{[-l]}=\nu_{0} + n_{k,[-l]}$
\For{$i\in[l]$}
\State Compute $R_{\phi_j}(i)=\rho_{\phi_{j}}(\mathcal{S}_{k,i},\mathcal{S}_{k,i})$ and $R_{\phi_j}(i,[-l])=\rho_{\phi_{j}}(\mathcal{S}_{k,i},\mathcal{S}_{k,[-l]})$
\State Calculate $M_{i}=R_{\phi_j}(i,[-l])R^{-1}_{\phi_j}([-l])$ and form $M_{y,i}=\begin{bmatrix} X_{k,i} & M_{i} \end{bmatrix}$
\State Calculate $\mu^{\star}_{i}=M_{y,i}\mu^{\star}_{\gamma,[-l]}$ and $V_{\Omega_i}=R_{\phi_j}(i)-M_{i}R_{\phi_j}([-l],i)$
\State Construct $V_{e,i}=V_{\Omega_i}+(\alpha_{j}^{-1}-1)$ and $V^{\star}_{i}=M_{y,i}V^{\star}_{\gamma,[-l]}M_{y,i}^{\T}+V_{e,i}$

\State Compute $pd_{k,j,i} = \operatorname{T}_{1,q}(Y_{k,i}\mid \nu^{\star}_{[-l]}, \mu^{\star}_{i}, V^{\star}_{i}, \Psi^{\star}_{[-l]})$.
\EndFor
\EndFor
\EndFor
\State Solve: $\max _{ z_k \in  S_1^J} \frac{1}{n_k} \sum_{i=1}^{n_k} \log \sum_{j=1}^J z_{k,j} pd_{k,j,i}$ such that $z_{k} \in[0,1]^J: \sum_{j=1}^J z_{k,j}=1$
\EndParFor
\State \textbf{return} $\{\hat{z}, \{pd_{k,j,i}\}, G_{all}\}$
\end{algorithmic}
\end{algorithm}

This tractability is possible if the range decay $\phi$ and $\alpha$ are fixed. While data can inform about these parameters, they are inconsistently estimable and lead to poorer convergence \citep{zhang_inconsistent_2004}. \citet{finley_efficient_2019} explored $K$-fold cross-validation, but inference is limited to only one set of values for the parameters. Instead, we pursue exact inference using \eqref{eq:MultivLatPost} and \eqref{eq:MultivLatPred} by stacking over different fixed values of $\{\alpha, \phi\}$ using \textsc{bps} of predictive densities as described in Section~\ref{sec:BPS}. This minimizes human intervention and enables automation.

For each subset of the data, we compute the stacking weights $z_{k}=\{z_{k,j}\}_{j=1,\ldots,J}$ as
\begin{equation}\label{eq:subsetOptProb}
   \max _{ z_k \in  S_1^J} \frac{1}{n_k} \sum_{i=1}^{n_k} \log \sum_{j=1}^J z_{k,j} p\left(Y_{k,i} \mid  \mathscr{D}_{k,[-l]}, \mathscr{M}_j \right),
\end{equation}
where $Y_{k,i}$ is the $i$-th of the $n_{k}$ rows of $Y_{k,[l]}\in\mathscr{D}_{k,[l]}$ (the $l$-th fold within the $k$-th dataset), and $\mathscr{D}_{k,[-l]}$ is the $k$-th dataset without the $l$-th fold, with $l=1,\dots,L$, and $L$ the number of folds for K-fold cross-validation estimates for the expected value of the score (see Section~\ref{sec:BPS}). Analogously, $X_{k,i}$ denotes the $i$-th row of $X_{k,[l]}$ and $\mathcal{S}_{k,i}$ contains the associated location, so that $\{Y_{k,i},X_{k,i},\mathcal{S}_{k,i}\}$ collects the $i$-th held-out observation in the $l$-th fold. The posterior predictive density, $p\left(Y_{k,i} \mid \mathscr{D}_{k,[-l]}, \mathscr{M}_j\right)$, is available in closed form as a matrix $t$ distribution, which makes the computation efficient. This leads us from \eqref{eq:subsetOptProb} to
\begin{equation}\label{eq:subsetOptMatrixT}
    \max _{ z_k \in  S_1^J} \frac{1}{n_k} \sum_{i=1}^{n_k} \log \sum_{j=1}^J z_{k,j} \operatorname{T}_{1,q}(Y_{k,i}\mid\nu^{\star}_{[-l]}, \mu^{\star}_{i}, V^{\star}_{i}, \Psi^{\star}_{[-l]}),
\end{equation}
where $\nu^{\star}_{[-l]}=\nu_{0} + n_{k,[-l]}$, $n_{k,[-l]}$ is the cardinality of $\mathcal{S}_{k,[-l]}$ (which is the set of locations in $\mathscr{D}_{k,[-l]}$), and $\Psi^{\star}_{[-l]}=\Psi_{0} + (\alpha_{j}^{-1}-1)^{-1}Y^{\T}_{k,[-l]}Y_{k,[-l]} + m_{0}^{\T}M_{0}m_{0} - \mu^{\T\star}_{\gamma,[-l]}V^{\star -1}_{\gamma,[-l]}\mu^{\star}_{\gamma,[-l]}$. The quantities $V^{\star}_{i}=M_{y,i}V_{\gamma}^{\star}M_{y,i}^{\T}+V_{\Omega_i}+(\alpha_{j}^{-1}-1)$ and $\mu^{\star}_{i}=M_{y, i}\mu^{\star}_{\gamma,[-l]}$ are computed from the following auxiliary quantities: $V^{\star -1}_{\gamma,[-l]}=\begin{bsmallmatrix}
    \frac{\alpha_{j}}{1-\alpha_{j}}X^{\T}_{k,[-l]}X_{k,[-l]}+M_{0}^{-1} & \frac{\alpha_{j}}{1-\alpha_{j}}X^{\T}_{k,[-l]} \\
    \frac{\alpha_{j}}{1-\alpha_{j}}X_{k,[-l]} & \rho^{-1}_{\phi_{j}}(\mathcal{S}_{k,[-l]},\mathcal{S}_{k,[-l]}) + \frac{\alpha_{j}}{1-\alpha_{j}}\mathbb{I}_{n_{k,[-l]}}
\end{bsmallmatrix}$, $\mu^{\star}_{\gamma,[-l]}=V^{\star}_{\gamma,[-l]}\begin{bsmallmatrix}
    \frac{\alpha_{j}}{1-\alpha_{j}}X^{\T}_{k,[-l]}Y_{k,[-l]}+m_{0}\\
    \frac{\alpha_{j}}{1-\alpha_{j}}Y_{k,[-l]} 
\end{bsmallmatrix}$, $M_{y,i}=\begin{bsmallmatrix} X_{k,i} & \quad \rho_{\phi_{j}}(\mathcal{S}_{k,i},\mathcal{S}_{k,[-l]})\rho^{-1}_{\phi_{j}}(\mathcal{S}_{k,[-l]},\mathcal{S}_{k,[-l]}) \end{bsmallmatrix}$, and $V_{\Omega_i}=\rho_{\phi_{j}}(\mathcal{S}_{k,i},\mathcal{S}_{k,i})-\rho_{\phi_{j}}(\mathcal{S}_{k,i},\mathcal{S}_{k,[-l]})\rho^{-1}_{\phi_{j}}(\mathcal{S}_{k,[-l]},\mathcal{S}_{k,[-l]})\rho_{\phi_{j}}(\mathcal{S}_{k,[-l]},\mathcal{S}_{k,i})$. Note that $\nu^{\star}_{[-l]}$ is scalar, $\mu^{\star}_{i}$ is $(1\times q)$ row vector, $V^{\star}_{i}$ is a scalar, and $\Psi^{\star}_{[-l]}$ is $(q\times q)$ matrix, so that \eqref{eq:subsetOptMatrixT} evaluates the density of a $(1\times q)$ matrix-variate $t$ distribution. 
Further details, including derivations and implementation, are supplied in Section~\ref{sec:matrixT} and Algorithm~\ref{alg:first_weights}.

% ALGO %%%%%%%%%%%%%%%%%%%%%%%%%%%%%%%%%%%%%%%%%
\begin{algorithm}[!t]
\caption{Calculating stacking weights between subsets using \textsc{bps}}\label{alg:second_weights}

\vspace{0.25em}
\textbf{Input:} $\hat{z} = \{\hat{z}_{k}=\{\hat{z}_{k,j}\}: k =1,\dots,K, j=1,\dots,J \}$: Stacking weights within subsets; $\{pd_{k,j,i}:k=1,\dots,K, j=1,\dots,J, i=1,\dots,n\}$:  point-wise predictive density of $Y$; $n, q, p$: Number of rows, number of outcomes, and number of predictors; $K, \{n_{k}:k=1,\dots,K\}, J$: Number of subsets, dimension of each subset, and number of competitive models in each subset.\\
\textbf{Output:} $\hat{w} = \{\hat{w}_{k}: k =1,\dots,K\}$: Stacking weights between subsets.

\begin{algorithmic}[1]
\State Construct $pd=[pd_{1}^{\T}:\cdots:pd_{K}^{\T}]^{\T}$ of dimension $(n\times J)$
\NoNumber{where $pd_{k}=\begin{bsmallmatrix} 
pd_{k,1,1} & \dots & pd_{k,J,1} \\
\vdots & pd_{k,j,i} & \vdots \\
pd_{k,1,n_{k}} & \dots & pd_{k,J,n_{k}} \\    
\end{bsmallmatrix}$ of dimension $(n_{k}\times J)$}

\For{$k=1,\dots,K$}
\State Compute $epd_{k} = pd\;\hat{z}_{k}$ of dimension $(n\times 1)$
\EndFor

\State Solve convex optimization problem:
\NoNumber{$\max _{ w \in  S_1^K} \frac{1}{n} \sum_{i=1}^{n} \log \sum_{k=1}^K w_{k} epd_{k,i} = \max _{ w \in  S_1^K} \frac{1}{n} \sum_{i=1}^{n} \log \sum_{k=1}^K w_{k} \sum_{j=1}^J \hat{z}_{k,j} pd_{k,j,i}$}
\NoNumber{where $pd_{k,j,i}=p\left(Y_{k,i} \mid  \mathscr{D}_{k,[-l]}, \mathscr{M}_j \right)=\operatorname{T}_{1,q}(Y_{k,i}\mid\nu^{\star}_{[-l]}, \mu^{\star}_{i}, V^{\star}_{i}, \Psi^{\star}_{[-l]})$} 
\NoNumber{for $\forall i\notin [-l],l\in\{1,\dots,L\}$ and $S_{1}^{K}=\{ w \in[0,1]^K: \sum_{k=1}^K w_{k}=1\}$}

\State \textbf{return} $\hat{w}=\{\hat{w}_{k}:k=1,\dots,K\}$
\end{algorithmic}
\end{algorithm}

For each dataset $\mathscr{D}_k$, \textsc{bps} computes: (i) an estimate of the posterior predictive $\hat{p}(\,\cdot \;;  \mathscr{D}_k) =\sum_{j=1}^J \hat{z}_{k,j} \; p\left(\,\cdot \mid  \mathscr{D}_k, \mathscr{M}_j\right)$ for $ k=1,\dots,K$ and $j=1,\dots,J$; and (ii) a set of stacking weights $\hat{z}_k=\{\hat{z}_{k,j}\}_{j=1,\dots,J}$.%, following Algorithm~\ref{alg:first_weights}.
Once these weights are available, we apply \textsc{bps} a second time to obtain a weighted average of  $\hat{p}(\,\cdot \;;  \mathscr{D}_k)$ over the $k$ subsets. This \textsc{double bps} (\textsc{dbps}) of predictive densities seeks weights $w = \{w_k\}_{k=1,\ldots,K}$ such that $\hat{w} = \max _{ w \in  S_1^K} \frac{1}{n} \sum_{i=1}^{n} \log \sum_{k=1}^K w_{k} \hat{p}\left(Y_{i} \;;  \mathscr{D}_{k}\right)$ (see Section~\ref{sec:BPSobjfun}).
Once stacking weights $ \hat{w}=\{\hat{w}_k\}_{k=1,\dots,K}$ are computed using Algorithm~\ref{alg:second_weights}, sampling from the posterior and posterior predictive distributions is obtained from
\begin{equation}\label{eq:doubleBPSpred}
    \hat{p}(\,\cdot \;;  \mathscr{D}) 
    =\sum_{k=1}^{K} \hat{w}_k \; \sum_{j=1}^J \hat{z}_{k,j} \; p\left(\,\cdot \mid  \mathscr{D}_k, \mathscr{M}_j\right)\;.
\end{equation}
Only posterior predictive distributions are considered to acquire the two sets of stacking weights.
Given the weights obtained from double stacking, the stacked posterior distribution is a mixture of finite mixtures. This makes sampling from \eqref{eq:doubleBPSpred} straightforward.

First, the set of stacking weights $\hat{ z}_k=\{\hat{z}_{k,j}\}_{j=1,\dots,J}$ obtained using \textsc{bps} within the subset of the data $\mathscr{D}_k$ is primarily used to approximate the subset posterior distribution $\hat{p}(\Theta \;;  \mathscr{D}_{k})=\sum_{j=1}^J \hat{z}_{k,j} p\left(\Theta \mid  \mathscr{D}_k, \mathscr{M}_j\right) $ for each subset $k=1,\dots,K $. By considering the second set of stacking weights $\hat{w}=\{\hat{w_k}\}_{k=1,\dots,K}$, the stacked full posterior distribution is $\hat{p}(\Theta \;;  \mathscr{D}) = \sum_{k=1}^K \hat{w}_k \; \hat{p}(\Theta \;;  \mathscr{D}_k) = \sum_{k=1}^K \hat{w}_k \; \sum_{j=1}^J \hat{z}_{k,j} \; p(\Theta \mid  \mathscr{D}_k, \mathscr{M}_j)$, where $\Theta=\{\gamma, \Sigma\}$. This is a substantial simplification over meta-kriging \citep{guhaniyogi_meta-kriging_2018, guhaniyogi_multivariate_2019} that does not require any empirical approximation of posterior or predictive distributions. Again, the predictive random variable $Y_{\mathcal{U}}$ is recovered from \eqref{eq:doubleBPSpred} as
\begin{equation}\label{eq:fullPredictiveYU}
    \hat{p}( Y_{\mathcal{U}} \;; \mathscr{D}) =\sum_{k=1}^{K} \hat{w}_k \; \sum_{j=1}^J \hat{z}_{k,j} \; p\left( Y_{\mathcal{U}} \mid  \mathscr{D}_k, \mathscr{M}_j\right).
\end{equation}
Each component $p\left( Y_{\mathcal{U}} \mid  \mathscr{D}_k, \mathscr{M}_j\right)$ is a matrix-variate $t$ distribution $\operatorname{T}_{n^{\prime},q}(\nu_{k}^{\star},\mu_{k,j}^{\star}, V_{k,j}^{\star}, \Psi_{k,j}^{\star})$. The mixture structure can inflate predictive covariances; disagreement tempering, described in Section~\ref{sec:disagree}, mitigates this effect. The latent spatial process %for $\Omega$ 
is learned by %estimating $\Omega_{\mathcal{U}}$ using 
sampling from
$\hat{p}(\Omega_{\mathcal{U}} \;;  \mathscr{D}) =\sum_{k=1}^{K} \hat{w}_k \;\hat{p}(\Omega_{\mathcal{U}} \;;  \mathscr{D}_k)$, where $\hat{p}(\Omega_{\mathcal{U}} \;;  \mathscr{D}_k)=\sum_{j=1}^J \hat{z}_{k,j} \; p\left(\Omega_{\mathcal{U}} \mid  \mathscr{D}_k, \mathscr{M}_j\right)$.

%%%%%%%%%%%%%%%%%%%%%%%%%%%%%%%%%%%%%%%%%%%%
\section{Approximation Insights}\label{sec:theory_dev}
 
We study the behavior of the \textsc{dbps} posterior predictive distribution by characterizing and bounding its divergence from the true posterior predictive distribution using the Kullback-Leibler (\textsc{kl}) divergence. We then introduce disagreement tempering as a principled approach to mitigate predictive variance inflation and oversmoothing that commonly arise in distributed settings. Together, these results establish approximation guaranties for the \textsc{dbps} posterior predictive and show that, while distributed settings may exhibit predictive variance inflation, disagreement tempering mitigates this over-dispersion.

%-------------------------------------------
\subsection{Kullback-Leibler Divergence from True Posterior Predictive}\label{sec:upperbound}

We investigate the behavior of \textsc{double bps} approximations to the posterior predictive distributions when the number of competitive models ($J$) and the number of partitions ($K$) grow in size. The reversed Kullback-Leibler divergence between the \textsc{double bps} posterior predictive in \eqref{eq:fullPredictiveYU} and the true predictive distribution $p_{t}(\;\cdot\mid\mathscr{D})$ in \eqref{eq:MultivLatPred}, where $\{\alpha,\phi\}$ are as in the data-generating process, offers analytical tractability. 
Writing the true predictive distribution as $p_{t} (y\mid\mathscr{D}) = p(y\mid\mathscr{D},\alpha_{t},\phi_{t}) = \operatorname{T}_{n,q}(y; \nu_{t},M_{t},V_{t},\Psi_{t})$ and the \textsc{double bps} approximation as $\hat{p}(y\;;\mathscr{D}) = \sum_{k=1}^{K}\hat{w}_{k}\sum_{j=1}^{J}\hat{z}_{k,j} \; \operatorname{T}_{n,q}(y\mid\nu_{k,j},M_{k,j},V_{k,j},\Psi_{k,j})$, and denoting $\hat{P}$ and $P_{t}$ as the probability distributions corresponding to the \textsc{double bps} approximation and the true predictive probability distributions, respectively, the reverse \textsc{kl} divergence is
\begin{equation*}
\begin{split}
    D_{\textsc{kl}}(\, \hat{P} \parallel P_{t}\,) %= \int_{y} \log\frac{\hat{P}(\dd y)}{P_{t}(\dd y)} \hat{P}(\dd y)
    &= \int_{y\in\mathbb{R}^{n\times q}} \sum_{k=1}^{K}\hat{w}_{k} \hat{p}_{k}(y\;; \mathscr{D}_{k}) \log\frac{\sum_{k=1}^{K}\hat{w}_{k} \hat{p}_{k}(y\;; \mathscr{D}_{k})}{p_{t}(y\mid \mathscr{D})}\dd y \\
    &= \sum_{k=1}^{K}\hat{w}_{k} \mathbb{E}_{\hat{p}_{k}}\Big[ \log\sum_{k=1}^{K}\hat{w}_{k} \hat{p}_{k}(y\;;\mathscr{D}_{k}) \Big] - \sum_{k=1}^{K}\hat{w}_{k} \mathbb{E}_{\hat{p}_{k}}\left[ \log p_{t}(y\mid\mathscr{D}) \right] \\
    &= \sum_{k=1}^{K}\hat{w}_{k} \mathbb{E}_{\hat{p}_{k}}\Bigg[\log\frac{\sum_{k=1}^{K}\hat{w}_{k} \hat{p}_{k}(y\;;\mathscr{D}_{k})}{p_{t}(y\mid\mathscr{D})}\Bigg] \leq \sum_{k=1}^{K}\hat{w}_{k} \log \mathbb{E}_{\hat{p}_{k}}\Bigg[\frac{\sum_{k=1}^{K}\hat{w}_{k} \hat{p}_{k}(y\;;\mathscr{D}_{k})}{p_{t}(y\mid\mathscr{D})}\Bigg],
\end{split}
\end{equation*}
where $\hat{p}_{k}(y\;;\mathscr{D}_{k})=\sum_{j=1}^{J}\hat{z}_{k,j} \; \operatorname{T}_{n,q}(y\mid\nu_{k,j},M_{k,j},V_{k,j},\Psi_{k,j})$ and the inequality follows from Jensen's inequality.
% Expanding algebraically, this term, we get the more convenient formulation in Equation~\eqref{eq:upperbound}.
Some further algebraic simplification yields
\begin{equation}\label{eq:upperbound}
    D_{KL}\big(\,\hat{P} \,\big|\!\big|\, P_{t} \,\big) \leq \operatorname{log}\prod_{k=1}^{K}\Bigg\{\sum_{k=1}^{K}\hat{w}_{k}\sum_{j=1}^{J}\hat{z}_{k,j} \;\mathbb{E}_{p_{k,j}}\Bigg[\frac{\sum_{j=1}^{J}\hat{z}_{k,j}\; p(y\mid\mathscr{D}_{k},\mathscr{M}_{j})}{p_{t}(y\mid\mathscr{D})}\Bigg]\Bigg\}^{\hat{w}_{k}}. %\;,\quad y\sim\hat{p}(y\mid\mathscr{D}_{k},\mathscr{M}_{j})
\end{equation}
The empirical behavior for the upper bound is studied using Monte Carlo experiments in Section~\ref{sec:upperbound_sim} of the Appendix. 

% ALGO %%%%%%%%%%%%%%%%%%%%%%%%%%%%%%%%%%%%%%%%%
\begin{algorithm}[!t]
\caption{Disagreement tempering of \textsc{dbps} predictions}\label{alg:disagreement}
\vspace{0.25em}
\textbf{Input:} $\hat{z} = \{\hat{z}_{k}=\{\hat{z}_{k,j}\}: k \in \{1,\dots,K\}, j \in\{1,\dots,J\} \}$: Stacking weights within subsets; $\hat{w} = \{\hat{w}_{k}: k =1,\dots,K\}$: stacking weights between subsets; $R$: Number of predictive samples.\\
\textbf{Output:} $\big\{Y_{\mathcal{U}}^{DT (r)}: r=1,\ldots,R\big\}$: Disagreement-tempered predictive sample.
\begin{algorithmic}[1]
\For{$r=1,\dots,R$}
\State Sample $k^{(r)}\sim\operatorname{Multinom}(1,\hat{w})$
\State Sample $j^{(r)}\sim\operatorname{Multinom}(1,\hat{z}_{k^{(r)}})$
\State Sample $Y_{\mathcal{U}}^{(r)}\sim p(Y_{\mathcal{U}}\mid \mathscr{D}_{k^{(r)}},\mathscr{M}_{j^{(r)}})$
\State Compute $\mu_{k^{(r)},j^{(r)}}^{\star}=\mathbb{E}[Y_{\mathcal{U}}\mid \mathscr{D}_{k^{(r)}},\mathscr{M}_{j^{(r)}}]$
\EndFor
\State Compute centered global mean $\mu^{\star}_{C}=\frac{1}{R}\sum_{r=1}^{R}\mu_{k^{(r)},j^{(r)}}^{\star}$
\For{$r=1,\dots,R$}
\State Compute $Y_{\mathcal{U}}^{DT (r)}= Y_{\mathcal{U}}^{(r)} - \mu_{k^{(r)},j^{(r)}}^{\star} + \mu^{\star}_{C}$
\EndFor
\State \textbf{return} $\big\{Y_{\mathcal{U}}^{DT (r)}: r=1,\ldots,R\big\}$
\end{algorithmic}
\end{algorithm}

%-------------------------------------------
\subsection{Disagreement Tempering}\label{sec:disagree}

Disagreement tempering (\textsc{dt}) aims to mitigate posterior predictive variability and predictive interval width when posterior predictive distributions are mixtures. A known issue with linear mixtures of predictive distributions is variance inflation due to a positive quantity called model disagreement. When component densities or forecasters differ in their means, the variance of the ensemble can be substantially larger than the variance of each model. Forecast combination theory shows that a linear pool tends to produce inflated credible intervals that may be too wide for practical purposes. In fact, the extra disagreement term in predictive variance can cause overdispersion, which has been studied in the literature on ``linear pooling'' \citep{knuppel_forecast_2022}. In double Bayesian predictive stacking, the predictive variance is
\begin{equation*}
    \operatorname{Var}(\tilde{Y}_{\mathcal{U}}\;;\;\mathscr{D})=\sum_{k=1}^{K}w_{k}\sum_{j=1}^{J}z_{k,j}\operatorname{Var}(\tilde{Y}_{\mathcal{U}}\mid \mathscr{D}_{k}, \mathscr{M}_{j}) + \operatorname{Dis}\left(\{m_{k,j}\}_{k,j}\right),
\end{equation*}
where $\tilde{Y}_{\mathcal{U}}=\operatorname{vec}(Y_{\mathcal{U}})$, $m_{k,j}=\operatorname{vec}(\mu^{\star}_{k,j})=\operatorname{vec}(\mathbb{E}[Y_{\mathcal{U}}\mid \mathscr{D}_{k}, \mathscr{M}_{j}])$. The disagreement term $\operatorname{Dis}\left(\{m_{k,j}\}_{k,j}\right) \succeq 0$ is derived in Section~\ref{sec:disagree_derivation}. A simple solution to mitigate the effect of $\operatorname{Dis}\left(\{m_{k,j}\}_{k,j}\right)$ is centering, or ``tempering'', the disagreement: subtract each model's predictive mean so that all component distributions share a common center; form the mixture; and re-add a global mean if desired. In ensemble forecasting, this is known as the centered linear pool \citep{knuppel_forecast_2022}. This procedure ensures that the disagreement term disappears, reducing the variability inflation and the width of the predictive intervals. Larger differences between means imply higher values of $\operatorname{Dis}\left(\{m_{k,j}\}_{k,j}\right)$ and therefore a greater impact of tempering disagreement (see Section~\ref{sec:disagree_derivation} for details).

In practice, we can apply a \textsc{dt} step without modifying the weights: we fit the stacking weights $w_k$ as usual, but when making the final predictions, we first recenter each distribution (by subtracting the posterior predictive mean of the model) and then combine them; optionally, we add the ensemble mean (see Algorithm~\ref{alg:disagreement}). This does not alter score-based weight optimization, but it can reduce the inflated uncertainty of the ensemble forecast. Hence, assimilation of Bayesian predictive distributions can suffer from extra variance due to model disagreement, and simple centering of predictions can mitigate this effect.

%-------------------------------------------
\subsection{Alternative Approximate Methods}\label{sec: alt_approx_methods}
Our approach is based on an exact model with analytically accessible posterior distributions. Alternative approaches would include approximate methods. These can be broadly classified as methods that (i) approximate the true posterior distribution; and (ii) build scalable Gaussian processes. Some distinctions are worth noting. Approximate posterior inference foregoes sampling from the exact posterior distribution \citep[as is done in MCMC; see, e.g.,][]{robertCasella2004monteCarloBook} and instead devises a faster algorithm to compute an approximate posterior (\citealp[such as INLA,][]{rue_approximate_2009}; \citealp[and variational Bayes,][]{blei2017vbreview}). However, these approximate methods do not necessarily scale to massive datasets in terms of floating-point operations or storage complexity. Hence, their only possible benefit in the context of divide and conquer is to use them for each subset of the data to achieve faster convergence than MCMC. Scalable Gaussian processes, on the other hand, can be implemented with substantially lower storage and computational complexity and avoid the need for dividing and conquering massive data sets. While a comprehensive comparison with these alternative approaches is beyond the scope of a single article, we briefly discuss variational Bayes and nearest neighbor Gaussian processes in the context of our model.  

Variational Bayes \citep[see, e.g.,][for an excellent review from a statistical perspective]{blei2017vbreview} is prominent in machine learning for seeking an optimal approximation to intractable posteriors. Given that our posteriors are available in closed form, there is no apparent benefit to using variational Bayes or, for that matter, any iterative algorithm that needs to converge to an optimal approximation. Furthermore, the mean field variational approximation induces biased inference, which can be explicitly quantified for the \textsc{mniw} model. We briefly discuss this in Section~\ref{sec: vb}. Nearest neighbor Gaussian processes \citep[introduced by][]{datta_hierarchical_2016} scale statistical learning to massive datasets by building likelihood approximations \citep{vecchia_estimation_1988} based on a sparse directed acyclic graphical model of the spatial topology. This results in a sparse precision matrix that enables fast computation. We briefly discuss this in Section~\ref{sec: scalable_gp}.

\subsubsection{Variational Inference}\label{sec: vb}
One advantage of the conjugate posterior distributions in our model is that we do not worry about biases resulting from approximate algorithms such as variational Bayes. It is well-known that mean-field variational approximations underestimate uncertainty in the posterior distribution \citep[see, e.g., Section~3 in][for an exposition using the Normal-Gamma Bayesian regression]{ren_variational_2011}. Here, we offer a brief overview of this bias in a mean-field approximation of the posterior distribution for $\theta = \{\beta,\Sigma\}$ derived from \eqref{eq:MatrixNormLik}.

Following \citet{blei2017vbreview}, we express our marginal distribution of $Y$ as
\begin{equation*}
    \label{eq:vb_basic}
    \log p(Y) = \log\left(\frac{p(Y,\theta)}{p(\theta\given Y)}\right) = \log p(Y,\theta) - \log p(\theta\given Y) = \log \left(\frac{ p(Y,\theta)}{q(\theta)}\right) + \log \left(\frac{q(\theta)}{p(\theta\given Y)}\right)\;. 
\end{equation*}
Let $q(\theta)$ be any valid probability density for $\theta$. Multiplying both sides by $q(\theta)$ and integrating with respect to $\theta$ and noting that $\log p(Y) = \int q(\theta) \log p(Y) \,d\theta$, we obtain
\begin{eqnarray*}
\log p(Y) %&=& \int q(\theta) \log p(Y) \,d\theta
  &=& \int q(\theta) \log\left(\frac{ p(Y,\theta)}{q(\theta)}\right) \,d\theta + \int q(\theta) \log\left(\frac{q(\theta)} {p(\theta\given Y)}\right) \,d\theta \\
    &=& \text{\textsc{elbo}}(q) + KL(q, p(\cdot\given Y)) \geq  \text{\textsc{elbo}}(q)  \; ,
\end{eqnarray*}
where $\mbox{\textsc{elbo}}(q) = \Exp[\log p(Y, \theta)] -  \Exp[\log q(\theta)]$ is the \emph{evidence lower bound}\index{evidence lower bound (ELBO)} with $\Exp[\cdot]$ denoting expectation with respect to $q$, and $KL(q, p(\cdot\given Y))$ is the Kullback-Leibler (KL) divergence between $q(\theta)$ to $p(\theta\given Y)$. The divergence $KL(q, p(\cdot\given Y))$ is always nonnegative, and $\log p(Y)$ is a constant for any given data $Y$, so the sum of $\mbox{\textsc{elbo}}(q)$ and $KL(q,p)$ is a constant. Hence, $\mbox{\textsc{elbo}}(q)$ acts as a lower bound for $\log p(Y)$. Given data $Y$, we can obtain the variational approximation of $p(\theta\given Y)$ by either maximizing $\mbox{\textsc{elbo}}(q)$ or minimizing $KL(q,p)$.

Focusing on maximizing $\mbox{\textsc{elbo}}(q)$, the mean-field approximation seeks an optimal $q$ within the class $q(\theta) = q_{\beta}(\beta)q_{\Sigma}(\Sigma)$. Therefore, to derive the optimal $q_{\beta}(\beta)$ we write
\begin{equation}
    \label{eq: elbo_coordinate_wise}
    \begin{split}
        \mbox{\textsc{elbo}}(q) &= \E_q[\log p(Y, \beta,\Sigma)] - \E_q[\log q(\beta) + \log(\Sigma)] \\
        &= \E_{q_\beta}\left\{ \E_{q_{\Sigma}}[\log p(Y, \beta,\Sigma)] - \E_{q_{\Sigma}}[\log q(\beta) + \log q(\Sigma)] \right\}\\
        &= \E_{q_\beta}\left\{ \E_{q_{\Sigma}}[\log p(Y, \beta,\Sigma)] - \log q(\beta)\right\} + \mbox{const} := \mbox{\textsc{elbo}}(q_{\beta})\;,
    \end{split}
\end{equation}
where ``$\mbox{const}$'' absorbs quantities free of $\beta$. Equation~\eqref{eq: elbo_coordinate_wise} reveals that $\mbox{\textsc{elbo}}(q_{\beta})$ is the negative Kullback-Leibler divergence between $q_{\beta}(\beta)$ and a density given by $\log q_{\beta}^*(\beta) = \Exp_{\Sigma}[\log p(Y,\beta, \Sigma)] + \mbox{constant}$. Therefore, we maximize $\mbox{\textsc{elbo}}(q_{\beta})$ when we set $q_{\beta}(\beta)$ as 
\begin{equation} \label{eq: vb_beta}
  q_{\beta}^*(\beta) %= \frac{\exp \left\{\Exp_{\Sigma}[\log p (Y,\beta,\Sigma)]\right\}}{\int \exp\left\{\Exp_{\Sigma}[\log p (Y,\beta, \Sigma)]\right\} \,d\Sigma}
 \propto \exp \left\{\E_{q_\Sigma}[\log p (Y,\beta,\Sigma)]\right\}
 \propto \exp \left\{\E_{q_\Sigma}[\log p (\beta \given Y, \Sigma)]\right\} \; ,
\end{equation}
where the last expression is a simple consequence of the fact that the full conditional distribution $p (\beta \given \Sigma, Y)$ is proportional to the joint distribution $p (Y,\beta, \Sigma)$. Analogously,
\begin{equation} \label{eq: vb_sigma}
 q_{\Sigma}^*(\Sigma)
 \propto \exp \left\{\E_{q_\beta}[\log p (Y,\beta,\Sigma)]\right\}
 \propto \exp \left\{\E_{q_\beta}[\log p (\Sigma \given Y, \beta)]\right\} \; 
\end{equation}
is the optimal mean-field approximation for the marginal posterior distribution of $\Sigma$. This draws a resemblance to the Gibbs sampler. While the Gibbs sampler draws samples from the full conditionals, the mean-field variational algorithm computes an optimal approximation based on the full conditional distribution.

The variational approximations in \eqref{eq: vb_beta}~and~\eqref{eq: vb_sigma} are available explicitly. Focusing on the Bayesian updating described below \eqref{eq:MatrixNormLik} in Section~\ref{sec:DivConBays} from the prior to a single dataset ${\cal D}_1 = \{Y, X\}$ with fixed row-covariance $V$, we easily see that $p(\beta \given Y, \Sigma) = \operatorname{MN}(\beta \given Mm, M, \Sigma)$, where $M^{-1} = M_0^{-1} + X^{\T}V^{-1}X$ and $m = m_0 + X^{\T}V^{-1}Y$. Therefore,
\begin{equation}
    \label{eq: vb_full_conditionals_beta}
    \log q^{\ast}_{\beta}(\beta) = \mbox{const} - \frac{1}{2}\mbox{tr}\left\{\zeta^{-1}\left(\beta - Mm\right)M^{-1}(\beta - Mm)\right\} = \log \operatorname{MN}(\beta \given Mm, M, \zeta)\;,
\end{equation}
where $\zeta^{-1} = \E_{q_{\Sigma}}[\Sigma^{-1}]$. %Hence, $q^{\ast}_{\beta}(\beta) = MN(\beta \given Mm, M, \zeta)$. 
Analogous calculations using \eqref{eq: vb_sigma} reveal that 
\begin{equation}
    \label{eq: vb_full_conditionals_sigma}
    \log q^{\ast}_{\Sigma}(\Sigma) = \mbox{const} - \frac{\nu_0 + q + p + n + 1}{2}\log \det(\Sigma) - \frac{1}{2}\mbox{tr}\left(\tilde{\Psi}\Sigma^{-1}\right) = \log \operatorname{IW}(\Sigma \given \tilde{\Psi}, \tilde{\nu})\;,
\end{equation}
where $\tilde{\nu} = \nu_0 + p + n$ and $\tilde{\Psi} = \Psi_0 + \E_{q_{\beta}}[S_\beta(\beta) + S_{Y}(\beta)]$, with $S_{\beta}(\beta) = (\beta - M_0m_0)^{\T}M_0(\beta-M_0m_0)$ and $S_{Y}(\beta) = (Y-X\beta)^{\T}V^{-1}(Y-X\beta)$ being the prior scatter and likelihood scatter, respectively. The expression for $\tilde{\Psi}$ simplifies, using elementary matrix algebra, to
\begin{equation}
    \label{eq: vb_full_conditionals_sigma_simplified_scale}
    \begin{split}
        \tilde{\Psi} &= \Psi_0 + m_0^{\T}M_0m_0 + Y^{\T}V^{-1}Y - m^{\T}Mm + \E_{q_{\beta}}\left[(\beta - Mm)^{\T}M^{-1}(\beta-Mm)\right] %\\
        %&= \Psi_0 + m_0^{\T}M_0m_0 + Y^{\T}V^{-1}Y - m^{\T}Mm + p\zeta
        \;.
    \end{split}
\end{equation}

The variational mean-field algorithm iterates between \eqref{eq: vb_full_conditionals_beta} and \eqref{eq: vb_full_conditionals_sigma} as follows. Given the variational approximations $q^{(t)}_{\beta}(\beta)$ and $q^{(t)}_{\sigma}(\Sigma)$ at iteration $t$, we compute
\begin{equation*}
    \label{eq: vb_iterations}
    q^{(t+1)}_{\beta}(\beta) = \operatorname{MN}\left(\beta \given Mm, M, \zeta^{(t)}\right)\quad \mbox{and}\quad q^{(t+1)}_{\Sigma}(\Sigma) = \operatorname{IW}\left(\Sigma \given \tilde{\Psi}^{(t)}, \tilde{\nu}\right)
\end{equation*}
where $\zeta^{(t)} = \left(\E_{q^{(t)}_{\Sigma}}[\Sigma^{-1}]\right)^{-1}$ and $\tilde{\Psi}^{(t)} = \Psi_0 + m_0^{\T}M_0m_0 + Y^{\T}V^{-1}Y - m^{\T}Mm + p\zeta^{(t)}$. The $\zeta^{(t)}$ appearing as the last term in the expression for $\tilde{\Psi}^{(t)}$ is obtained by simplifying the expectation in the last term of \eqref{eq: vb_full_conditionals_sigma_simplified_scale} with respect to $q^{(t+1)}_{\beta}(\beta)$. The sequence $\{\zeta^{(t)}\}$ is updated using recursion,
\begin{equation}
    \label{eq: vb_recursive_update_zeta}
    \zeta^{(t+1)} = \left(\E_{q^{(t+1)}_{\Sigma}}[\Sigma^{-1}]\right)^{-1} = \frac{\tilde{\Psi}^{(t)}}{\tilde{\nu}} = \frac{\Psi^{\ast} + p\zeta^{(t)}}{\tilde{\nu}}\;, %\;.
\end{equation}
where $\Psi^{\ast} = \Psi_0 + m_0^{\T}M_0m_0 + Y^{\T}V^{-1}Y - m^{\T}Mm$ does not depend on $t$. 

Denoting $\lim_{t\to\infty}\zeta^{(t)} = \zeta$, we find $\zeta$ by taking the limits of both sides of \eqref{eq: vb_recursive_update_zeta} as $t\to\infty$. This yields $\zeta = \frac{\Psi^{\ast}}{\tilde{\nu}-p}$ and, therefore, $\lim_{t\to\infty} \tilde{\Psi}^{(t)} = \Psi^{\ast} + p\zeta = \frac{\tilde{\nu}\Psi^{\ast}}{\tilde{\nu}-p}$. While the exact marginal posterior distribution of $\Sigma$ is $p(\Sigma \given Y) = \operatorname{IW}(\Sigma \given \Psi^{\ast},\nu)$ with $\nu = \nu_0 + n$, its variational approximation, upon convergence, yields $\operatorname{IW}\left(\Sigma \given \frac{\tilde{\nu}\Psi^{\ast}}{\tilde{\nu}-p}, \tilde{\nu}\right)$. The true marginal posterior distribution $p(\beta \given Y)$ is a matrix-variate $\mbox{\textsc{T}}$ distribution, while the variational approximation converges to the matrix-normal $\operatorname{MN}\left(\beta \given Mm, M, \frac{\Psi^{\ast}}{\tilde{\nu}-p}\right)$. These distinctions between the true marginal posteriors and their variational approximations typically lead to underestimations of the variability. While these biases are mitigated for large sample sizes (under broad regularity conditions), applying VB to subsets of the data will induce discrepancies that are unnecessary since we have full analytical tractability for the joint posterior distributions.  

\subsubsection{Scalable Gaussian Process Models}\label{sec: scalable_gp}
Scalable Gaussian processes have been a prominent field of research in spatial statistics and are too vast to be comprehensively reviewed here \citep[see, e.g.,][and discussions therein]{banerjee_high-dimensional_2017, heaton_case_2017}. Rather than dividing and conquering massive datasets or building faster algorithms to diminish iteration complexity, this approach builds approximate Gaussian processes that ensure feasible computational and storage requirements. For subsequent comparisons with our approach, we briefly outline one particular class of models that builds a sparse approximation of the spatial precision matrix.    

To see how sparsity accrues computational benefits, consider the model in \eqref{eq:MatrixNormHier} with the posterior distribution given in \eqref{eq:MultivLatPost}. Sampling from this posterior distribution requires the Cholesky decomposition of the $n\times n$ spatial covariance matrix $\rho_{\phi}(\mathcal{S},\mathcal{S})$ or its inverse, which is computationally unfeasible for large $n$. \citet{zhang_high-dimensional_2021} investigates computational strategies for the \textsc{mniw} family using the nearest neighbor Gaussian process \citep[\textsc{nngp},][]{datta_hierarchical_2016, zhang_practical_2019}. The \textsc{nngp} specifies a sparse spatial precision matrix $\tilde{\rho}_{\phi}(\mathcal{S},\mathcal{S})^{-1}$ using the likelihood approximation developed by \citet{vecchia_estimation_1988} and extends it to a well-defined spatial process over the entire domain. We provide a brief discussion of this construction and leave the details to Section~2.2.2 of \citet{zhang_high-dimensional_2021}.     

The sparse precision matrix is $\tilde{\rho}_{\phi}(\mathcal{S},\mathcal{S})^{-1} = H_{\rho}^{\T}H_{\rho}$, where $H_{\rho} = D_{\rho}^{-1/2}(I-B_{\rho})$%. The matrices $D_{\rho}$ and $B_{\rho}$ are 
is %
constructed using a directed acyclic graph based on a fixed topological order of the $n$ spatial locations, say $(s_1,\ldots, s_n)$. We construct neighbor sets $\mathcal{N}(s_i)$ for each $s_i$ to comprise at most $m$ nearest neighbors of $s_i$ from the set of locations preceding it in the topological order. We construct $B_{\rho}$ to be a sparse strictly lower-triangular matrix with the $(i,j)$-th entry $0$ if $s_j \notin \mathcal{N}(s_i)$ for $j > i$. All entries in the first row of $B_{\rho}$ are $0$ because $\mathcal{N}(s_1)$ is empty, as are all diagonal entries since $s_j$ is not included in $\mathcal{N}(s_j)$. This ensures that there are at most $m$ non-zero entries in $B_{\rho}$. Let $i_1 < i_2 < \cdots < i_m$ be the column indices in the $i$-th row of $B_{\rho}$ that have non-zero entries. These non-zero entries are obtained by solving $m\times m$ linear systems $\rho_{\phi}(\mathcal{N}(s_i),\mathcal{N}(s_i))b_i = \rho_{\phi}(\mathcal{N}(s_i), s_i)$ for each $b_i$, $i=2,\ldots,m$, where $\rho_{\phi}(\mathcal{N}(s_i),\mathcal{N}(s_i))$ is the $m\times m$ spatial correlation matrix of $\omega(s)$ constructed using a spatial correlation function $\rho_\phi(s,s')$ over the neighbor sets $\mathcal{N}(s_i)$ and $\rho_{\phi}(\mathcal{N}(s_i), s_i)$ is the $m\times 1$ vector of spatial correlations between $\omega(s_j)$ and $\omega(s_i)$ for each $s_j \in \mathcal{N}(s_i)$. The matrix $D_{\rho}^{-1/2}$ is diagonal with elements given by $\left(1 - b_i^{\T}\rho_{\phi}(\mathcal{N}(s_i), s_i)\right)^{-1/2}$. This completes the specification for $H_{\rho}$.     

The above construction exempts us from computing the Cholesky decomposition. We are essentially \emph{modeling} the Cholesky decomposition rather than \emph{computing} it. Furthermore, the $V_{\gamma}^{\ast-1} = \begin{bmatrix} \frac{\alpha}{1-\alpha}X^{\T}X + M_0^{-1} & \frac{\alpha}{1-\alpha}X^{\T} \\ \frac{\alpha}{1-\alpha}X & H_{\rho}^{\T}H_{\rho} + \frac{\alpha}{1-\alpha}\mathbb{I}_n\end{bmatrix}$ in \eqref{eq:MultivLatPost} is easily factored as
\begin{equation*}
    \label{eq: mniw_nngp_precision_decomposition} 
    V_{\gamma}^{\ast -1} %= \begin{bmatrix} \frac{\alpha}{1-\alpha}X^{\T}X + M_0^{-1} & \frac{\alpha}{1-\alpha}X^{\T} \\ \frac{\alpha}{1-\alpha}X & H_{\rho}^{\T}H_{\rho} + \frac{\alpha}{1-\alpha}\mathbb{I}_n\end{bmatrix} 
    = \underbrace{\begin{bmatrix}
        \sqrt{\frac{\alpha}{1-\alpha}}X^{\T} & M_0^{-1/2} & O \\
        \sqrt{\frac{\alpha}{1-\alpha}}\mathbb{I}_n & O & H_{\rho}^{\T}
    \end{bmatrix}}_{H_{\gamma}^{\ast\T}}\underbrace{ 
    \begin{bmatrix}
        \sqrt{\frac{\alpha}{1-\alpha}}X & \sqrt{\frac{\alpha}{1-\alpha}}\mathbb{I}_n \\
        M_0^{-1/2} & O \\
        O & H_{\rho}
    \end{bmatrix}}_{H_{\gamma}^{\ast}} = H_{\gamma}^{\ast\T}H_{\gamma}^{\ast}\;.
\end{equation*}
Sampling from \eqref{eq:MultivLatPost} is achieved by sampling $\Sigma \sim \text{IW}(\Psi^{\ast},\nu^{\ast})$, computing the Cholesky decomposition of the $q\times q$ matrix (cheaply) $\Sigma = R_{\Sigma}^{\T}R_{\Sigma}$, forming an $n\times q$ matrix $Z$ with elements drawn from $Z_{ij} \overset{iid}{\sim} N(0,1)$, solving the matrix equation $H_{\gamma}^{\ast}A = ZR_{\Sigma}$ for $A$, and setting $\gamma = \mu_{\gamma}^{\ast} + A$. Computational feasibility is ensured by avoiding the Cholesky decomposition of any $n\times n$ spatial correlation matrices. \citet{zhang_high-dimensional_2021} develops and analyzes more efficient algorithms that reformulate sampling from \eqref{eq:MultivLatPost} as a sparse least-squares problem solved using the LSMR algorithm of \citet{fong2011lsmr}.

Predictive inference follows from the marginal posterior predictive distribution in \eqref{eq:MultivLatPred}, but with %$\text{MN}_{n',q}\left(\Omega_{\mathcal{U}}\given M_{\mathcal{U}}\Omega, V_{\Omega_{\mathcal{U}}},\Sigma\right)$ 
$M_{\mathcal{U}}$ and $V_{\Omega_{\mathcal{U}}}$ %
derived from the \textsc{nngp} model. We define the neighbor set $\mathcal{N}(u)$ of an arbitrary location $u \in \mathcal{U}$ as comprising the $m$ closest neighbors of $u$ from $\{s_1,\ldots,s_n\}$. That is, any arbitrary location in $\mathcal{U}$ where predictions are sought is deemed to appear after $s_n$ in the topological order of locations. Using this neighbor set, we construct the $n'\times n$ matrix $M_{\mathcal{U}}$ similar to $B_{\rho}$. Each row of $M_{\mathcal{U}}$ comprises at most $m$ non-zero elements occupying column indices $i_1 < i_2 < \cdots < i_m$ in row $i$. These are obtained by solving the $m\times m$ system  $\rho_{\phi}(\mathcal{N}(u_i),\mathcal{N}(u_i))b_i = \rho_{\phi}(\mathcal{N}(u_i), u_i)$ for each $b_i$, $i=1,\ldots,n'$. The $n'\times n'$ covariance matrix $V_{\Omega_{\mathcal{U}}} = {D}_{\mathcal{U}}$ is diagonal with entries $d_i = 1-b_i^{\T}\rho_{\phi}(\mathcal{N}(u_i), u_i)$. Posterior predictive sampling now proceeds exactly as described below \eqref{eq:MultivLatPred}.

%%%%%%%%%%%%%%%%%%%%%%%%%%%%%%%%%%%%%%%%%%%%
\section{Computer Programs and Resources}\label{sec: comp_programs}
All our subsequent analyses are implemented in native \texttt{R} and \texttt{c++} using the \texttt{spBPS} package. All programs required to reproduce the analysis are publicly accessible from the GitHub repository
\href{https://github.com/lucapresicce/Bayesian-Transfer-Learning-for-GeoAI}{lucapresicce/Bayesian-Transfer-Learning-for-GeoAI }
that links the \texttt{Rcpp}-based \href{https://github.com/lucapresicce/spBPS}{\texttt{spBPS}} package. The reported results are from a standard laptop running an Intel Core I7-$8750$H CPU with $5$ cores for parallel computation and $16$ GB of \textsc{RAM}.

We fit a linear model of coregionalization \citep[\textsc{lmc},][]{banerjee_hierarchical_2025} and multivariate seemingly unrelated Bayesian additive regression trees \citep[multivariate \textsc{bart},][]{esser_seemingly_2025} using \texttt{spBayes} and \texttt{suBART} packages, respectively. We also compare with machine learning methods and AI systems using a scalable platform for parallelized supervised and unsupervised machine learning algorithms offered by \texttt{h2o} \citep{fryda_h2o_2024}. We specifically fit distributed random forest (\textsc{drf}), gradient boosting (\textsc{gbm}), deep neural network (\textsc{dnn}), and a fully automatic machine learning algorithm (\textsc{automl}).
For parallel implementations of \textsc{dbps}, we employ \texttt{R} packages \texttt{doParallel}, and \texttt{foreach} \citep{corporation_doparallel_2022,microsoft_foreach_2022}. We map the interpolated spatial surfaces using \texttt{MBA} \citep{finley_mba_2022}, while sampling from the matrix-variate normal and t distributions is achieved using \texttt{mvnfast} \citep{fasiolo_introduction_2014}. Section~\ref{sec:compdetails} specifies computational considerations and sensitivity to the number of data shards, $K$, for spatial ``BIG'' data analysis.

We build a Bayesian transfer learning engine to conduct amortized Bayesian inference \citep{zammit-mangion_neural_2024} using \textsc{dbps}. We implement a residual neural network (ResNet) \citep{he_deep_2015} using the \texttt{R} interfaces supplied by \texttt{tensorflow} \citep{allaire_tensorflow_2024} and \texttt{keras} \citep{kalinowski_keras_2024} for native \texttt{Python}.

%%%%%%%%%%%%%%%%%%%%%%%%%%%%%%%%%%%%%%%%%%%%
\section{Simulation Experiments}\label{sec:sim}

We evaluate computational and inferential performance of \textsc{double bps}, while comparing against multiple alternative methodologies.
We present a selection of simulation results for multivariate models here and refer the reader to Section~\ref{sec:sim_suppl_M} for further experiments.

% ------------------------------------------
\subsection{Predictive Coverage Performance}\label{sec:sim_precover}

We evaluate the predictive coverage performance and computational efficiency of our proposed framework using a synthetic spatial data set comprising $2,250$: $n=2,000$ locations used for training and $u=250$ held for predictive evaluations. We also include a design matrix $X$ with $p=2$ comprising an intercept and a single predictor whose values were sampled independently from a uniform distribution on $[0,1]$, and a univariate ($q = 1$) response $Y$. The spatial coordinates are sampled uniformly over the unit square $[0,1]^2$ and the spatial correlation matrix $n \times n$ over these coordinates using $\rho_{\phi}(s_i,s_j) = \exp(-\phi \|s_i-s_j\|)$ with $\phi=4$. The response is generated according to a Gaussian process model with regression coefficients $\beta = (1.0, 0.5)^\top$, spatial variance $\sigma^{2} = 1$, and nugget variance $\tau^2 = 0.25$, which corresponds to $\alpha=0.8$.

Although our methodology is inherently multivariate, we restricted this study to a univariate response and a moderate sample size. This design allows direct comparison with established gold-standard approaches: full Gaussian process models (Full \textsc{gp}) and nearest-neighbor Gaussian process (\textsc{nngp}) models. Larger or multivariate datasets would render repeated full \textsc{gp} analyses infeasible, yet these settings are sufficient to evaluate predictive coverage, \textsc{mspe}, and interval width across competing methods.

We assess distributed learning via \textsc{dbps}, varying the number of subsets $K \in \{5,10,20\}$, with candidate models defined over a grid of hyperparameters $\alpha \in \{0.7,0.8,0.9\}$ and $\phi \in \{3,4,5\}$. We evaluate the impact of the disagreement term on the predictive interval width by applying the disagreement tempering technique described in Section~\ref{sec:disagree} to double Bayesian predictive stacking predictions. For comparison, we fit \textsc{nngp} models with neighbor numbers ranging from $m \in \{5,10,20\}$ and a full GP model as a comparison. For both simulation-based methods, we specify the same non-informative default priors and $2,000$ \textsc{mcmc} samples. Each method is applied to $B = 50$ replicated datasets to obtain a frequentist evaluation for each method and assess variability in predictive performance.  

Table~\ref{tab:sim_predcover} summarizes the results in terms of the average predictive interval width, the mean squared prediction error (\textsc{mspe}), empirical coverage, and computational time in seconds for the complete Gaussian process (\textsc{full gp}), nearest neighbor Gaussian process (\textsc{nngp}), and double predictive stacking models. To be precise, we present summaries of the distributions of these quantities over the $B=50$ replicated datasets. Hence, we present the means and the $2.5$th and $97.5$th percentiles (in parentheses) of these quantities over the $B=50$ replicated values. In Table~\ref{tab:sim_predcover}, we denote the application of disagreement tempering to double predictive stacking, or not, using the acronyms \textsc{dbps-dt} and \textsc{dbps}, respectively.

From Table~\ref{tab:sim_predcover}, we see that \textsc{dbps}-based models tend to achieve predictive accuracy and coverage comparable to the \textsc{nngp} and full \textsc{gp} models while dramatically reducing the computation time with respect to the latter. In particular, while most of the $95\%$ intervals presented in Table~\ref{tab:sim_predcover} show large overlap and, therefore, not statistically significant differences, we note that disagreement tempering (\textsc{dt}) provides substantial improvements to \textsc{dbps}. Indeed, \textsc{dt} introduces a net advantage in terms of smaller predictive interval widths while also retaining a plausible empirical coverage (just one percent lower than the nominal $95\%$). The computational time for the double predictive stacking decreases as the number of subsets increases, highlighting the scalability of the framework without compromising inferential quality. These findings demonstrate that a distributed approach can efficiently replicate the predictive performance of full Gaussian process models, making it a practical alternative for large spatial datasets. These results also prove that \textsc{double bps} achieves substantial computational savings relative to full \textsc{gp} and \textsc{nngp} models while maintaining comparable predictive accuracy and coverage. This simulation experiment illustrates the scalability of the framework without compromising inference quality when compared with nearest-neighbor Gaussian process models.

\begin{table}[t!]
\centering
\scriptsize
\begin{tabularx}{\linewidth}{M Y Z{2.75cm} Z{2.75cm} Z{2.75cm}}
\toprule
\textbf{Model} & \textbf{Time (sec)} & \textbf{Pred. Int. Width} & \textbf{MSPE} & \textbf{Emp. Coverage} \\
\midrule
\rowcolor{gray!10}
\textsc{full gp} & 1654 (1626, 1683) & 2.00 (1.96, 2.05) & \textbf{0.264} (0.222, 0.293) & 0.945 (0.925, 0.969) \\
\textsc{nngp} ($m=20$) & 120 (116, 130) & 2.01 (1.97, 2.08) & \textbf{0.264} (0.220, 0.290) & 0.947 (0.923, 0.971) \\
\rowcolor{gray!10}
\textsc{nngp} ($m=10$) & 39 (32, 40) & 2.01 (1.97, 2.07) & \textbf{0.264} (0.221, 0.288) & \textbf{0.949} (0.931, 0.969) \\
\textsc{nngp} ($m=5$) & 17 (16, 18) & 2.01 (1.98, 2.07) & \textbf{0.264} (0.223, 0.291) & 0.948 (0.929, 0.968) \\
\rowcolor{gray!10}
\textsc{dbps} ($K=5$) & 4.8 (4.7, 5) & 2.03 (1.97, 2.10) & 0.265 (0.222, 0.283) & 0.946 (0.926, 0.966) \\
\textsc{dbps-dt} ($K=5$) & 4.8 (4.7, 5) & \textbf{1.96} (1.89, 2.03) & \textbf{0.264} (0.222, 0.281) & 0.939 (0.918, 0.959) \\
\rowcolor{gray!10}
\textsc{dbps} ($K=10$) & 2 (1.7, 2.1) & 2.06 (1.96, 2.11) & 0.268 (0.247, 0.305) & 0.955 (0.938, 0.974) \\
\textsc{dbps-dt} ($K=10$) & 2 (1.7, 2.1) & 1.99 (1.91, 2.06) & 0.267 (0.248, 0.304) & 0.942 (0.913, 0.963) \\
\rowcolor{gray!10}
\textsc{dbps} ($K=20$) & \textbf{1.6} (1.5, 1.7) & 2.19 (2.11, 2.27) & 0.271 (0.243, 0.306) & 0.960 (0.941, 0.978) \\
\textsc{dbps-dt} ($K=20$) & \textbf{1.6} (1.5, 1.7) & 2.04 (1.99, 2.14) & 0.271 (0.242, 0.307) & 0.947 (0.922, 0.965) \\
\bottomrule
\end{tabularx}
\caption{Average predictive interval width, \textsc{mspe}, empirical coverage at $95\%$, and computation time (in seconds) for different specifications of \textsc{nngp}, \textsc{dbps}, and full Gaussian process models. These quantities are computed for each of our 50 replicated datasets. We present their means and the $2.5\%$ and $97.5\%$ quantiles calculated over the 50 replicates. Boldface values indicate the best performance with respect to each metric.}
\label{tab:sim_predcover}
\end{table}

Distributed approaches \citep[e.g.,][]{guhaniyogi_meta-kriging_2018} can lead to prediction oversmoothing, but this phenomenon is not manifested strongly in double \textsc{bps}. However, they clearly inflate predictive variability as the number of subsets grows. Consequently, increasing the number of subsets induces wider predictive intervals and some extra coverage.
The origin of this extra width of the predictive interval may arise from a disagreement term, often associated with linear pooling forecasts \citep{knuppel_forecast_2022}. However, the results in Table~\ref{tab:sim_predcover} confirm that the effect is negligible.

%-------------------------------------------
\subsection{Transfer Learning in \texorpdfstring{$\mathscr{M}$}{M}-closed \texorpdfstring{\&}{and} \texorpdfstring{$\mathscr{M}$}{M}-open Settings}\label{sec:sim4}

This simulation study serves a dual purpose: it assesses the performance of \textsc{dbps} under $\mathscr{M}$-closed and $\mathscr{M}$-open settings and, crucially, it investigates the sensitivity of predictive inference to the choice of candidate values for $\{\alpha,\phi\}$.
To this end, we consider both the $\mathscr{M}$-closed scenario, in which the true spatial parameters lie in the candidate grid, and the $\mathscr{M}$-open scenario, in which they do not. This directly addresses the practical question of what happens when the prior elicitation or automatic selection of the $\{\alpha,\phi\}$ grid is imperfect or coarse.
While exploring how \textsc{double bps} behaves in the $\mathscr{M}$-closed and $\mathscr{M}$-open settings, we compare with the exact transfer learning framework that we devise in Section~\ref{sec:DivConBays}.
In the latter, the model specification is characterized by different values of $\alpha$ and $\phi$ representing (i) a well-specified (\textsc{ws}) setting with the data generating values $\{\alpha=0.8, \phi=4\}$; (ii) a moderately misspecified (\textsc{ms}) setting with  $\{\alpha=0.45, \phi=6.63\}$; and (iii) a highly misspecified (\textsc{hms}) setting with $\{\alpha=0.25, \phi=50\}$.

\begin{figure}[t]
    \centering
    \includegraphics[scale=0.425]{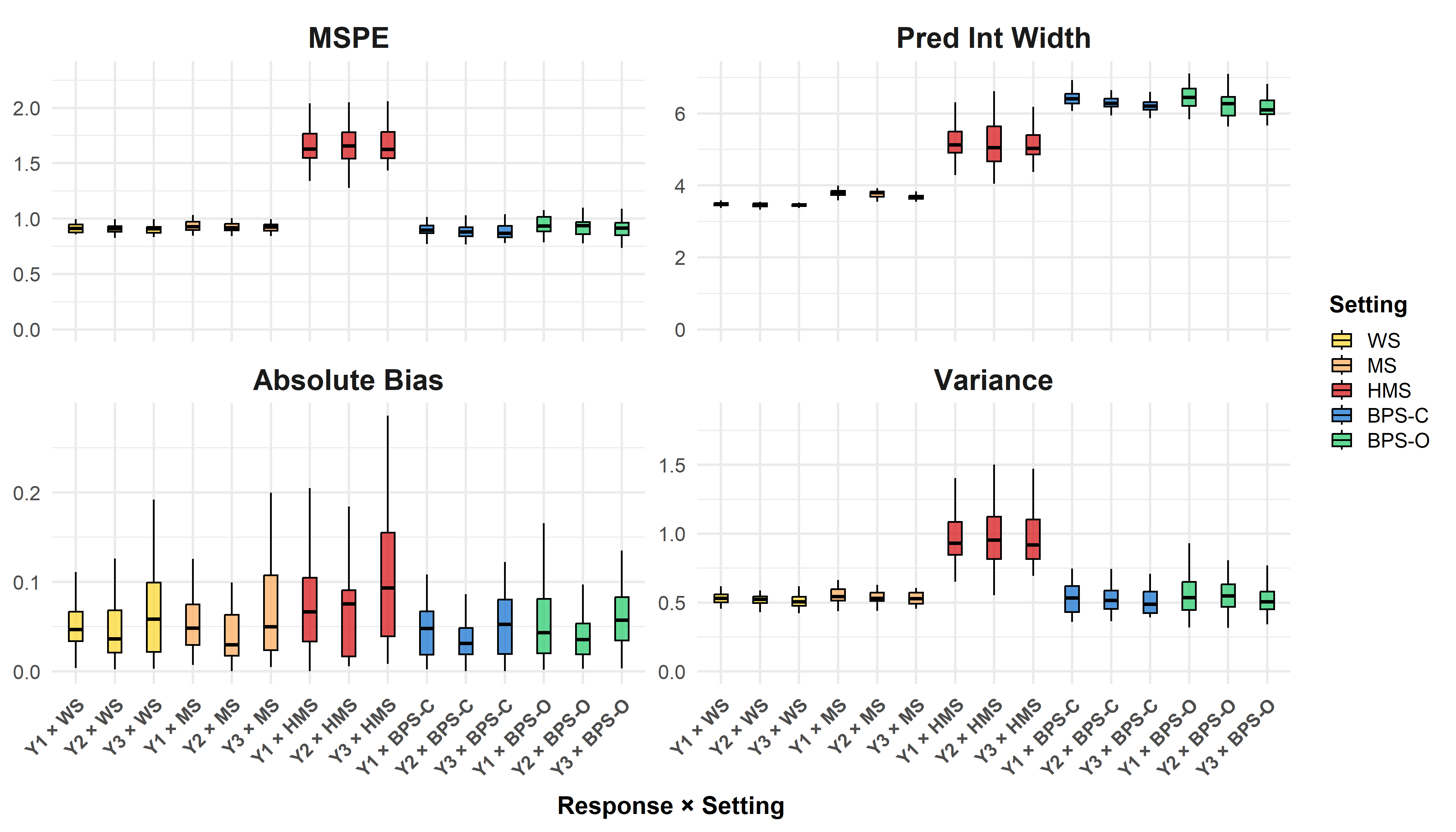}
    \caption{ Predictive MSPE, interval width, absolute bias, and variance boxplot across responses and settings from 50 replications.}
    \label{fig:seq_pred_bpsM}
\end{figure}

Double \textsc{bps} was tested on the $\mathscr{M}$-closed and $\mathscr{M}$-open settings. The former considers situations where the true model exists and is identified within a finite set of considered models. Here, the ``true'' model is the one such that $\{\alpha=0.8,\phi=4\}$. Then, for \textsc{double bps} under closed setting (\textsc{bps-c}) we specify $J=9$ competitive models with $\alpha \in \{0.75, 0.80, 0.85\}$ and $\phi \in \{2, 4, 6\}$ that yield effective spatial ranges in the percentage of maximum point inter-distance of $105\%, 53\%, 35\%$ respectively, including the true model as one of the possible candidates. Conversely, in the $\mathscr{M}$-open setting, even though the true model exists, it cannot be fully specified. Thus, for \textsc{double bps} under open settings (\textsc{bps-o}), we randomly define $J=9$ candidate models. In particular, we uniformly sample 3 values for $\alpha\in(0,1)$ and 3 values for $\phi\in(0,50)$. We perform the experiment using 50 replications.
Each replicate consists of values of the $n\times q$ outcome $Y$ generated from \eqref{eq:MatrixNormHier} with $n=5,000$, $q=3$, and $p=2$, the matrix $X$ includes an intercept as its first column and a predictor generated from a standard uniform distribution over $[0,1]$, $\beta= \begin{bsmallmatrix}
    -0.75 & 1.05 & -0.35 \\ 2.20 & -1.10 & 0.45
\end{bsmallmatrix}$ and $\Sigma = \begin{bsmallmatrix}
    2.00 & 0.80 & 0.20 \\ 0.80 & 2.00 & -0.45 \\ 0.20 & -0.45 & 2.00
\end{bsmallmatrix}$. The $n\times n$ spatial correlation matrix $V$ is specified using an exponential correlation function with $\phi=4$ and $\alpha=0.8$.   

\begin{figure}[t]
    \centering
    \includegraphics[scale=0.425]{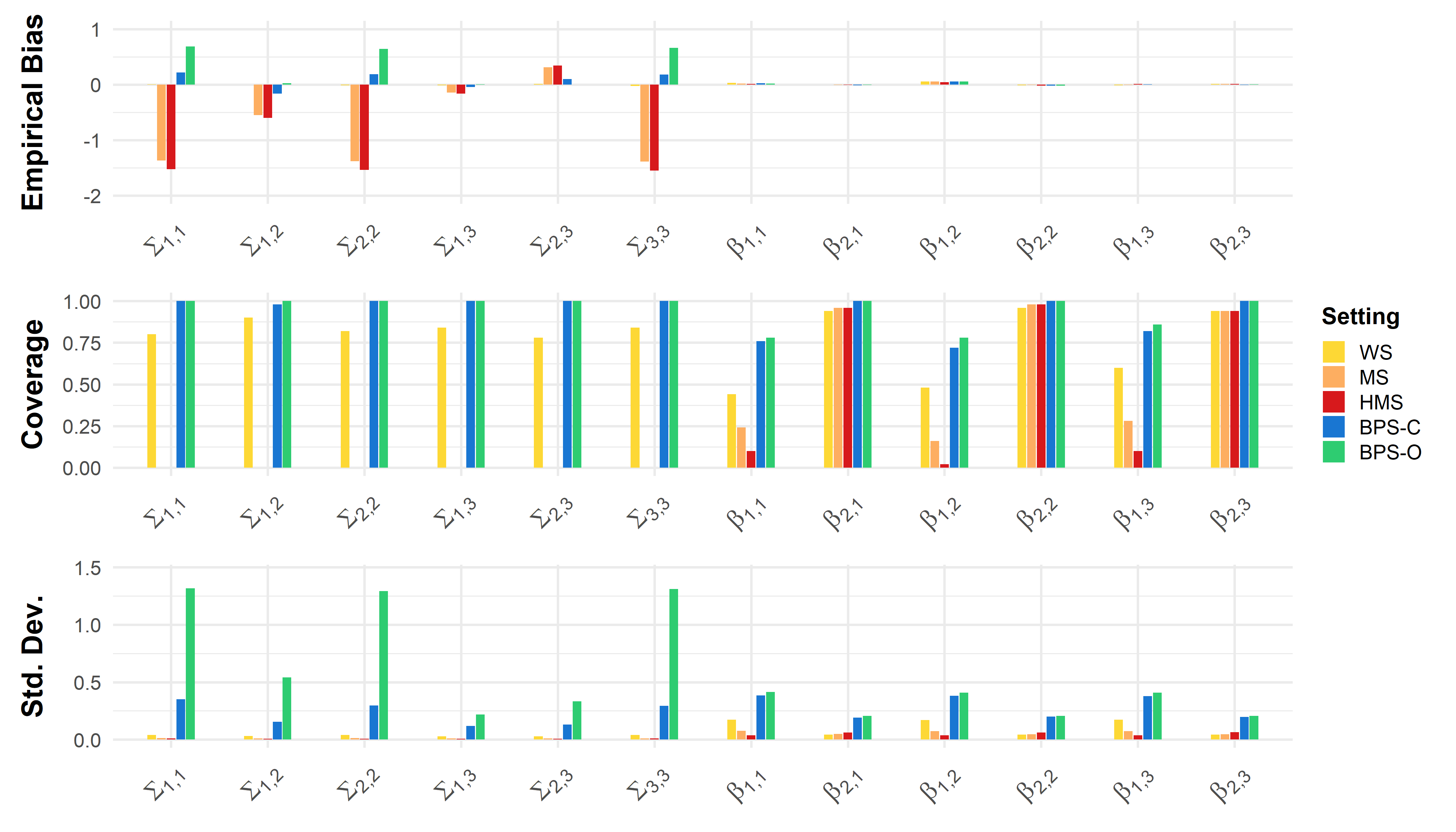}
    \caption{ Average posterior bias, coverage, and standard deviation across parameters and settings from 50 replications.}
    \label{fig:seq_post_bpsM}
\end{figure}

Figure~\ref{fig:seq_pred_bpsM} presents (i) mean square prediction error (\textsc{mspe}); (ii) predictive interval width; (iii) absolute bias; and (iv) variance. We present boxplots for the distribution of each metric over the 50 replicates. This is made for each response and for each setting. In terms of predictive \textsc{mspe}, absolute bias, and variance, the settings \textsc{bps-c} and \textsc{bps-o} exhibit slightly better performance. However, this seems to compromise the predictive interval width, as the uncertainty in the predictions is much higher. Approaches that estimate $\{\alpha,\phi\}$, rather than fix them, are somehow expected to introduce more uncertainty. In addition, we find no evidence of any difference between \textsc{bps-c} and \textsc{bps-o} for any metric. This is surprising, as it suggests the reliability of \textsc{dbps} even in the transfer learning setting we devised in Section~\ref{sec:AccelerMultivarMod}. Finally, irrespective of $\mathscr{M}$-closed or $\mathscr{M}$-open settings, it is more convenient to specify a set of candidate models using \textsc{double bps} instead of trying to fix $\{\alpha,\phi\}$.

Figure~\ref{fig:seq_post_bpsM} presents posterior inference for (i) average empirical bias; (ii) average coverage; and (iii) average standard deviation, where the average is taken over the 50 replications.
As expected, misspecification induces empirical bias in posterior estimates. The top panel reports how higher levels of bias are associated with models that are farther away from the truth. Here, \textsc{double bps} is placed in the middle for both $\mathscr{M}$ settings.
The middle plot in Figure~\ref{fig:seq_post_bpsM} shows that $\Sigma$ is the parameter most affected by misspecifications. Its elements are well captured only by \textsc{double bps}, with coverage close to nominal, followed by \textsc{ws} specification, which performs worse. A similar pattern holds for $\beta$, where only \textsc{double bps} ensures adequate coverage. This reflects greater posterior variability in stacking approaches, as shown in the bottom panel reporting posterior standard deviations between settings. Interestingly, the inferential performance is weakest for spatial variance, which is not identifiable from the data obtained \citep{zhang_inconsistent_2004}. \textsc{double bps} behaves very similarly among $\mathscr{M}$-closed and $\mathscr{M}$-open.
See Section~\ref{sec:sim_suppl_M} for additional simulation experiments. 

Although \textsc{dbps} outperforms exact transfer learning (Section~\ref{sec:smk}) in the $\mathscr{M}$-closed and $\mathscr{M}$-open settings, the latter performs competitively with improved predictive performance in the \textsc{ws} and \textsc{ms} settings than in the \textsc{hms} settings. 
This empirically demonstrates the robustness of \textsc{dbps} with respect to the choice of the candidate grid for spatial parameters, confirming that stacking is resilient to hyperparameter misspecification, adapting to concentrate predictive mass on the closest approximating models.
Our overall findings appear consistent with theoretical insights that Gaussian processes tend to deliver good predictive performance even for misspecified covariance functions in fixed domains \citep{stein_asymptotically_1988, stein_asymptotic_1989}.

%-------------------------------------------
\subsection{Amortized Bayesian Inference}\label{sec:sim_abi}

% setting explanation
We perform transfer learning by supervising a neural network using the output of \textsc{double bps} to deliver amortized Bayesian inference. We generate $N=100$ instances of $Y$ from \eqref{eq:MatrixNormHier} using a fixed realization of $\Omega$ for $q=2$ correlated outcomes, $n=500$ spatial locations that remain fixed across the datasets, and a fixed design matrix $X$ with $p=2$ comprising an intercept and a single predictor whose values were sampled independently from a uniform distribution over $[0,1]$. The true regression coefficients are fixed at $\beta = \begin{bsmallmatrix} -0.75 & 1.85 \\ 0.9 & -1.10 \end{bsmallmatrix}$, with $\Sigma = \begin{bsmallmatrix} 1 & -0.3 \\ -0.3 & 1 \end{bsmallmatrix}$, $\alpha = 0.8$, and $\rho_{\phi}(s_i,s_j) = \exp(-\phi\|s_i-s_j\|)$ with $\phi = 4$.

\begin{figure}[t]
    \centering
    \includegraphics[scale=0.5]{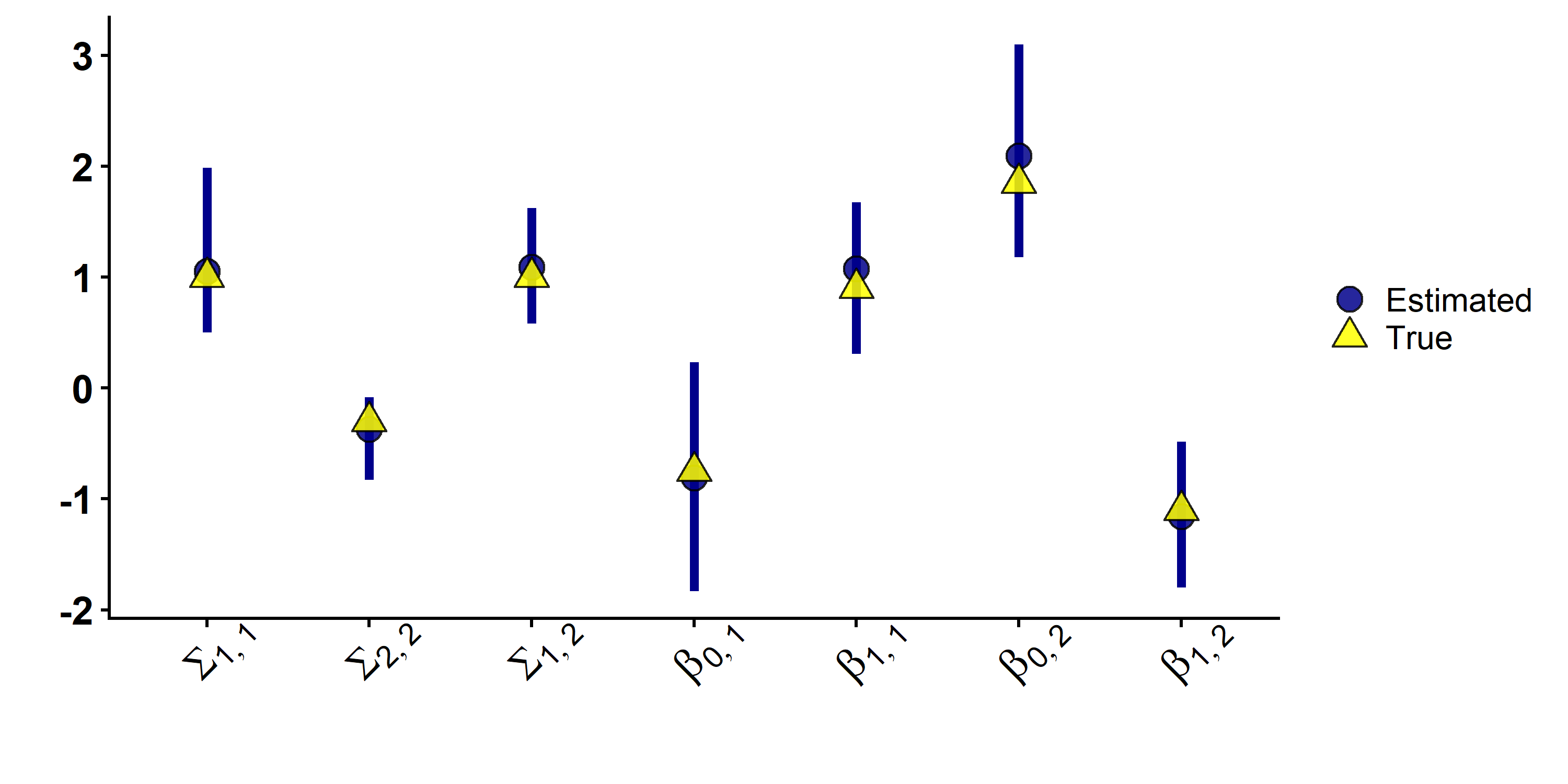}
    \caption{ Amortized posterior credible intervals for parameters. True parameters in yellow.}
    \label{fig:postint_AB}
\end{figure}

We train the neural network in a supervised learning framework using $R=250$ posterior samples. We apply \textsc{double bps} to each generated dataset with $K=5$ subsets, $\alpha \in \{0.7, 0.8, 0.9\}$, and $\phi \in \{3, 4, 5\}$. These yield $100$ instances of $\{Z, \Theta\}$, where $Z = [Y:X] \in \mathbb{R}^{n\times(q+p)}$, while $\Theta\in\mathbb{R}^{[(qp)+(q(q+1)/2)+(nq)]\times 3}$ comprises the posterior quantiles ($\{2.5\%, 50\%, 97.5\%\}$) for the distinct elements of $\{\beta,\Sigma,\Omega\}$.

We supervise a deep neural network comprising 3 hidden layers with 128, 256, and 512 nodes, with ReLU activations. The residual network is trained over 50 epochs (with 24 batches per epoch) by minimizing $\hat{\psi}=\arg\min_{\psi} 1/N \sum_{i=1}^{N} L\left(g_{\psi}(Z_{i}),\, \Theta_{i}\right)$, where $g_{\psi}(Z_i)$ is the neural output from $Z_i$ trained by the $\Theta_i$ obtained using \textsc{double bps} for each of the $N=100$ generated data sets. This training is implemented in \texttt{Keras} using the Adam optimizer with mean squared error (\textsc{mse}) loss $L(\cdot\,,\cdot)$. For evaluation, we apply the trained model to unseen datasets with the same dimensions. 

\begin{figure}[t]
    \centering
    \includegraphics[scale=0.35]{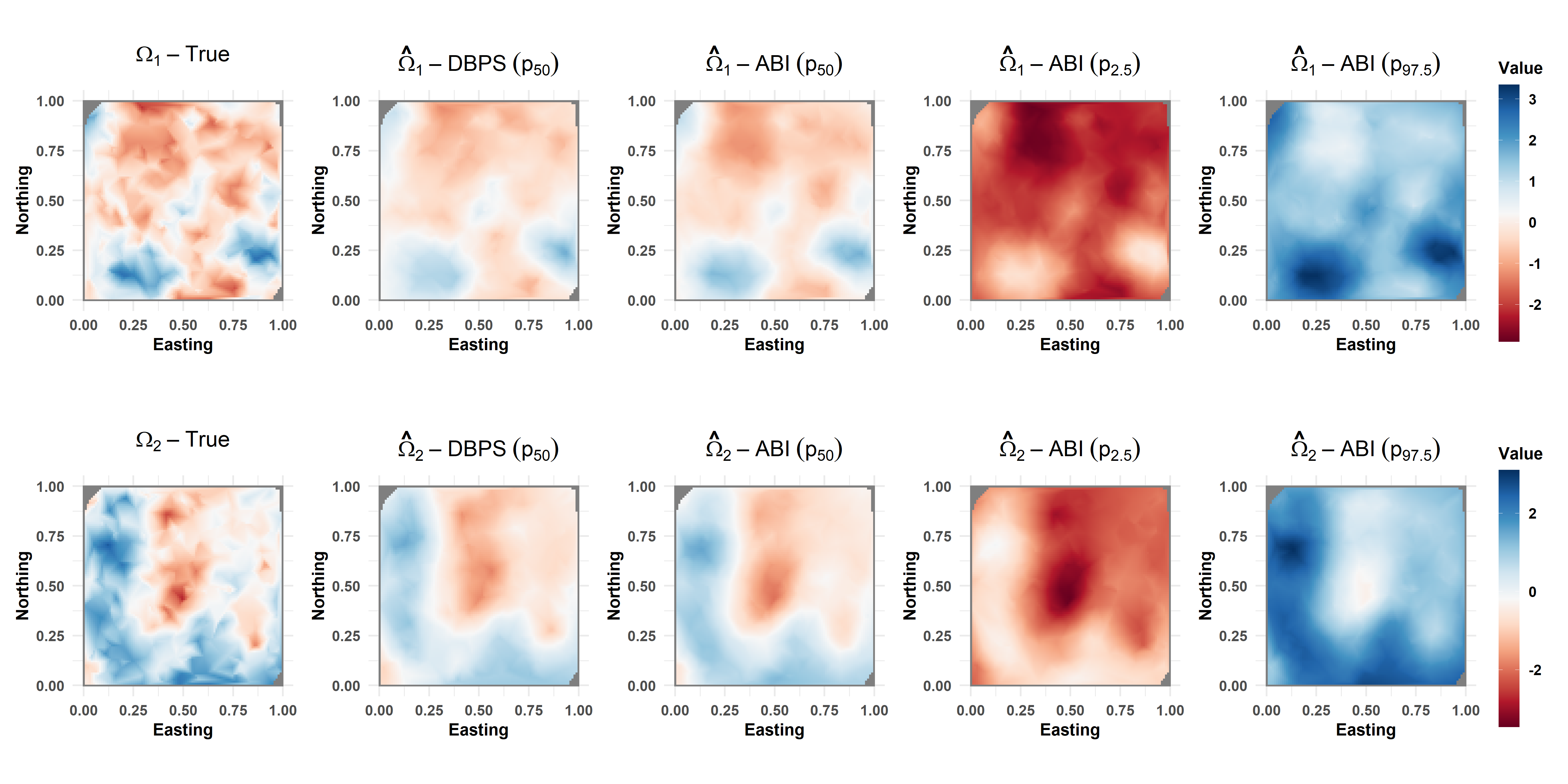}
    \caption{ Surface interpolations for true spatial process, BPS prediction (50-th quantile), and amortized prediction of $\{50, 2.5, 97.5\}$ quantiles. Each row corresponds to an outcome.}
    \label{fig:heatmap_AB}
\end{figure}

Figure~\ref{fig:postint_AB} displays the amortized posterior credible intervals (blue bars) for $\{\beta,\Sigma\}$, along with the true values of the parameters (yellow triangles). The amortized $95\%$ credible intervals capture all true values and the medians ($50\%$) align closely with the true values, highlighting the effectiveness of the deep network in recovering posterior summaries. Figure~\ref{fig:heatmap_AB} displays amortized inference for $\Omega$. We compare the results from amortized inference with the true values of $\Omega$, and the \textsc{double bps} prediction for the $50$th quantile presented in the first and second columns of Figure~\ref{fig:heatmap_AB}, respectively.

These results illustrate the strengths of amortized inference and transfer learning. Once trained, the deep network provides instantaneous posterior quantile estimates for new datasets, without requiring us to rerun \textsc{double bps}. This amortizes the computational cost for future tasks. Additionally, the model generalizes across a range of data-generating conditions, effectively enabling posterior transfer learning to new but structurally similar problems. This makes the approach especially useful in large-scale or resource-constrained applications where repeated full Bayesian inference is prohibitively expensive.

%%%%%%%%%%%%%%%%%%%%%%%%%%%%%%%%%%%%%%%%%%%%
\section{Data Analysis}\label{sec:dataappl}

%-------------------------------------------
\subsection{Vegetation Index Data}\label{sec:dataset}
 
Statisticians and machine learners face growing demands to analyze and study global warming datasets \citep[see, e.g.,][]{data_sst_fisher1958, data_sst_neville1989, data_sst_friehe1991, data_sst_carroll2019}. The sheer volume of such datasets and the process-driven models required for their analysis have naturally raised the question of migrating such analysis to AI platforms. Our current application analyzes vegetation index data from the Moderate Resolution Imaging Spectroradiometer (\textsc{modis}) developed by \textsc{nasa}. We specifically focus on ``\textsc{mod13c1}.061 - Terra Vegetation Indices 16-Day L3 Global 0.05 Deg Climate Modeling Grid'' \citep{earth_science_data_systems_modisterra_2025}, which provides vegetation indices per pixel on a 0.05-degree climate modeling grid (3600 rows by 7200 columns of 5600-meter pixels). It contains 16-day global composites, cloud-free, with additional reflectance and angular information. 

Modeling the Normalized Difference Vegetation Index (\textsc{ndvi}) and Red Reflectance (\textsc{rr}) jointly is scientifically important, as it separates and defines the specific biophysical factors influencing vegetation. \textsc{ndvi} is built on the fundamental principle that healthy vegetation absorbs red light and reflects near-infrared (NIR) light. By examining both the calculated \textsc{ndvi} and its red reflectance component, scientists can gain deeper insights into the structural and biochemical properties of vegetation. The solar zenith angle (\textsc{sza}), which is the angle between the Sun and the point directly overhead, is a shared predictor that influences how much solar irradiation reaches the surface and is, therefore, crucial when assessing biomass and vegetation indices. 
\textsc{ndvi} and \textsc{rr} reflect vegetation's capacity to absorb photosynthetically active radiation, helping scientists understand the underlying mechanisms of climate change \citep[][]{tucker_red_1979,sellers_canopy_1985,justice_analysis_1985,haque_effects_2024}.

\begin{table}[t!]
\centering
\small
\begin{tabularx}{\textwidth}{L{3.5cm} Y Y Y Y L{2.5cm} L{2.5cm}}
\toprule
\textbf{Vegetation index} & \textbf{Mean} & \textbf{Std.Dev} & \textbf{Min.} & \textbf{Max.} & \textbf{Histogram} & \textbf{Boxplot} \\
\midrule

\cellcolor{gray!10}\textbf{NDVI}
& \cellcolor{gray!10} 8.593 
& \cellcolor{gray!10} 0.517 
& \cellcolor{gray!10} 6.909 
& \cellcolor{gray!10} 9.469 
& \cellcolor{gray!10}\raisebox{-0.2\totalheight}{\includegraphics[width=0.8\linewidth]{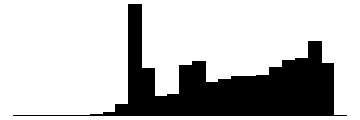}} 
& \cellcolor{gray!10}\raisebox{-0.2\totalheight}{\includegraphics[width=0.8\linewidth]{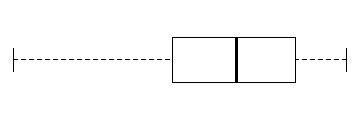}} \\

\textbf{Red Reflectance} 
& 8.563 
& 0.447 
& 8.007 
& 9.472 
& \raisebox{-0.2\totalheight}{\includegraphics[width=0.8\linewidth]{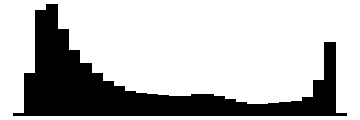}} 
& \raisebox{-0.2\totalheight}{\includegraphics[width=0.8\linewidth]{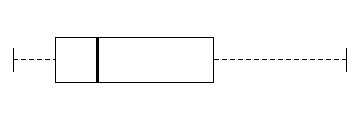}} \\

\bottomrule
\end{tabularx}
\caption{Summary statistics and visual representation of response variables.}
\label{tab:eda_summary}
\end{table}

The data set comprises $1,500,000$ locations. 
We randomly selected $n=1,000,000$ sites for model fitting. To assess predictive performance and its stability with respect to the choice of evaluation locations, we constructed $5$ disjoint held-out sets by randomly sampling $u=250,000$ locations from the remaining sites and partitioning them into $5$ non-overlapping subsets of $50,000$ test locations each. All competing methods were fitted once on the full training set of $n$ observations; predictive evaluation was then carried out independently on each of the $5$ held-out sets. Reported metrics in Table~\ref{tab:analysis2_ai} are averages over the $5$ replicates.
This protocol extends standard practice in large-scale spatial data benchmarking, where a single held-out set is typically reserved for predictive evaluation \citep[see, e.g.,][]{zhang_practical_2019}. Rather than relying on a single split, we evaluate all competing methods on multiple independent, disjoint held-out sets, allowing us to assess not only predictive accuracy but also its stability. This yields a comprehensive out-of-sample assessment over a total of $250,000$ held-out multivariate observations while keeping fitting fixed, as repeated refitting and predictions at this scale would be prohibitive for competing methods.
Following \citet{zhang_spatial_2022}, outcomes were logarithmically transformed and we labeled $\log(\textsc{ndvi} + 1)$ as \textsc{ndvi} and $\log(\textsc{rr} + 1)$ as \textsc{rr}. All variables, including \textsc{sza}, were averaged over a 16-day window in May 2024. Table~\ref{tab:eda_summary} summarizes the spatial response distribution. The maximum distance between sites is approximately $42,909$ kilometers.

Our central scientific objective is to jointly predict \textsc{ndvi} and \textsc{rr} from massive, globally distributed datasets of this scale using \textsc{GeoAI}. We assess whether multivariate statistical models can deliver accurate, scalable, and timely predictions across millions of spatial locations while accounting for dependence structures in the solar zenith angle. 
These capabilities are crucial for their application in ecology, agriculture, and climate policy.

%-------------------------------------------
\subsection{Results Using DBPS}\label{sec:results_dbps}

\begin{figure}[t]
    \centering
    \includegraphics[scale=0.3]{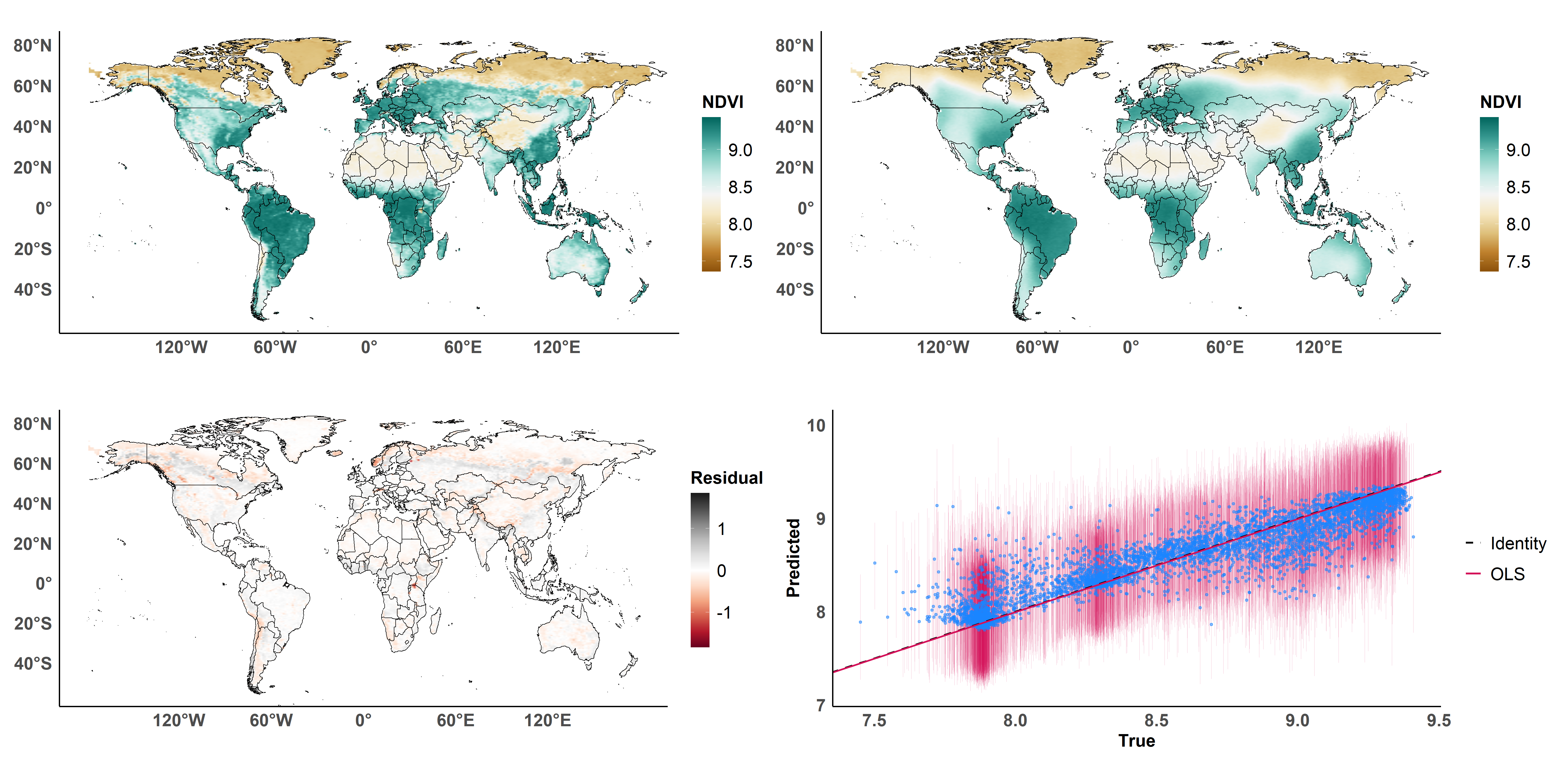}
    \caption{ Left to right: Maps for test data (top left), predicted data (top right), and residual surface (bottom left) for \textsc{ndvi}. Predicted values scatterplot against the truth, with error bars for held-out values, is in the bottom right, alongside identity and \textsc{ols} lines. Results correspond to $K=2,000$.}
    \label{fig:res_analysis2_ndvi}
\end{figure}

To analyze Vegetation Index (\textsc{vi}) data, we conduct machine-generated \textsc{eda} in Appendix~\ref{sec:eda}. %to guide grid settings for ${\alpha,\phi}$.
The predictors comprise an intercept and the solar zenith angle for that location ($p=2$).
Based on the variogram analysis in Appendix~\ref{sec:eda}, we set $\alpha\in\{0.825, 0.909\}$, and $\phi\in\{0.049, 0.067\}$ respectively. We specify $\{\gamma,\Sigma\}$ in \eqref{eq:MultivJoinPrior} using $m_{0}=0_{p\times q}$, $M_{0}=10\mathbb{I}_{p}$, $\Psi_{0}=\mathbb{I}_{q}$, $\nu_{0}=3$, and finally opt for an exponential spatial covariance function.
Following insights from Section~\ref{sec:sim5_sens}, we fix the subset size at $n_{k}\in\{250, 500\}$, leading to a number of subsets $K\in\{4,000 \;,\; 2,000\}$ respectively. We use a random scheme to form the partition of the data set and present the results when $K=2,000$ and in Table~\ref{tab:analysis2} for $K=4,000$.

\begin{figure}[t]
    \centering
    \includegraphics[scale=0.3]{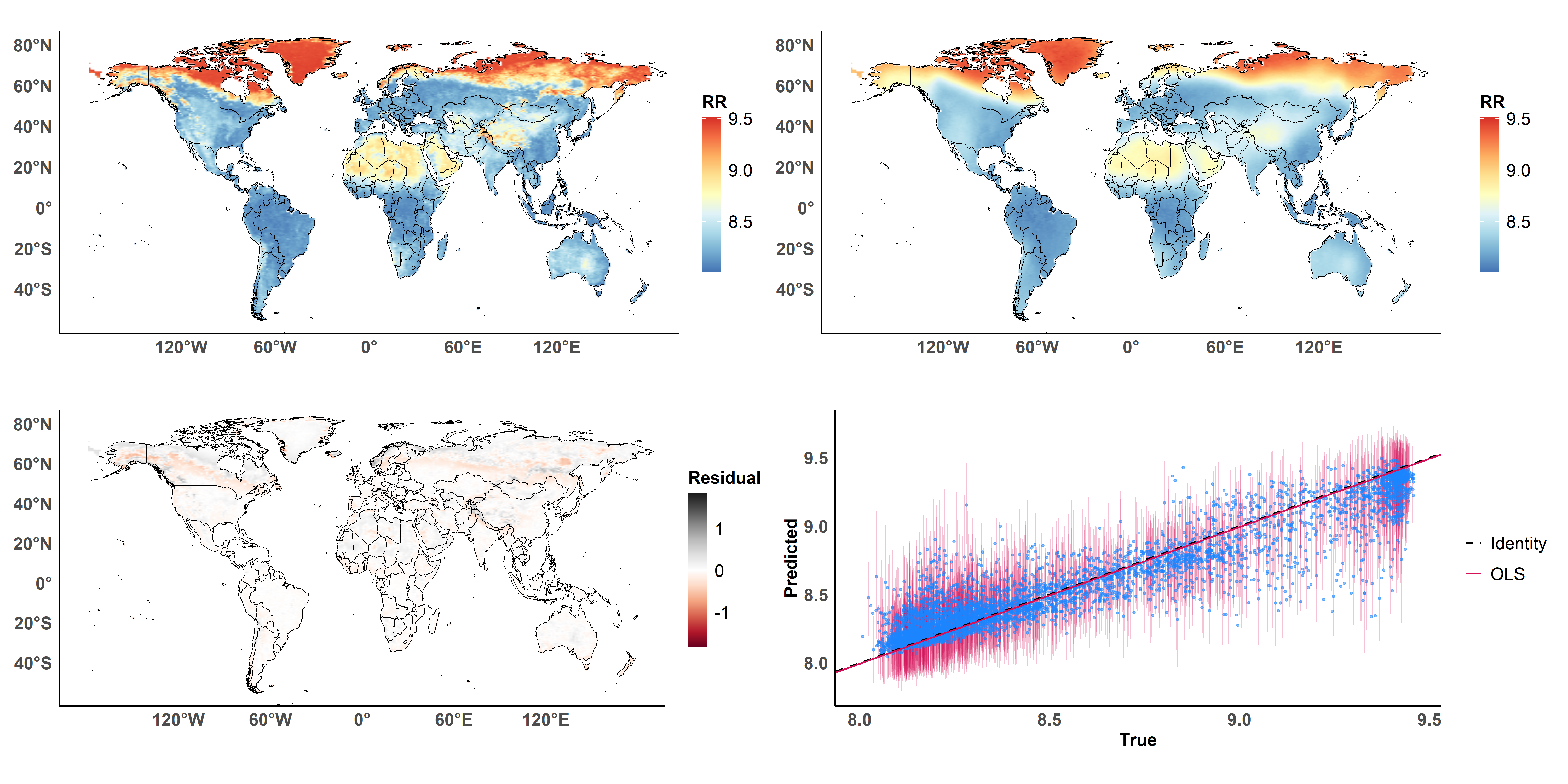}
    \caption{ Left to right: Maps for test data (top left), predicted data (top right), and residual surface (bottom left) for \textsc{rr}. Predicted values scatterplot against the truth, with error bars for held-out values, is in the bottom right, alongside identity and \textsc{ols} lines. Results correspond to $K=2,000$.}
    \label{fig:res_analysis2_rr}
\end{figure}

We fit the multivariate model developed in Section~\ref{sec:AccelerMultivarMod} with \textsc{ndvi} and \textsc{rr} comprising the $q=2$ columns of $Y$. 
Figures~\ref{fig:res_analysis2_ndvi}~and~\ref{fig:res_analysis2_rr} illustrate maps and predictive diagnostic corresponding to \textsc{ndvi} and \textsc{rr}, respectively, using $K=2,000$ subsets. The top left panel in each figure presents the spatially interpolated map of the held-out test observations for the respective responses, revealing pronounced spatial variation: darker shades of green in \textsc{ndvi} represent higher values of detected biomass, while lighter shades of brown represent low biomass. In contrast, for \textsc{rr}, warmer colors in the red spectrum represent higher reflectance.

The top right panel of figures~\ref{fig:res_analysis2_ndvi}~and~\ref{fig:res_analysis2_rr} shows the \textsc{dbps} interpolated posterior mean surfaces, which are nearly indistinguishable from the observed test surfaces (top left). This suggests that automated \textsc{dbps} effectively, perhaps even strikingly, recovers spatial patterns despite simplifications of the modeling over more elaborate statistical models \citep{banerjee_modeling_2020, zhangEtAl2025jasa}. 
The bottom left panel of each figure shows the map of pointwise prediction residuals at the held-out locations. The absence of systematic spatial structure in the residuals confirms that \textsc{dbps}  does not exhibit localized bias and adequately captures the global spatial dependence in both responses. The bottom right panel presents scatterplots of observed versus posterior mean predicted values at the held-out observations, with the identity line and the \textsc{ols} regression line overlaid. The perfect overlap between the two lines confirms the absence of systematic predictive bias, and the tight concentration around them corroborates the quantitative accuracy reported in Table~\ref{tab:analysis2_ai} for \textsc{ndvi} and \textsc{rr}, respectively.

We attempted comparisons with other Bayesian models, which represent the benchmark for spatial data analysis: the \textsc{lmc}, \textsc{nngp}, and \textsc{bart} models.
We were unable to fit any of them into our multivariate data because they exceeded memory. 
However, since \textsc{nngp} is the gold standard in large-scale analysis, the restricted computational resources we use to fit the data set do not allow us to allocate the output. This highlights a main advantage of using \textsc{dbps} when working on limited computational frameworks.

Table~\ref{tab:analysis2} compares the posterior estimates of the model parameters for \textsc{dbps} with $K\in\{4,000\,,\;2,000\}$, and the Bayesian conjugate linear model (\textsc{blm}), which does not take into account spatial variability.
The notably higher magnitudes of the intercepts in the non-spatial linear model are unsurprising, as the spatial random effects absorb much of the variation in the intercepts. The solar zenith angle is positively associated with \textsc{rr} and negatively associated with \textsc{ndvi} in the non-spatial model. This is also expected since higher levels of solar irradiation are associated with higher red reflectance and with more arid regions with less vegetation. However, the spatial models reveal that the slope for \textsc{ndvi} is significantly positive with solar irradiation after the spatial effects have absorbed previously unaccounted for latent or lurking factors that might have contributed to the negative slopes in the non-spatial model.

\begin{table}[t!]
\centering
\scriptsize
\begin{tabularx}{\linewidth}{Y Y Y Y}
\toprule
\textbf{Parameter} & \textbf{BLM} & \textbf{DBPS ($K=4{,}000$)} & \textbf{DBPS ($K=2{,}000$)} \\
\midrule
\rowcolor{gray!10}
$\beta_{0,\textsc{ndvi}}$             & 34.495 (34.392, 34.601) & 1.670 (-0.570, 4.159) & 1.814 (-0.787, 4.312) \\
\addlinespace[2pt]
$\beta_{1,\textsc{ndvi}}$             & -2.708 (-2.719, -2.697) & 0.716 (0.453, 0.955) & 0.700 (0.431, 0.979) \\
\addlinespace[2pt]
\rowcolor{gray!10}
$\beta_{0,\textsc{rr}}$               & -16.924 (-17.014, -16.837) & -0.586 (-2.512, 1.498) & -0.633 (-2.408, 1.407) \\
\addlinespace[2pt]
$\beta_{1,\textsc{rr}}$               & 2.664 (2.655, 2.674) & 0.962 (0.739, 1.162) & 0.966 (0.749, 1.156) \\
\addlinespace[2pt]
\rowcolor{gray!10}
$\Sigma_{\textsc{ndvi}}$              & 0.221 (0.220, 0.221) & 0.182 (0.133, 0.245) & 0.147 (0.114, 0.211) \\
\addlinespace[2pt]
$\Sigma_{\textsc{ndvi},\textsc{rr}}$  & -0.167 (-0.168, -0.166) & -0.116 (-0.167, -0.081) & -0.093 (-0.132, -0.072) \\
\addlinespace[2pt]
\rowcolor{gray!10}
$\Sigma_{\textsc{rr}}$                & 0.155 (0.154, 0.155) & 0.118 (0.082, 0.165) & 0.093 (0.073, 0.134) \\
\bottomrule
\end{tabularx}
\caption{Vegetation Index data analysis parameter estimates for Bayesian conjugate linear model (\textsc{blm}) and \textsc{dbps} models. Parameter posterior summary 50 (2.5, 97.5) percentiles.}
\label{tab:analysis2}
\end{table}

In addition to \textsc{double bps} (\textsc{dbps}), \textsc{lmc}, \textsc{nngp}, and Bayesian multivariate linear regression, we expanded the analysis by including four competitive
algorithms: distributed random forest (\textsc{drf}), gradient boosting (\textsc{gbm}), deep neural network (\textsc{dnn}), and a fully automatic machine learning algorithm (\textsc{automl}).
As detailed in Section~\ref{sec:compdetails}, we implement and perform the analysis using the \texttt{h2o} \texttt{R} package \citep{fryda_h2o_2024}.
As no method for multivariate models is available within the \texttt{h2o} framework, we fit these algorithms separately for \textsc{ndvi}, and \textsc{rr}, considering the same explanatory variables. Then, we compute the empirical correlation among the predictions.
Table~\ref{tab:analysis2_ai} compares the different methods with regard to computational costs, predictive performances, and empirical correlation, averaged over the $5$ disjoint held-out sets. These metrics are computed for each held-out set and summarized in terms of their means and standard deviations in the row below each model.

Predictive accuracy is assessed through three complementary criteria. The root mean square prediction error (\textsc{rmspe}), reported separately for \textsc{ndvi}, \textsc{rr}, and their average, measures pointwise accuracy of the posterior mean predictions. The continuous ranked probability score (\textsc{crps}) is a strictly proper scoring rule \citep{gneiting_strictly_2007} that evaluates the full predictive distribution against each scalar observation, rewarding both sharpness and calibration; lower values indicate more accurate probabilistic predictions. The energy score (\textsc{es}) extends this evaluation to the multivariate setting, assessing the joint predictive distribution over both responses simultaneously; particularly informative here with multivariate outcomes.
The \textsc{dbps} dominates both running time and predictive performances for all three metrics.
With $K=2,000$, it achieves \textsc{rmspe} of $(0.208, 0.163)$, \textsc{crps} of $(0.116,0.080)$, and \textsc{es} of $0.150$, substantially outperforming all competitors. Extremely close performance was achieved with $K=4,000$, still dominating other models. The gap is particularly pronounced relative to machine learning methods, which achieve comparable values among themselves, but consistently yield \textsc{rmspe},\textsc{crps}, and \textsc{es} values roughly two to three times larger than \textsc{dbps}. The Bayesian conjugate multivariate linear model (\textsc{blm}), while offering a proper predictive distribution, shows significantly lower (by almost 100\%) pointwise accuracy, confirming the importance of spatial dependence modeling. 
For \textsc{dbps} and \textsc{blm}, Table~\ref{tab:analysis2_ai} also reports posterior estimates of $\varrho_{\textsc{ndvi}}$ and $\varrho_{\textsc{rr}}$. The results reveal a well-documented negative association between the two responses. Indeed, the spatial patterns in these indices are almost the reverse of each other as revealed in Figures~\ref{fig:res_analysis2_ndvi}~and~\ref{fig:res_analysis2_rr}.
The conjugate Bayesian linear model estimates a higher negative correlation, while \textsc{dbps} tends to underestimate it, considerably so as $n_{k}$ decreases.

Recalling that all the analyses are produced on a personal laptop (with just 5 physical cores) with minimal human intervention, the total run time of only 4.5 minutes with $K=4,000$ and 15 minutes with $K=2,000$ for \textsc{dbps} is impressive and confirms the quadratic dependence of the partition size discussed in Section~\ref{sec:sim5_sens}. Moreover, the strong dependence on the number of $J$ competitive models is worth noting. With $J=4$, this suggests a marginal computational burden for each competitive model.

\begin{table}[t!]
\centering
\scriptsize
\begin{tabularx}{\linewidth}{Z{2.35cm} Z{1.00cm} Z{3.20cm} Z{2.35cm} Z{0.80cm} Z{3.00cm}}
\toprule
\textbf{Model} & \textbf{Time (min)} & \textbf{RMSPE} & \textbf{CRPS} & \textbf{ES} & {\large$\bm{\varrho_{\textsc{ndvi},\textsc{rr}}}$} \\
\midrule
\rowcolor{gray!10}
\textsc{dbps} ($K=2{,}000$) & 15 & [0.208, 0.163, 0.185] & [0.116, 0.080] & 0.150 & -0.905 (-0.918, -0.885) \\
\rowcolor{gray!10}
\textcolor{gray}{\textit{\ \ \textsc{sd}}} & -- & \textcolor{gray}{\textit{[0.0005, 0.0004, 0.0003]}} & \textcolor{gray}{\textit{[0.0003, 0.0001]}} & \textcolor{gray}{\textit{0.0002}} & \textcolor{gray}{0.002 (0.001, 0.002)} \\
\addlinespace[2pt]
\textsc{dbps} ($K=4{,}000$) & 4.5 & [0.233, 0.184, 0.209] & [0.133, 0.094] & 0.173 & -0.900 (-0.920, -0.863) \\
\textcolor{gray}{\textit{\ \ \textsc{sd}}} & -- & \textcolor{gray}{\textit{[0.0005, 0.0010, 0.0007]}} & \textcolor{gray}{\textit{[0.0001, 0.0004]}} & \textcolor{gray}{\textit{0.0003}} & \textcolor{gray}{0.003 (0.002, 0.005)} \\
\addlinespace[2pt]
\rowcolor{gray!10}
\textsc{blm} & 3 & [0.471, 0.394, 0.432] & [0.276, 0.229] & 0.370 & -0.921 (-0.927, -0.916) \\
\rowcolor{gray!10}
\textcolor{gray}{\textit{\ \ \textsc{sd}}} & -- & \textcolor{gray}{\textit{[0.0010, 0.0009, 0.0009]}} & \textcolor{gray}{\textit{[0.0007, 0.0006]}} & \textcolor{gray}{\textit{0.0008}} & \textcolor{gray}{0.0001 (0.0001, 0.0001)} \\
\addlinespace[2pt]
\textsc{lmc} & -- & [--, --, --] & [--, --] & -- &-- (--, --) \\
\textcolor{gray}{\textit{\ \ \textsc{sd}}} & -- & \textcolor{gray}{[--, --, --]} & \textcolor{gray}{[--, --]} & -- & \textcolor{gray}{-- (--, --)} \\
\addlinespace[2pt]
\rowcolor{gray!10}
\textsc{nngp} (m = 5) & -- & [--, --, --] & [--, --] & -- & -- (--, --) \\
\rowcolor{gray!10}
\textcolor{gray}{\textit{\ \ \textsc{sd}}} & -- & \textcolor{gray}{[--, --, --]} & \textcolor{gray}{[--, --]} & -- & \textcolor{gray}{-- (--, --)} \\
\addlinespace[2pt]
\textsc{bart}  & -- & [--, --, --] & [--, --] & -- & -- (--, --) \\
\textcolor{gray}{\textit{\ \ \textsc{sd}}} & -- & \textcolor{gray}{[--, --, --]} & \textcolor{gray}{[--, --]} & -- & \textcolor{gray}{-- (--, --)} \\
\addlinespace[2pt]
\rowcolor{gray!10}
\textsc{gbm} & 3 & [0.420, 0.346, 0.383] & [0.353, 0.275] & 0.463 & -- (--,--) \\
\rowcolor{gray!10}
\textcolor{gray}{\textit{\ \ \textsc{sd}}} & -- & \textcolor{gray}{\textit{[0.0015, 0.0012, 0.0014]}} & \textcolor{gray}{\textit{[0.0011, 0.0009]}} & \textcolor{gray}{\textit{0.0013}} & \textcolor{gray}{-- (--, --)} \\
\addlinespace[2pt]
\textsc{drf} & 10 & [0.421, 0.347, 0.384] & [0.353, 0.275] & 0.463 & -- (--,--)\\
\textcolor{gray}{\textit{\ \ \textsc{sd}}} & -- & \textcolor{gray}{\textit{[0.0016, 0.0013, 0.0014]}} & \textcolor{gray}{\textit{[0.0011, 0.0010]}} & \textcolor{gray}{\textit{0.0014}} & \textcolor{gray}{-- (--, --)} \\
\addlinespace[2pt]
\rowcolor{gray!10}
\textsc{dnn} & 26 & [0.420, 0.347, 0.383] & [0.354, 0.274] & 0.463 &-- (--,--)\\
\rowcolor{gray!10}
\textcolor{gray}{\textit{\ \ \textsc{sd}}} & -- & \textcolor{gray}{\textit{[0.0017, 0.0013, 0.0014]}} & \textcolor{gray}{\textit{[0.0012, 0.0009]}} & \textcolor{gray}{\textit{0.0014}} & \textcolor{gray}{-- (--, --)} \\
\addlinespace[2pt]
\textsc{automl} & 44 & [0.420, 0.346, 0.383] & [0.353, 0.275] & 0.463 & -- (--,--)\\
\textcolor{gray}{\textit{\ \ \textsc{sd}}} & -- & \textcolor{gray}{\textit{[0.0015, 0.0013, 0.0014]}} & \textcolor{gray}{\textit{[0.0011, 0.0009]}} & \textcolor{gray}{\textit{0.0013}} & \textcolor{gray}{-- (--, --)} \\
\bottomrule
\end{tabularx}
\caption{Vegetation Index data analysis computing time in minutes, \textsc{rmspe}, \textsc{crps}, \textsc{es}, and empirical correlation ($\varrho$) for candidate models. Root mean square prediction error(s) presentation [\textsc{ndvi}, red reflectance, average], continuous ranked probability score (s) presentation [\textsc{ndvi}, red reflectance]. Metrics are averaged over 5 different holdout sets and the row below each model reports their standard deviation across the 5 holdout sets.}
\label{tab:analysis2_ai}
\end{table}

%%%%%%%%%%%%%%%%%%%%%%%%%%%%%%%%%%%%%%%%%%%%
\section{Discussion}\label{sec:discussion}
This manuscript devises a statistical modeling component for an artificially intelligent geospatial system (\textsc{GeoAI}) geared to analyze massive data with minimal human intervention. The contribution harnesses analytically accessible multivariate statistical distributions in conjunction with Bayesian predictive stacking to deliver rapid inference by circumventing iterative algorithms that often require extensive human tuning. Our proposed GeoSpatial AI system relies on Bayesian transfer learning using double Bayesian predictive stacking to process massive amounts of streaming spatial data on high-performance \textsc{cpu} architectures.    

Some additional remarks are warranted. The development here has been elucidated with a hierarchical matrix-variate spatial process framework. Although modeling simplifications have been introduced to minimize human intervention, we emphasize that \textsc{dbps} seamlessly applies to more versatile but analytically intractable models. For spatially misaligned multivariate data, where not all variables have been observed in the same set of locations, we encounter missing entries in $Y$ using analytically tractable closed-form distribution theory \citep{zhang_spatial_2022}. Should users wish to explore more complex models with unknown $\{\alpha,\phi\}$ or work with alternate multivariate models, we can implement Bayesian inference (using \textsc{mcmc}, Variational Bayes, or Laplace approximations) on each subset and invoke \textsc{bps} to pool across the subsets of the data. The time for analyzing each subset increases due to model complexity, but scaling inference to massive datasets is still achieved by \textsc{bps}---and this is achieved with considerable efficiency compared to divide-and-conquer approaches such as the geometric median, or \textsc{gm}, of posteriors \citep[see meta-kriging in][]{guhaniyogi_meta-kriging_2018}. 

Future research can build upon recent work by \citet{cabel_bayesian_2025} to enrich spatial-temporal dependence structures within \textsc{dbps} and further accelerate amortized inference.
Section~\ref{sec:AccelerMultivarMod} devises posterior inference as a ``mixture of mixtures''. Following \citet{yao_using_2018}, we prefer stacking to fitting a mixture model because the former is numerically robust and requires almost no human adjustment. Moreover, the structure of ``mixture of mixtures'' conceptually resembles the ``Mixture of Experts'' (\textsc{MoE}) that are adopted by AI platforms such as \textsc{gpt}-4 and Mistral. Disseminating our proposed methodology with our accompanying software, which is currently being migrated to \texttt{R}, is expected to significantly boost \textsc{Geospatial AI} systems.  Future directions will also explore perceived potential of \textsc{dbps} as a feeder for emerging amortized inference methods \citep{ganguly_amortized_2023, zammit-mangion_neural_2024, sainsbury-dale_neural_2024} to achieve Bayesian inference. Rapid delivery of posterior estimates of the entire spatial process from \textsc{dbps} will amount to more training data for amortized neural learners that can result in accelerated tuning for subsequent Bayesian inference. We do not see our proposed approach as a competitor to, but rather as supplementary to, amortized neural inference. See Section~\ref{sec:sim_abi} for an effective representation.
Such developments will be pursued as future research. We also seek to expand and fully investigate automated \textsc{double bps} using Markovian graphical structures across data subsets to further expedite and improve \textsc{Geospatial AI} systems. 

%%%%%%%%%%%%%%%%%%%%%%%%%%%%%%%%%%%%%%%%%%%%
% \section*{Supplementary Materials}
% The online supplement includes exploratory data analyses, and automated settings for model hyperparameters without human intervention (Section~\ref{sec:eda}), additional derivations and details on matrix-variate distributions (Section~\ref{sec:distributiontheory}),
% some theoretical insights (Section~\ref{sec:upperbound}), computational details and algorithms (Section~\ref{sec:compdetails}), and additional simulation experiment results (Section~\ref{sec:sim_suppl_M}). Computer programs to reproduce all our analysis are publicly accessible from the GitHub repository \href{https://github.com/lucapresicce/Bayesian-Transfer-Learning-for-GeoAI}{lucapresicce/Bayesian-Transfer-Learning-for-\textsc{GeoAI}}.

%%%%%%%%%%%%%%%%%%%%%%%%%%%%%%%%%%%%%%%%%%%%
% Acknowledgements and Disclosure of Funding should go at the end, before appendices and references

% \acks{All acknowledgements go at the end of the paper before appendices and references.
% Moreover, you are required to declare funding (financial activities supporting the
% submitted work) and competing interests (related financial activities outside the submitted work).
% More information about this disclosure can be found on the JMLR website.}

\acks{Sudipto Banerjee acknowledges funding from the National Science Foundation through the Division of Mathematical Sciences (DMS grants 2113778 and 2515898); and funding from the National Institute of Health through the National Institute of General Medical Sciences (NIGMS grant R01GM148761).}

%%%%%%%%%%%%%%%%%%%%%%%%%%%%%%%%%%%%%%%%%%%%
\newpage


\appendix

%%%%%%%%%%%%%%%%%%%%%%%%%%%%%%%%%%%%%%%%%
\section{Theoretical Derivations}\label{sec:theory_derivations}

%-------------------------------------------
\subsection{Posterior and Predictive Matrix-Variate T Distributions}\label{sec:matrixT}

The joint posterior predictive for $Y_\mathcal{U}$ and the unobserved latent process $\Omega_{\mathcal{U}}$, can be recast by integrating out $\{\gamma,\Sigma\}$ from the joint conditional posterior predictive, that is
\begin{align*}
 p\left(Y_{\mathcal{U}}, \Omega_{\mathcal{U}} \mid\mathscr{D}\right)= \int& \mathrm{MN}_{{n^{\prime}}, q}\left(Y_{\mathcal{U}} \mid X_{\mathcal{U}} \beta+\Omega_{\mathcal{U}},\left(\alpha^{-1}-1\right) \mathbb{I}_{{n^{\prime}}}, \Sigma\right) \times \operatorname{MN}_{{n^{\prime}}, q}\left(\Omega_{\mathcal{U}} \mid M_{\mathcal{U}} \Omega,  V_{\Omega_{\mathcal{U}}}, \Sigma\right)
 \\ &\times \operatorname{MNIW}\left(\gamma, \Sigma \mid \mu_{\gamma}^{\star} ,  V^{\star}_{\gamma}, \Psi^{\star}, \nu^{\star}\right) d \gamma d \Sigma,   
\end{align*}
where $M_{\mathcal{U}}=\rho_\phi(\mathcal{U}, \mathcal{S}) \rho_\phi^{-1} (\mathcal{S}, \mathcal{S})$ and $ V_{\Omega_{\mathcal{U}}}=\rho_\phi(\mathcal{U}, \mathcal{U})-\rho_\phi(\mathcal{U}, \mathcal{S}) \rho_\phi^{-1}(\mathcal{S}, \mathcal{S}) \rho_\phi(\mathcal{S}, \mathcal{U})$. % The integration can be avoided by exploiting augmented linear systems and distribution theory results. 
We derive $p(\Omega_{\mathcal{U}}, Y_{\mathcal{U}} \mid \Sigma, \mathscr{D})$ by avoiding direct integration with respect to $\gamma$ using the following augmented linear system
\begin{equation*}\label{eqn:augmented_system_suppl}
    \underbrace{\begin{bmatrix}
        \Omega_{\mathcal{U}} \\
        Y_\mathcal{U} 
    \end{bmatrix}}_{\Upsilon} =
    \underbrace{\begin{bmatrix}
        0_{n^{\prime} \times p} & M_{\mathcal{U}} \\
        X_\mathcal{U} & M_{\mathcal{U}}
    \end{bmatrix}}_{M}
    \underbrace{\begin{bmatrix}
        \beta \\
        \Omega
    \end{bmatrix}}_{\gamma}
    + \underbrace{\begin{bmatrix}
        E_{\Omega u} \\
        E_{Yu} 
    \end{bmatrix}}_{E}, \;\;
    E \sim \operatorname{MN}_{2{n^{\prime}},q}\left(0_{2n^{\prime} \times q}, V_{E}, \Sigma\right)\;,
\end{equation*}
where $V_{E} = \begin{bmatrix}
        V_{\Omega_{\mathcal{U}}} & V_{\Omega_{\mathcal{U}}} \\
        V_{\Omega_{\mathcal{U}}} & (\alpha^{-1}-1)\mathbb{I}_{n^{\prime}} + V_{\Omega_{\mathcal{U}}} 
    \end{bmatrix}$.
We write the posterior distribution $p(\gamma\mid\mathscr{D})$ in Equation~\eqref{eq:MultivLatPost} as a linear equation, $\gamma = \mu_{\gamma}^{\star} + E_{\gamma}$, with $E_{\gamma} \sim \operatorname{MN}_{p+n,q}(O, V^{\star}_{\gamma}, \Sigma)$, where $E$ and $E_{\gamma}$ are independent of each other. Then,
\begin{equation}\label{eq:JointCondPred}
    \Upsilon = M \mu_{\gamma}^{\star} + M E_{\gamma} + E \sim \operatorname{MN}_{2n^{\prime}, q}\left(M \mu_{\gamma}^{\star}, V^{\star},\Sigma\right),
\end{equation}
where $V^{\star} = MV^{\star}_{\gamma} M^\top + V_E$. This yields $p(\Upsilon\mid\Sigma,\mathscr{D})=p(\Omega_{\mathcal{U}},Y_{\mathcal{U}}\mid\Sigma,\mathscr{D})$ as the closed-form joint predictive distribution by integrating out $\Sigma$ from $p(\Omega_{\mathcal{U}},Y_{\mathcal{U}}\mid\Sigma,\mathscr{D}) p(\Sigma\mid\mathscr{D})$ to get
\begin{equation*}
    \int \operatorname{MNIW}(\Omega_{\mathcal{U}},Y_{\mathcal{U}},\Sigma\mid M\mu_{\gamma}^{\star}, V^{\star}, \Psi^{\star}, \nu^{\star}) \; d\Sigma = \operatorname{T}_{2n^{\prime},q}(\nu^{\star}, M\mu_{\gamma}^{\star}, V^{\star}, \Psi^{\star})\,
\end{equation*}
which is a matrix-variate %Student's 
$\operatorname{t}$ random variable.
Defining $\Upsilon=[\Omega_{\mathcal{U}}^{\T}, Y_{\mathcal{U}}^{\T}]^{\T}$ as a matrix of dimension $m\times q$, where $m=2n^{\prime}$, the predictive distribution is
\begin{equation*}\label{eq:integralpredictive}
    p(\Upsilon\mid Y) = \int P(\Upsilon,\Sigma\mid Y) d\Sigma\;.
\end{equation*}
This matrix-variate integral can be avoided by simply writing
\begin{equation}\label{eq:alterDeriv}
    p(\Upsilon\mid Y) = \frac{p(\Upsilon,\Sigma\mid Y)}{p(\Sigma\mid\Upsilon,Y)}.
\end{equation}
The density $p(\Upsilon,\Sigma\mid Y)$ comes from Equation~\eqref{eq:JointCondPred}, while the denominator is obtained as
\begin{equation*}
\begin{aligned}
    p(\Sigma\mid\Upsilon,Y&) = \frac{p(\Sigma\mid Y) p(\Upsilon\mid\Sigma,Y)}{p(\Upsilon\mid Y)}\\
    &\propto \frac{ \left|\Psi^{\star}\right|^{\frac{\nu^{\star}}{2}} \left|V_{\gamma}^{\star}\right|^{-\frac{q}{2}} \left|\Sigma\right|^{-\frac{\nu^{\star}+m+q+1}{2}} }{ 2^{\frac{(\nu^{\star}+m)q}{2}} (\pi)^{\frac{mq}{2}} \Gamma_{q}\left(\frac{\nu^{\star}}{2}\right)} \exp\left\{ -\frac{1}{2} \operatorname{tr}\left[ \Sigma^{-1}\left(\Psi^{\star} + (\Upsilon - \mu^{\star})^{\T}V_{\gamma}^{\star -1}(\Upsilon - \mu^{\star})\right) \right]\right\} \\
    &\propto \left|\Sigma\right|^{-\frac{\nu^{\star}+m+q+1}{2}} \exp\left\{ -\frac{1}{2} \operatorname{tr}\left[ \Sigma^{-1}\left(\Psi^{\star} + (\Upsilon - \mu^{\star})^{\T}V_{\gamma}^{\star -1}(\Upsilon - \mu^{\star})\right) \right]\right\},
\end{aligned}
\end{equation*}
where $\mu^{\star}=M\mu_{\gamma}^{\star}$. Hence, $\Sigma\mid\Upsilon,Y \sim \operatorname{IW}(\hat{\Psi},\hat{\nu})$ with $\hat{\Psi}=\left(\Psi^{\star} + (\Upsilon - \mu^{\star})^{\T}V_{\gamma}^{\star -1}(\Upsilon - \mu^{\star})\right)$, and $\hat{\nu}=\nu^{\star}+m$. The joint posterior predictive density then follows from Equation~\eqref{eq:alterDeriv}
\begin{equation*}
\begin{aligned}
    p(\Upsilon\mid Y) &= \frac{\operatorname{MNIW}(\Upsilon,\Sigma\mid\mu^{\star}, V_{\gamma}^{\star}, \Psi^{\star}, \nu^{\star})}{\operatorname{IW}(\Sigma\mid\hat{\Psi},\hat{\nu})} \\
    &= K(\Upsilon) \left|\Psi^{\star}\right|^{-\frac{\nu^{\star}+m}{2}} 
    \left| \mathbb{I}_{m} +  V_{\gamma}^{\star -1}(\Upsilon - \mu^{\star}) \Psi^{\star -1} (\Upsilon - \mu^{\star})^{\T}  \right|^{-\frac{\nu^{\star}+m}{2}}\;, \\
\end{aligned}
\end{equation*}
where $K(\Upsilon)= \frac{\Gamma_{q}\left(\frac{\hat{\nu}}{2}\right) \left|\Psi^{\star}\right|^{\frac{\nu^{\star}}{2}} \left|V_{\gamma}^{\star}\right|^{-\frac{q}{2}}}{\Gamma_{q}\left(\frac{\nu^{\star}}{2}\right)(\pi)^{\frac{mq}{2}} } = \frac{\Gamma_{q}\left(\frac{\nu^{\star}+m}{2}\right) \left|\Psi^{\star}\right|^{\frac{\nu^{\star}}{2}} \left|V_{\gamma}^{\star}\right|^{-\frac{q}{2}}}{ \Gamma_{q}\left(\frac{\nu^{\star}}{2}\right) (\pi)^{\frac{mq}{2}} }$, since $\hat{\nu}=\nu^{\star}+m$. 

This is a matrix-variate T density, which we denote as $\Upsilon\mid Y \sim \operatorname{T}_{m,q}(\nu^{\star}, \mu^{\star}, V^{\star}, \Psi^{\star})$. We recover the same result, without needing to integrate out $\Sigma$ \citep{iranmanesh_conditional_2010,gupta_matrix_2000}, by using only Bayes theorem and related distribution theory. Finally, the marginal predictive distributions $\Omega_{\mathcal{U}}\mid\mathscr{D}$ and $Y_{\mathcal{U}}\mid\mathscr{D}$ are also available in analytic form as matrix T distributions for any set of predictive points $\mathcal{U}$.

%-------------------------------------------
\subsection{Details on Disagreement Tempering}\label{sec:disagree_derivation}

Double Bayesian predictive stacking produces posterior predictive distributions that are expressed as mixtures. We can then take advantage of linear pooling \citep[see, e.g.,][]{knuppel_forecast_2022} to mitigate inflated predictive variances due to disagreements between the mixture and the true posterior. Unlike standard linear pooling, \textsc{dbps} posterior predictive distributions are mixtures of a finite mixture distribution that includes a second layer of mixing. We adapt to this setting and show how disagreement tempering (\textsc{dt}) can mitigate over-dispersion in our model. 

The \textsc{dbps} posterior predictive distribution $\hat{p}(Y_{\mathcal{U}}\,;\,\mathscr{D})$ in \eqref{eq:fullPredictiveYU} can also be written as 
\begin{equation*}
    \hat{p}(Y_{\mathcal{U}}\,;\,\mathscr{D})=\sum_{k=1}^{K}w_{k}\;\hat{p}(Y_{\mathcal{U}}\,;\,\mathscr{D}_{k}) = \sum_{k=1}^{K}w_{k}\;\sum_{j=1}^{J}z_{k,j}\;p(Y_{\mathcal{U}}\mid \mathscr{D}_{k}, \mathscr{M}_{j}).
\end{equation*}
Each component in the mixture is a matrix-variate $\operatorname{T}$ distribution $p(Y_{\mathcal{U}}\mid \mathscr{D}_{k}, \mathscr{M}_{j})=\operatorname{T}_{n^{\prime},q}(\nu_{k}^{\star}, \mu_{k,j}^{\star}, V_{k,j}^{\star}, \Psi_{k,j}^{\star})$ with degrees of freedom $\nu_{k}^{\star}=\nu_{0}+n_{k}$, an $n^{\prime}\times q$ location matrix $\mu_{k,j}^{\star}=M_{y,k,j}\mu_{\gamma,k,j}$, an $n^{\prime}\times n^{\prime}$ row-scale matrix $V_{k,j}^{\star}$, and a column-scale matrix $\Psi_{k,j}^{\star}$ of dimension $q\times q$, where all the parameters are defined in \eqref{eq:MultivLatPost},~\eqref{eq:MultivLatPred}, and \eqref{eq:subsetOptMatrixT}.

Consider now the vectorization of $Y_{\mathcal{U}}$, i.e., $\tilde{Y}_{\mathcal{U}}=\operatorname{vec}(Y_{\mathcal{U}})$, and denote $m_{k,j}=\operatorname{vec}(\mu_{k,j}^{\star})$, $\bar{m}_{k}=\sum_{j=1}^{J}z_{k,j}\;m_{k,j}$ and $\bar{m}=\sum_{k=1}^{K}w_{k}\;\bar{m}_{k}$. Under the properties of the matrix-variate $\operatorname{t}$ distribution, $\tilde{Y}_{\mathcal{U}}$ is a multivariate %Student's 
$\operatorname{t}$, and given that $\nu_{k}^{\star}=\nu_{0}+n_{k}>2$ is satisfied under our prior choice $\nu_{0}=3$, its conditional covariance matrix is
\begin{equation*}
    \operatorname{Var}(\tilde{Y_{\mathcal{U}}}\mid \mathscr{D}_{k}, \mathscr{M}_{j}) = \frac{\nu_{k}^{\star}}{\nu_{k}^{\star}-2}\;\Psi_{k,j}^{\star}\otimes V_{k,j}^{\star}.
\end{equation*}

The posterior predictive covariance can then be derived as
{\footnotesize
\begin{align*}
    \operatorname{Var}(\tilde{Y}_{\mathcal{U}}\;;\;\mathscr{D})&=\sum_{k=1}^{K}w_{k}\operatorname{Var}(\tilde{Y}_{\mathcal{U}}\;;\;\mathscr{D}_{k}) + \sum_{k=1}^{K}w_{k}\bar{m}_{k}\bar{m}_{k}^{\top} - \bar{m}\bar{m}^{\top}\\
    &= \sum_{k=1}^{K}w_{k} \left[ \sum_{j=1}^{J}z_{k,j}\operatorname{Var}(\tilde{Y}_{\mathcal{U}}\mid \mathscr{D}_{k}, \mathscr{M}_{j}) + \sum_{j=1}^{J}z_{k,j}m_{k,j}m_{k,j}^{\top} - \bar{m}_{k}\bar{m}_{k}^{\top} \right] + \sum_{k=1}^{K}w_{k}\bar{m}_{k}\bar{m}_{k}^{\top} - \bar{m}\bar{m}^{\top}.
\end{align*}
}
Expanding all terms and simplifying some of them, we get the complete formulation
{\footnotesize
\begin{align*}
    &= \sum_{k=1}^{K}w_{k}\sum_{j=1}^{J}z_{k,j}\operatorname{Var}(\tilde{Y}_{\mathcal{U}}\mid \mathscr{D}_{k}, \mathscr{M}_{j}) + \sum_{k=1}^{K}w_{k}\sum_{j=1}^{J}z_{k,j}m_{k,j}m_{k,j}^{\top} - \sum_{k=1}^{K}w_{k}\bar{m}_{k}\bar{m}_{k}^{\top} \\
    &+ \sum_{k=1}^{K}w_{k}\bar{m}_{k}\bar{m}_{k}^{\top} - \bar{m}\bar{m}^{\top} \\
    &= \sum_{k=1}^{K}w_{k}\sum_{j=1}^{J}z_{k,j}\operatorname{Var}(\tilde{Y}_{\mathcal{U}}\mid \mathscr{D}_{k}, \mathscr{M}_{j}) + \sum_{k=1}^{K}w_{k}\sum_{j=1}^{J}z_{k,j}m_{k,j}m_{k,j}^{\top} - \bar{m}\bar{m}^{\top} \\
    &= \sum_{k=1}^{K}w_{k}\sum_{j=1}^{J}z_{k,j}\frac{\nu_{k}^{\star}}{\nu_{k}^{\star}-2}\;\Psi_{k,j}^{\star}\otimes V_{k,j}^{\star} + \underbrace{\sum_{k=1}^{K}w_{k}\sum_{j=1}^{J}z_{k,j}m_{k,j}m_{k,j}^{\top} - \bar{m}\bar{m}^{\top}}_{\operatorname{Dis}\left(\{m_{k,j}\}_{k,j}\right)} \\
    &= \sum_{k=1}^{K}w_{k}\sum_{j=1}^{J}z_{k,j}\frac{\nu_{k}^{\star}}{\nu_{k}^{\star}-2}\;\Psi_{k,j}^{\star}\otimes V_{k,j}^{\star} + \operatorname{Dis}\left(\{m_{k,j}\}_{k,j}\right)\;.
\end{align*}
} 
The disagreement term is the covariance matrix of the discrete random vector that takes value $m_{k,j}=\operatorname{vec}(\mu_{k,j}^{\star})$ with probability $w_{k}z_{k,j}$: it measures how much the location matrices $\mu_{k,j}^{\star}$ vary across the components of the mixture. As a covariance matrix, it satisfies $\operatorname{Dis}\left(\{m_{k,j}\}_{k,j}\right) \succeq 0$, where $A \succeq 0$ denotes that a matrix $A$ is positive semidefinite.

The idea of disagreement tempering, which can be considered a heuristic adjustment in machine learning, is to consider the individual posterior predictive distributions so that the new posterior predictive means are all equal to a constant, i.e., $\mathbb{E}_{\text{adjusted}}[Y_{\mathcal{U}}\mid \mathscr{D}_{k}, \mathscr{M}_{j}] = C$ for all $k,j$. If we are able to do this, then this will lead to $\operatorname{Dis}\left(\{m_{k,j}\}_{k,j}\right)=0$ and deflate the covariance matrix from the mixing. This ``disagreement tempered'' predictive covariance is
\begin{equation*}
\sum_{k=1}^{K}w_{k}\sum_{j=1}^{J}z_{k,j}\frac{\nu_{k}^{\star}}{\nu_{k}^{\star}-2}\;\Psi_{k,j}^{\star}\otimes V_{k,j}^{\star} = \operatorname{Var}(\tilde{Y}_{\mathcal{U}}\;;\;\mathscr{D}) - \operatorname{Dis}\left(\{m_{k,j}\}_{k,j}\right).
\end{equation*}
In practice, the centering is implemented as follows. For each $r=1,\ldots,R$, we sample %$(k^{(r)}, j^{(r)})$ according to $(\hat w_k, \hat z_{k,j})$ and draw 
$Y_{\mathcal{U}}^{(r)}\sim\operatorname{T}_{n^{\prime},q}(\nu_{k^{(r)}}^{\star}, \mu_{k^{(r)},j^{(r)}}^{\star}, V_{k^{(r)},j^{(r)}}^{\star}, \Psi_{k^{(r)},j^{(r)}}^{\star})$ and record the corresponding posterior predictive mean $\mu_{k^{(r)},j^{(r)}}^{\star} = \mathbb{E}\left[ {Y}_{\mathcal{U}} \mid \mathscr{D}_{k^{(r)}}, \mathscr{M}_{j^{(r)}} \right]$ as the mean of the sampled matrix-variate $t$ component. Each predictive draw is then centered by subtracting $\mu_{k^{(r)},j^{(r)}}^{\star}$, and a common global mean $\mu_{C}^{\star} = \frac{1}{R}\sum_{r=1}^R \mu_{k^{(r)},j^{(r)}}^{\star}$ is added back to obtain the disagreement-tempered predictive samples. Algorithm~\ref{alg:disagreement} formalizes this procedure. Note that no explicit averaging over $K$ and $J$ is required, since the sampling at each iteration already accounts for the mixture weights, as is standard sampling from mixture models.

%%%%%%%%%%%%%%%%%%%%%%%%%%%%%%%%%%%%%%%%%%%%
\section{Computational Details}\label{sec:compdetails}

Key computational aspects of the proposed method involve two points: comparing its theoretical complexity to state-of-the-art approaches and addressing memory constraints. The complexity comparison evaluates the method’s time and space efficiency, particularly its scalability with larger datasets. This includes insights from the explicit objective function, highlighting its computational impact. Memory constraints are equally critical, as limitations can hinder performance despite powerful processors. The proposed method addresses these challenges, ensuring both scalability and efficient resource use.

In summary, this section will examine both the theoretical complexity, including the explicit derivation of the objective function for the optimization problems detailed in Equations~\eqref{eq:subsetOptProb} and \eqref{eq:combOptProb}, as well as the memory management strategies, offering a comprehensive view of the computational feasibility in practical applications.

%-------------------------------------------
\subsection{Objective Function for Double Bayesian Predictive Stacking}\label{sec:BPSobjfun}

We expound the double Bayesian predictive stacking in Section~\ref{sec:AccelerMultivarMod}. The optimization problem used to compute the stacking weights in Equation~\eqref{eq:doubleBPSpred} is formally defined as: %Equation~\eqref{eq:combOptProb}.
\begin{equation}\label{eq:combOptProb}
    \max _{ w \in  S_1^K} \frac{1}{n} \sum_{i=1}^{n} \log \sum_{k=1}^K w_{k} \hat{p}\left(Y_{i} \;;  \mathscr{D}_{k}\right) = \max _{ w \in  S_1^K} \frac{1}{n} \sum_{i=1}^{n} \log \sum_{k=1}^K w_{k} \sum_{j=1}^{J} \hat{z}_{k,j} p\left(Y_{k,i} \mid  \mathscr{D}_{k,[-l]}, \mathscr{M}_j \right),
\end{equation}
as $\hat{p}\left(Y_{i} \;;  \mathscr{D}_{k}\right)=\sum_{j=1}^{J} \hat{z}_{k,j} p\left(Y_{k,i} \mid  \mathscr{D}_{k,[-l]}, \mathscr{M}_j \right)$.
In this \textsc{double bps} framework, we focus exclusively on $\mathscr{D}_{k}$, which is treated equivalently to $\mathscr{M}_{j}$ in the first step. It is crucial to discriminate the predictive performance induced by each $\mathscr{D}_{k}$. We must use a common set of $Y$ across all $\mathscr{D}_{k}$---namely, $Y$ itself. This stems from the construction of \textsc{double bps}. Specifically, for \textsc{double bps} to be effective, it necessitates predictive assessments over a common set of points for each model in the competition. Otherwise, the predictive performances cannot be directly compared, and the weights cannot be optimized to distinguish predictive capabilities across models, as different points would be used for different models.

To illustrate, consider the first stacking step. Here, we compute $p\left(Y_{k,i} \mid \mathscr{D}_{k},\mathscr{M}_{j}\right)$ for each subset, where $i=1,\dots,n_{k}$ and $j=1,\dots,J$. This allows us to evaluate $Y_{k}$ with respect to the predictive density of all $J$ models under consideration. Similarly, in \textsc{double bps}, the goal is to evaluate $Y_{i}$ with respect to the predictive density across all $K$ subsets (acting as competing models) for comparison.
The weights $\{\hat{z}_{k,j}\}$, which are derived from the optimization problem specified in Equation~\eqref{eq:subsetOptProb} also appear in Equation~\eqref{eq:combOptProb}. However, comparing the right-hand sides of \eqref{eq:combOptProb}~and~\eqref{eq:subsetOptProb}, we observe that the objective functions are almost identical, with the only difference being the second convex linear combination governed by the weights $\{w_{k}\}$. Therefore, the predictive distributions in both optimization problems refer to the same quantity. To summarize, the objective function in Equation~\eqref{eq:subsetOptProb} can be derived by substituting each $\{z_{k,j}\}$ with its optimized counterpart $\{\hat{z}_{k,j}\}$ and incorporating the weights $\{w_{k}\}$. This leads to the maximization objective in \eqref{eq:combOptProb}. 

Next, we consider the optimization problem in \eqref{eq:subsetOptProb} with the objective function,
\begin{equation*}\label{eq:subsetobjfunDefinition}
    \max _{ z_k \in  S_1^J} \frac{1}{n_k} \sum_{i=1}^{n_k} \log \sum_{j=1}^J z_{k,j} p\left(Y_{k,i} \mid  \mathscr{D}_{k,[-l]}, \mathscr{M}_j \right) = \max _{ z_k \in  S_1^J}  f\left( z_{k} \right),
\end{equation*}
where $f\left( z_{k} \right)=\frac{1}{n_k} \sum_{i=1}^{n_k} \log \sum_{j=1}^J z_{k,j} p\left(Y_{k,i} \mid  \mathscr{D}_{k,[-l]}, \mathscr{M}_j \right)$. An explicit form of $f(z_k)$ is
\begin{equation*}\label{eq:subsetobjfunDerivation}
\begin{aligned}
    f(z_{k} ) &= f\left( z_{k,1},\ldots, z_{k,J}\right) = \frac{1}{n_k} \sum_{i=1}^{n_k} \log \sum_{j=1}^J z_{k,j} p\left(Y_{k,i} \mid  \mathscr{D}_{k,[-l]}, \mathscr{M}_j \right) \\
    &= \frac{1}{n_k} \sum_{i=1}^{n_k} \log \sum_{j=1}^J z_{k,j} \operatorname{T}_{1,q}(Y_{k,i}\mid\nu^{\star}_{[-l]}, \mu^{\star}_{i}, V^{\star}_{i}, \Psi^{\star}_{[-l]}) \\
    &= \frac{1}{n_k} \sum_{i=1}^{n_k} \log \sum_{j=1}^J z_{k,j} K(Y_{k,i}) \left| 1 +  V^{\star -1}_{i}(Y_{k,i} - \mu^{\star}_{i}) \Psi^{\star -1}_{[-l]} (Y_{k,i} - \mu^{\star}_{i})^{\T}  \right|^{-\frac{\nu^{\star}_{[-l]}+1}{2}}, \\
\end{aligned}
\end{equation*}
where $K(Y_{k,i}) = \frac{\left|\Psi^{\star}_{[-l]}\right|^{-\frac{1}{2}} \left|V^{\star}_{i}\right|^{-\frac{q}{2}}\Gamma_{q}\left(\frac{\nu^{\star}_{[-l]}+1}{2}\right) }{(\pi)^{\frac{1q}{2}} \Gamma_{q}\left(\frac{\nu^{\star}_{[-l]}}{2}\right)}$. The logarithm of a linear combination precludes further accessibility, but $f(z_{k})$ is computed easily by evaluating the matrix-T density. This is standard convex optimization \citep{yao_using_2018}; see Section~\ref{sec: comp_programs} for further details.

The objective function in Equation~\eqref{eq:combOptProb} is related to Equation~\eqref{eq:subsetOptProb} as
\begin{equation*}\label{eq:objfunDefinition}
    \max _{ w \in  S_1^K} \frac{1}{n} \sum_{i=1}^{n} \log \sum_{k=1}^K w_{k} \sum_{j=1}^{J} \hat{z}_{k,j} p\left(Y_{k,i} \mid  \mathscr{D}_{k,[-l]}, \mathscr{M}_j \right) = \max _{ w \in  S_1^K} g(w),
\end{equation*}
where $w = (w_{1},\ldots, w_{K})^{\T}$ and
\begin{equation*}\label{eq:objfunDerivation}
\begin{aligned}
    g(w ) &= \frac{1}{n} \sum_{i=1}^{n} \log \sum_{k=1}^K w_{k} \sum_{j=1}^{J} \hat{z}_{k,j} p\left(Y_{k,i} \mid  \mathscr{D}_{k,[-l]}, \mathscr{M}_j \right) \\
    &= \frac{1}{n} \sum_{i=1}^{n} \log \sum_{k=1}^K w_{k} \sum_{j=1}^{J} \hat{z}_{k,j} \operatorname{T}_{1,q}(Y_{k,i}\mid\nu^{\star}_{[-l]}, \mu^{\star}_{i}, V^{\star}_{i}, \Psi^{\star}_{[-l]}) \\
    &= \frac{1}{n} \sum_{i=1}^{n} \log \sum_{k=1}^K w_{k} \sum_{j=1}^{J} \hat{z}_{k,j} K(Y_{k,i}) \left| 1 +  V^{\star -1}_{i}(Y_{k,i} - \mu^{\star}_{i}) \Psi^{\star -1}_{[-l]} (Y_{k,i} - \mu^{\star}_{i})^{\T}  \right|^{-\frac{\nu^{\star}_{[-l]}+1}{2}}. \\
\end{aligned}
\end{equation*}

In practical applications, we address \eqref{eq:combOptProb} separately from \eqref{eq:subsetOptProb}, which is defined for each subset. To evaluate the latter, we perform $K$ separate maximizations in \eqref{eq:subsetOptProb}, one for each subset. Once we obtain all $K$ sets of $\{\hat{z}_{k,j}\}$, we can recover the weights $\{\hat{w}_k\}$ across subsets by solving the convex optimization problem in \eqref{eq:combOptProb}.

Setting aside the potential for parallel computation of the subset stacking weights $\{z_{k,j}\}$, this method offers a significant computational advantage: no additional quantities are required to solve the problem in Equation~\eqref{eq:combOptProb}. All necessary components are already available from the independent computations performed within each subset. Specifically, all terms in Equations~\eqref{eq:subsetOptProb}~and~\eqref{eq:combOptProb} are known and identical except for the weights $w$, which remain to be optimized. Consequently, there is no need to recompute the cross-validated predictive distributions or $\{\hat{z}_{k,j}\}$.

%-------------------------------------------
\subsection{Theoretical Complexity}\label{sec:theocomp}

In terms of theoretical computational complexity, we provide a comparison between spatial meta-kriging \citep[\textsc{smk},][]{guhaniyogi_meta-kriging_2018} and the \textsc{double bps}. Given a dataset $\mathscr{D}_n$, let $n$ denote the total number of observations, $K$ the number of subsets, and $M$ the number of target posterior samples. The approach developed by \citet{guhaniyogi_meta-kriging_2018} generates each posterior sample through a parallel implementation of \textsc{smk} over $K$ cores, which requires $\mathcal{O}\left((\frac{n}{K})^3\right)$ operations. Thus, obtaining $M$ draws from each subset posterior yields a complexity of $\mathcal{O}\left(M(\frac{n}{K})^3\right)$ per computational core. The cost of computing the geometric median must also be added. As stated in \citet{minsker_robust_2017}, Weiszfeld’s algorithm has a complexity of $\mathcal{O}\left(M^2\right)$ per step and requires at most $\mathcal{O}\left(1/\epsilon\right)$ steps to achieve an accuracy within $\epsilon$. However, especially in large-scale applications, allocating a computational core to each subset is not always possible. If we let $m$ denote the number of available cores, where generally $m << K$, the total computational complexity of \textsc{smk} across $K$ partitions distributed over $m$ cores becomes $\mathcal{O}\left(\frac{K}{m}\left[M(\frac{n}{K})^3\right] + \frac{(KM)^2}{\epsilon}\right)$.

% For the theoretical complexity of double Bayesian predictive stacking, we have to specify $J$ as the number of competitive models, and $L$ as the number of folds used for cross-validation. Equivalently to \textsc{smk}, model fitting within subsets is dominated by Cholesky decompositions implying costs in the order of $\mathcal{O}\Big((\frac{n}{K})^3\Big)$. Nevertheless, in \textsc{double bps}, we perform $J$ Cholesky decompositions, and for each of them, we refit the model $L$ times. Hence, the theoretical complexity boils down to $\mathcal{O}\Big(\frac{K}{m} JL (\frac{n}{K})^3\Big)$. 
% In addition, we use the package \textsc{cvxr} \citet{fu_cvxr_2020} in the R statistical computing environment by applying disciplined convex programming \citet{grant_graph_2008, cvx_research_cvx_2012} to find the stacking weights in polynomial time using an interior-point algorithm. We used the solvers \textsc{scs} \citep{odonoghue_conic_2016} and ECOSolveR \citep{fu_ecosolver_2023} to obtain the stacking weights. This introduces the discipline convex problems into the theoretical complexity, turning out to be $\mathcal{O}\Big(\frac{K}{m}[JL(\frac{n}{K})^3 + J^p] + K^p \Big)$, for $K$ subsets over $m$ cores, and a polynomial degree $p$.
% The portion in square brackets pertains to model fitting within each subset, consisting of a term related to cross-validation and the polynomial cost of \textsc{double bps} across $J$ models. Finally, we account for the complexity introduced by the second stacking process across the $K$ subsets.

For the theoretical complexity of double Bayesian predictive stacking, we let $J$ denote the number of competing models and $L$ the number of folds used for cross-validation. Equivalent to \textsc{smk}, model fitting within subsets is dominated by Cholesky decompositions, implying a cost on the order of $\mathcal{O}\left((\frac{n}{K})^3\right)$. However, in \textsc{double bps}, we perform $J$ Cholesky decompositions and refit the model $L$ times for each. Distributed over $m$ cores, the theoretical complexity for this step is $\mathcal{O}\left(\frac{K}{m} JL (\frac{n}{K})^3\right)$. 

In addition, we find the stacking weights in polynomial time using an interior-point algorithm via the \textsc{cvxr} package \citep{fu_cvxr_2020} in the R statistical computing environment, which applies disciplined convex programming \citep{grant_graph_2008, cvx_research_cvx_2012}. We use the solvers \textsc{scs} \citep{odonoghue_conic_2016} and ECOSolveR \citep{fu_ecosolver_2023} to obtain these weights. Incorporating the disciplined convex programming steps into the total computational complexity yields $\mathcal{O}\left(\frac{K}{m}[JL(\frac{n}{K})^3 + J^p] + K^p \right)$ for $K$ subsets distributed over $m$ cores, where $p$ represents the polynomial degree of the solver.

The term in square brackets corresponds to the local operations within each subset, combining the cross-validation model fitting and the polynomial optimization cost across $J$ models. The final $K^p$ term accounts for the complexity introduced by the second stacking process across the $K$ subsets.

Next, we compare the computational complexities of the two approaches. We will separately examine the terms associated with subset modeling and global inference combination. Thus, for \textsc{smk} and \textsc{double bps}, respectively, the computational complexities are as follows:
\begin{equation}\label{eq:theorcomporders}
\mathcal{O}\left(\frac{K}{m}\underbrace{ \left[M\left(\frac{n}{K}\right)^3\right]}_{\mathclap{\text{subset modeling}}} \;\;+\;\;\underbrace{\vphantom{\left[M\left(\frac{n}{K}\right)^3\right]}\frac{(KM)^2}{\epsilon} }_{\mathclap{\text{combination}}}\;\right), \qquad \mathcal{O}\left(\frac{K}{m}\underbrace{\vphantom{\left[M\left(\frac{n}{K}\right)^3\right]}\left[JL\left(\frac{n}{K}\right)^3 + J^p\right]}_{\mathclap{\text{subset modeling}}} \;+\; \underbrace{\vphantom{\left[JL\left(\frac{ n}{K}\right)^3 + J^p\right]} K^p}_{\text{combination}}\;\right)
\end{equation}

Focusing on the subset modeling component, as in \eqref{eq:theorcomporders}, two specifications stand out. First, consider the difference in magnitude between $M$ and the product $JL$. In this context, \textsc{double bps} offers a theoretical advantage when $JL < M$, a quite common condition in practice. This is because $M$ represents the number of posterior samples required for convergence across all the Markov chains involved, and it typically needs to be at least on the order of $10^3$. In contrast, the product $JL$ consists of relatively small terms, making it highly likely that this inequality will hold. Second, due to the significant difference in scale, the term $J^p$ is absorbed by $(n/K)^3$

When comparing the combination phase, the analysis reduces to a comparison between the geometric median approximation and Bayesian predictive stacking. Since a discrete number of posterior samples $M$ is required by \textsc{smk} for each of the $K$ partitions, we generally find that $K^p < (KM)^2 /\epsilon$. Thus, while empirical computational times are significantly lower for double Bayesian predictive stacking compared to \textsc{smk}, there are some modest theoretical differences between the two methods. The major advantage lies in avoiding simulation-based methods, such as \textsc{mcmc} while achieving local inferences through exact approaches.

Like Weiszfeld’s algorithm, modern disciplined convex programming encounters computational challenges in high-dimensional contexts, particularly in managing random memory allocation. In Section~\ref{sec:pBMA}, we present a feasible strategy for approximating the \textsc{double bps} weights, tailored for very large-scale memory problems.

%-------------------------------------------
\subsection{Memory Management and Pseudo-BMA}\label{sec:pBMA}

When modeling GeoAI systems, as the number of locations exceeds the order of millions, managing storage space becomes crucial. Timing issues may arise depending on the available optimizer. While open-source solvers theoretically offer faster solutions compared to iterative algorithms, e.g., geometric median, they often face practical challenges when the problem size considerably exceeds dimensions of $10^2$.
In contrast, commercial optimizers behave slightly better, even if these approaches are not exempt from random allocation memory constraints. We emphasize working with portable approaches, i.e., with open-source solvers, that can effectively handle large-scale problems. 

We present a computationally cheaper alternative that facilitates better management of available \textsc{ram}. The subsequent contents, including Algorithm~\ref{alg:pseudoBMA_weights}, were implemented in data analyses involving millions ($10^6$) of locations in Section~\ref{sec:dataappl}. When addressing optimization problems of significant dimensions, \textsc{aic}-based alternatives could be considered.

\begin{algorithm}[!t]
\caption{Calculating stacking weights between subsets using pseudo-\textsc{bma}}\label{alg:pseudoBMA_weights}

\vspace{0.25em}
\textbf{Input:} $\hat{z} = \{\hat{z}_{k}=\{\hat{z}_{k,j}\}: k \in \{1,\dots,K\}, j \in\{1,\dots,J\} \}$: Stacking weights within subsets; $\{pd_{k,j,i}= \operatorname{T}_{1,q}(Y_{k,i}\mid \nu^{\star}_{[-l]}, \mu^{\star}_{i}, V^{\star}_{i}, \Psi^{\star}_{[-l]}):k=1,\dots,K, j=1,\dots,J, i=1,\dots,n\}$:  point-wise predictive density of $Y$; $n, q, p$: Number of rows, number of outcomes, and number of predictors; $K, \{n_{k}:k\in\{1,\dots,K\}\}, J$: Number of subsets, dimension of each subset, and number of competitive models in each subset.\\
\textbf{Output:} $\hat{w} = \{\hat{w}_{k}: k =1,\dots,K\}$: Stacking weights between subsets.

\begin{algorithmic}[1]
\State Construct $pd=\underbrace{[pd_{1}^{\T}:\cdots:pd_{K}^{\T}]^{\T}}_{n\times J}$, $\;pd_{k}=\underbrace{\begin{bmatrix} 
pd_{k,1,1} & \dots & pd_{k,J,1} \\
\vdots & pd_{k,j,i} & \vdots \\
pd_{k,1,n_{k}} & \dots & pd_{k,J,n_{k}} \\    
\end{bmatrix}}_{n_{k}\times J}$

\For{$k=1,\dots,K$}
\State Compute $\widehat{elpd}^{k} = \sum_{i=1}^{n}\log\big(pd\;\hat{z}_{k}\big)$
\EndFor

\For{$k=1,\dots,K$}
\State Compute $\hat{w}_{k} = \exp\big(\widehat{elpd}^{k}\big)\big/ \sum_{k=1}^{K}\exp\big(\widehat{elpd}^{k}\big)$
\EndFor

\State \textbf{return} $\hat{w}=\{\hat{w}_{k}:k\in\{1,\dots,K\}\}$
\end{algorithmic}
\end{algorithm}

To facilitate model stacking, various methodologies exist within the Bayesian model averaging (\textsc{bma}) framework. In particular, we present an approach based on information criteria, which was formerly introduced in \citet{yao_using_2018}. To ensure comparability between datasets and enhance interpretability, we estimate the expected log point-wise predictive density (as done in \textsc{double bps}). The expected log pointwise predictive density ($elpd$) for each partition $k$ is defined as 
\begin{equation*}\label{eq:elpd_def}
    \widehat{\text{elpd}}^{k} = \sum_{i=1}^{n} \widehat{\text{elpd}}_{i}^{k} = \sum_{i=1}^{n} \log\; \hat{p}(Y_{k,i}\;;\mathscr{D}_{k,[-l]})
\end{equation*}
Importantly, each $elpd$ term need not be computed individually, as these values are generated during the first \textsc{bps} procedure within each subset for all model configurations, significantly reducing memory storage requirements and the total computational burden. 
Given the set $\{\widehat{\text{elpd}}^{k}\}_{k=1,\dots,K}$, the pseudo Bayesian model averaging (pseudo-\textsc{bma}) weights are computed as $\displaystyle \hat{w}_k=\frac{\exp\left(\widehat{\text{elpd}}^{k}\right)}{\sum_{k=1}^{K}\exp\left(\widehat{\text{elpd}}^{k}\right)}$.
\begin{comment}
\begin{equation*}\label{eq:pBMAweights_def}
    \hat{w}_k=\frac{\exp\left(\widehat{\text{elpd}}^{k}\right)}{\sum_{k=1}^{K}\exp\left(\widehat{\text{elpd}}^{k}\right)}.    
\end{equation*}
\end{comment}

This formulation, introduced by \citet{yao_using_2018}, simplifies the computation of the stacking weights, significantly reducing computational costs in terms of complexity and storage while maintaining the Bayesian predictive stacking framework. Thus, it serves as a viable alternative to \textsc{bps} for predictive densities in challenging scenarios. When dealing with datasets comprising millions of instances and a substantial number of partitions, optimization solvers may fail or produce errors due to memory constraints, and the iterative processes involving large matrices can lead to increased procedure times.

Based on empirical experience, we primarily use pseudo-\textsc{bma} as a Bayesian predictive stacking approach when datasets require excessive memory storage, particularly when $n >> 10^5$ and $K >> 10^2$. When feasible, we generally prefer convex optimization using \textsc{bps} of predictive densities without reservations. Simulations highlighting potential differences in posterior predictive and posterior inference performances between these two model stacking approaches can be found in \citet{yao_using_2018}, where several alternatives to Bayesian stacking approaches are discussed. We opted for pseudo-\textsc{bma} due to its simpler analytical formulation, which enables matrix algebra to mitigate the computational burden of both random allocation memory and runtime.

%-------------------------------------------
\subsection{Memory-Efficient Posterior Sampling}\label{sec:smartsampling}

In the matrix-variate conjugate Bayesian linear regression model presented in Equation~\eqref{eq:MatrixNormLik}, the Bayesian updating process may become costly when several data shards arrive, even more so when the dimensions involved are large. Computational problems are often related to the available \textsc{ram}, especially when working with sizable datasets. We present a memory-efficient posterior sampling scheme for the regression coefficient $\beta$ to reduce the computational burden in such contexts. We consider the model in Equation~\eqref{eq:MatrixNormLik} of Section~\ref{sec:DivConBays},
\begin{equation}
\begin{split}
Y &= X\beta + E\;, \quad E\mid\Sigma \sim \operatorname{MN}(O, V, \Sigma)\;;\\
\beta &= M_0m_0 + E_{\beta}\;, \quad E_{\beta}\mid\Sigma \sim \operatorname{MN}(O, M_0, \Sigma)\;; \quad \Sigma \sim \operatorname{IW}(\Psi_0,\nu_0)\;, 
\end{split}
\end{equation}
where $\Sigma$ is assumed to be known hereafter. Then, by matrix normal distribution theory, we know the exact form of the posterior distribution
\begin{equation}
    \beta\mid \mathscr{D},\Sigma\sim\operatorname{MN}(M_{n}m_{n},M_{n},\Sigma),
\end{equation}
where $M_{n}^{-1}=M_{0}^{-1} + X^\top V^{-1} X$, $m_{n}=m_{0} + X^\top V^{-1} Y$. We provide a memory-efficient way to sample from this distribution and reduce its computational burden. We define random variables $Y_{rep}\sim\operatorname{MN}(Y,V,\Sigma)$ and $Z\sim\operatorname{MN}(M_{0}m_{0},M_{0},\Sigma)$. Expressing the relation between $Y_{rep}$, $Z$, and $B$ as
\begin{equation}
    M_{n}^{-1}B=A_{1}Z+A_{2}Y_{rep}.
\end{equation}
We seek matrices $A_{1},A_{2}$ such that $B\overset{d}{=}\beta\mid \mathscr{D},\Sigma$.
Since $\beta\mid \mathscr{D},\Sigma$ is distributed as a Gaussian random variable, it is fully characterized by its mean and variance (in such cases, with both row and column covariance matrices). Then, all we need are $A_{1}$ and $A_{2}$ so that the first two moments of $B$ match those of $\beta\mid \mathscr{D},\Sigma$.

For $X\sim\operatorname{MN}(m,v,s)$, we have $DXC\sim\operatorname{MN}(DmC, DvD^\top, CsC^\top)$. Moreover, if $X$ is $n\times q$, the row-variance matrix is defined as $v = \mathbb{V}_{row}(X) = \mathbb{E}[(X - m )(X - m )^{\T}]\operatorname{tr}(s)^{-1}$ of dimension $(n\times n)$ and its elements are defined as the variance computed on each row, while the $q\times q$ column covariance matrix is denoted by $s = \mathbb{V}_{col}(X) = \mathbb{E}[(X - m )^{\T}(X - m )]\operatorname{tr}(v)^{-1}$ \citep[see, e.g.,][for further details]{gupta_matrix_2000}. Hence, we compute the row covariance matrix for $B$ (the column covariance matrix $\Sigma$ is given). Note that
\begin{equation}
\begin{split}
    M_{n}^{-1}\mathbb{V}_{row}(B)M_{n}^{-1} &= A_{1}\mathbb{V}_{row}(Z)A_{1}^\top + A_{2}\mathbb{V}_{row}(Y_{rep})A_{2}^\top = A_{1}M_{0}A_{1}^\top + A_{2}VA_{2}^\top
\end{split}    
\end{equation}
Setting these matrices as $A_{1}=M_{0}^{-1}$, and $A_{2}=X^\top V^{-1}$, we have
\begin{equation}
\begin{split}
    M_{n}^{-1}\mathbb{V}_{row}(B)M_{n}^{-1} &= M_{0}^{-1}M_{0}M_{0}^{-1} + X^\top V^{-1}VV^{-1}X = M_{0}^{-1} + X^\top V^{-1}X = M_{n}^{-1}.
\end{split}
\end{equation}
This implies $\mathbb{V}_{row}(B)=M_{n}$. The mean follows from
\begin{equation}
\begin{split}
    M_{n}^{-1}\mathbb{E}[B] &= A_{1}\mathbb{E}[Z] + A_{2}\mathbb{E}[Y_{rep}] = M_{0}^{-1}M_{0}m_{0} + X^\top V^{-1}Y = m_{n},
\end{split}
\end{equation}
and we obtain $\mathbb{E}[B]=M_{n}m_{n}$. Therefore, we can derive the next equality in distribution between $B$ and the posterior distribution of the regression coefficient $\beta$ as
\begin{equation}
    B\overset{d}{=}\beta\mid \mathscr{D},\Sigma\sim\operatorname{MN}(M_{n}m_{n},M_{n},\Sigma).
\end{equation}
This implies we can sample from $\beta \mid \mathscr{D}, \Sigma$ by solving a linear system. Specifically, by simply drawing samples from $Z$ and $Y_{\text{rep}}$, we obtain a sample from $\beta \mid \mathscr{D}, \Sigma$ by solving the system $(M_{0}^{-1} + X^\top V^{-1} X)B = (M_{0}^{-1}Z + X^\top V^{-1}Y_{\text{rep}})$ for $B$.
This approach is particularly advantageous for Bayesian transfer learning, as it avoids storing several large matrices when computing the posterior of $\beta \mid \mathscr{D}, \Sigma$. Instead, only the prior precision matrix for $\beta$, $M_{0}^{-1}$, and the product matrix $X^\top V^{-1}$ need to be stored, significantly reducing the memory footprint.

%%%%%%%%%%%%%%%%%%%%%%%%%%%%%%%%%%%%%%%%%%%%
\section{Simulations for Multivariate Models}\label{sec:sim_suppl_M}

We supplement empirical results from Section~\ref{sec:sim}; for further details, we refer to the main article. The current section follows this structure: we begin by comparing the computational performance of the double Bayesian predictive stacking (\textsc{dbps}) approach with the multivariate spatial meta-kriging \citep[\textsc{msmk},][]{guhaniyogi_multivariate_2019}, the linear model of coregionalization \citep[\textsc{lmc},][]{finley_spbayes_2015}, and the seemingly unrelated \textsc{bart} model \citep[su\textsc{bart},][]{esser_seemingly_2025}, and conclude with empirical investigations for data partition dimension sensitivity for the \textsc{double bps}.

% %%%%%%%%%%%%%%%%%%%%%%%%%%%%%%%%%%%%%%%%%%%%%%%%%%%%%%%%%%%%%%%

%-------------------------------------------
\subsection{Computational Performance}\label{sec:sim3}

We investigate the running times of our framework on two synthetic datasets with similar structures but different sizes. Both datasets consist of $p=2$ predictors and $q=2$ response variables, but differ in the number of spatial locations, $n=5,000$ and $n=10,000$, respectively. We generate and fix spatial coordinates from a uniform distribution on the unit square ($[0,1]^2$). We build the $n\times n$ spatial correlation matrix $V$ over these coordinates using $\rho_{\phi}(s_i,s_j) = \exp(-\phi \|s_i-s_j\|)$ with $\phi=4$ and specify $\Sigma = \mathbb{I}_q$. From these specifications we generate the $n\times q$ matrix $Y$ from the first equation of \eqref{eq:MatrixNormHier} with fixed $p \times q$ matrix $\beta = \begin{bsmallmatrix}
    -0.75 & 1.85 \\ 0.90 & -1.10
\end{bsmallmatrix}$, a fixed $n\times p$ matrix $X$ with a first column of ones, representing the intercept, and $p-1$ columns of values randomly simulated from a uniform distribution on $[0,1]$ (emulating standardized predictors), and the proportion of spatial variability $\alpha=0.8$.

For distributed learning approaches, i.e., \textsc{double bps} and \textsc{msmk}, we perform the analyses
in two settings: (i) $K = 10$ and $n=5,000$; (ii) $K = 10$ and $n=10,000$; (iii) $K=5$ and $n=5,000$; and (iv) $K = 20$ and $n=10,000$. These settings produce subsets of size $n/K \in \{500, 1000\}$. We implement \textsc{double bps} using $J=9$ candidate models $\mathscr{M}_j, j=1,\dots,J$, where each model is specified by a set of candidate values for the hyperparameters ${\alpha_{j},\phi_{j}}$ in \eqref{eq:MatrixNormHier}. These hyperparameters represent the proportion of spatial variability and the parameter(s) of the spatial correlation function, respectively. The set of candidate models is constructed as the set of all possible combinations of values for these hyperparameters. In the subsequent experiments, the grid of models was built using $\alpha \in \{0.70, 0.80, 0.90\}$ and $\phi \in \{3, 4, 5\}$. These values resemble an effective spatial range of $\{0.99, 0.75, 0.60\}$ units, corresponding to $70\%, 53\%, 42\%$ of the maximum inter-site distance inside the unit square, beyond which the spatial correlation drops below $0.05$.
% priors
Equation~\eqref{eq:MultivJoinPrior} follows choices in \citet{zhangEtAl2025jasa}. Specifically, we set $m_{0}=0_{p\times  q}$, $M_{0}= 10\mathbb{I}_p$, $\Psi_{0}=\mathbb{I}_{q}$, and $\nu_{0}=3$ in the $\operatorname{MNIW}$ joint prior for $\{\gamma, \Sigma\}$ in \ref{eq:MultivJoinPrior}. For data analysis, we use an exponential spatial correlation function for $\rho_{\phi}(\,\cdot\,,\cdot\,)$, which is completely defined conditionally on $\mathscr{M}_j$, since it specifies a value for $\phi$. Again, in the conjugate framework, we draw $R=250$ posterior samples used for inference.

\begin{table}[t!]
\centering
\scriptsize
\begin{tabularx}{\textwidth}{L{3cm} L{1cm} Y Y L{5cm}}
\toprule
\textbf{Setting} & \textbf{Model} & \textbf{Time (min)} & \textbf{Relative to BPS} & \textbf{Visual} \\
\midrule

\multirow{2}{*}{\textbf{$n=5000, K=10$}}
& \cellcolor{gray!10} \textsc{dbps}       & \cellcolor{gray!10} 0.24   & \cellcolor{gray!10} 1.0$\times$       & \cellcolor{gray!10} \barwidth{0.01} \\
& \textsc{msmk}       & 51.41  & 220.3$\times$      & \barwidth{2.2} \\

\midrule
\multirow{2}{*}{\textbf{$n=5000, K=5$}}
& \cellcolor{gray!10} \textsc{dbps}       & \cellcolor{gray!10} 0.98   & \cellcolor{gray!10} 1.0$\times$       & \cellcolor{gray!10} \barwidth{0.01} \\
& \textsc{msmk}       & 237.23 & 241.2$\times$      & \barwidth{2.4} \\

\midrule
\multirow{3}{*}{\textbf{$n=5000$}}
& \cellcolor{gray!10} \textsc{dbps}       & \cellcolor{gray!10} 0.23 -- 0.98 & \cellcolor{gray!10} 1.0 -- 1.0$\times$   & \cellcolor{gray!10} \barwidth{0.001} -- \barwidth{0.01} \\
& \textsc{lmc}       & 8975.31     & $>>10000\times$       & \barwidth{5} \\
& \cellcolor{gray!10} su\textsc{bart}    & \cellcolor{gray!10} 215.67      & \cellcolor{gray!10} 924 -- 219$\times$    & \cellcolor{gray!10} \barwidth{3} -- \barwidth{0.9} \\

\midrule
\multirow{2}{*}{\textbf{$n=10000, K=20$}}
& \cellcolor{gray!10} \textsc{dbps}       & \cellcolor{gray!10} 0.42    & \cellcolor{gray!10} 1.0$\times$       & \cellcolor{gray!10} \barwidth{0.01} \\
& \textsc{msmk}       & 103.36  & 248.1$\times$      & \barwidth{2.5} \\

\midrule
\multirow{2}{*}{\textbf{$n=10000, K=10$}}
& \cellcolor{gray!10} \textsc{dbps}       & \cellcolor{gray!10} 1.92   & \cellcolor{gray!10} 1.0$\times$       & \cellcolor{gray!10} \barwidth{0.01} \\
& \textsc{msmk}       & 446.01  & 232.7$\times$      & \barwidth{2.4} \\

\midrule
\multirow{3}{*}{\textbf{$n=10000$}}
& \cellcolor{gray!10} \textsc{dbps}       & \cellcolor{gray!10} 0.42 -- 1.92 & \cellcolor{gray!10} 1.0 -- 1.0$\times$   & \cellcolor{gray!10} \barwidth{0.0005} -- \barwidth{0.0005} \\
& \textsc{lmc}       & --           & --                & -- \\
& \cellcolor{gray!10} su\textsc{bart}    & \cellcolor{gray!10} 557.60       & \cellcolor{gray!10} 1338 -- 290$\times$    & \cellcolor{gray!10} \barwidth{3.5} -- \barwidth{1} \\

\bottomrule
\end{tabularx}
\caption{Running times (in minutes), relative to \textsc{double bps}. Bars give a visual impression of time cost (where applicable).}
\label{tab:sim3}
\end{table}

We also apply \textsc{msmk} to the two simulated datasets with the same combinations of $n$ and $K$ as \textsc{double bps}. Unlike \textsc{double bps}, where we stack analytically tractable posteriors over a range of fixed values of spatial covariance kernel parameters, the \textsc{msmk} implementation attempts full Bayesian inference using prior distributions on spatial covariance kernel parameters. We fit the linear model of coregionalization described in \citet{finley_spbayes_2015} for each subset of the multivariate spatial data using \textsc{mcmc}. The posterior samples from the $K$ subsets are combined using Weiszfeld's iterative algorithm \citep[][]{minsker_robust_2017} to produce an estimate of the geometric median of the posterior distributions.
For both experimental settings, we then fit \textsc{lmc} and su\textsc{bart} on the full dataset, using default prior settings, following \citet{finley_spbayes_2015}, and \citet{esser_seemingly_2025}, respectively. 

Table~\ref{tab:sim3}, which compares the computational speed of \textsc{double bps} with other considered approaches, reveals massive computational gains accrued from \textsc{double bps}. The computational advantage evident from the relative ratio becomes more pronounced as the size of the data increases, despite the larger subsets. This is explained by the fact that fitting the Gaussian process regression dominates the computation relative to assimilating inference from the subsets. If the number of locations explodes, then the geometric median of posteriors required by \textsc{msmk} is computationally unfeasible. While \textsc{msmk} offers Bayesian estimates using \textsc{mcmc} for each subset, \textsc{double bps} avoids \textsc{mcmc} and, hence, issues of convergence.
Similar arguments follow for \textsc{lmc} and su\textsc{bart}. As expected, the linear model of coregionalization, when fitted on the entire set of locations, gives a disastrous performance, taking almost a week of computation in the lighter simulation settings, and makes it infeasible to record results for $n=10,000$. Notwithstanding the scalability offered by this multivariate extension of the Bayesian additive regression model, the su\textsc{bart} does not have any chance to provide inference in a comparable time.

\begin{figure}[t]
    \centering
    \includegraphics[scale=0.385]{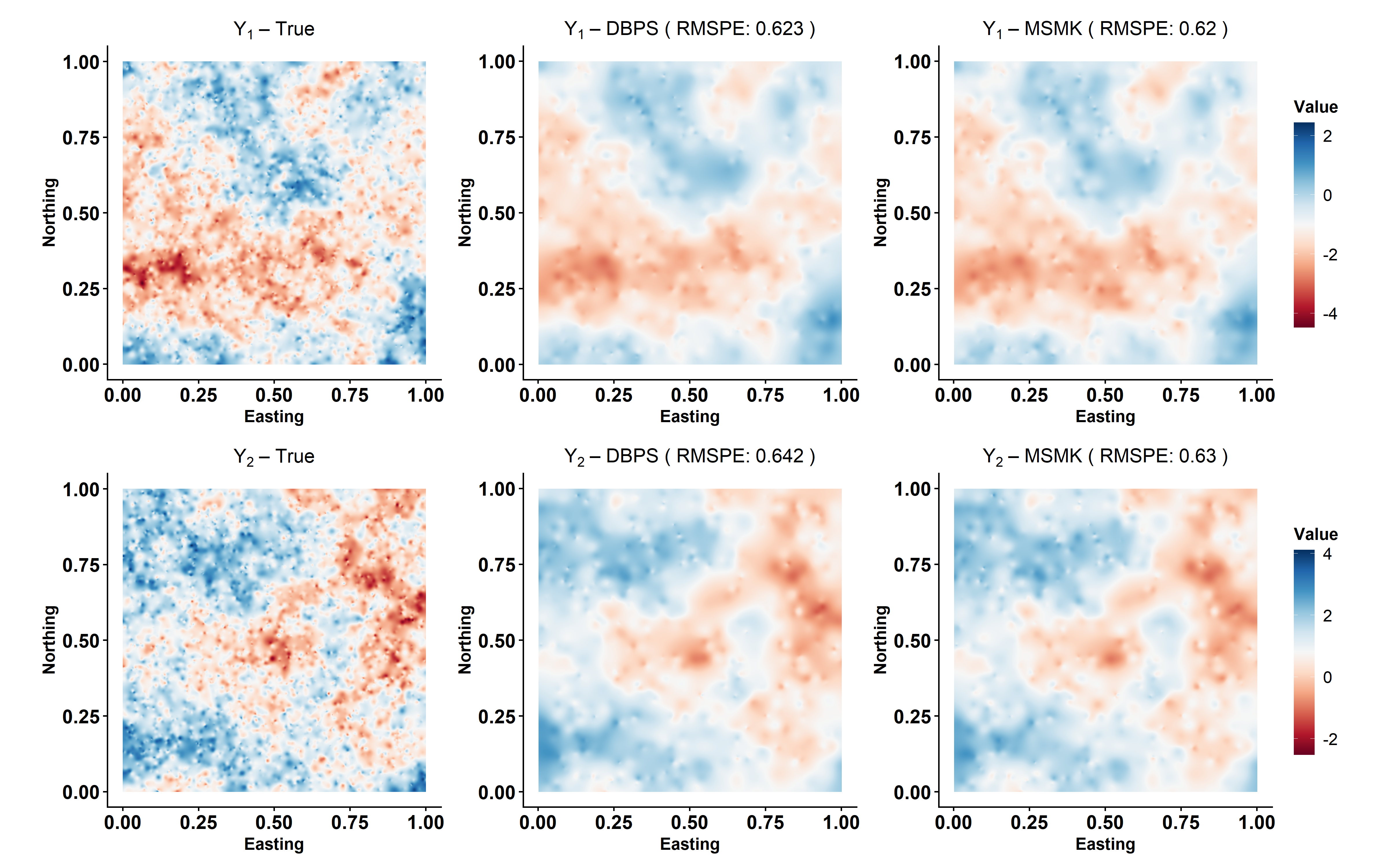}
    \caption{ From left to right: comparison between the true generated response surfaces, the surfaces predicted from \textsc{double bps} and \textsc{msmk} (posterior mean), with \textsc{rmspe}. For $n=5000$, $K=10$.}
    \label{fig:resp_5_500m}
\end{figure}

Figure~\ref{fig:resp_5_500m}, which depicts estimated response surfaces using \textsc{double bps} and \textsc{msmk} corresponding to $n=5000$ and $K=10$, shows that inferences are practically indistinguishable. The root mean squared prediction error (\textsc{rmspe}), reported in Figure~\ref{fig:resp_5_500m}, denotes the average prediction error and reveals minor discrepancies between \textsc{double bps} and \textsc{msmk}. Section~\ref{sec:sim_suppl_M} presents results for the other configurations of $n$ and $K$, each of which reveals that \textsc{double bps} offers practically indistinguishable spatial interpolation from \textsc{msmk} at a fraction of the computational cost.  

\begin{figure}[t]
    \centering
    \includegraphics[scale=0.385]{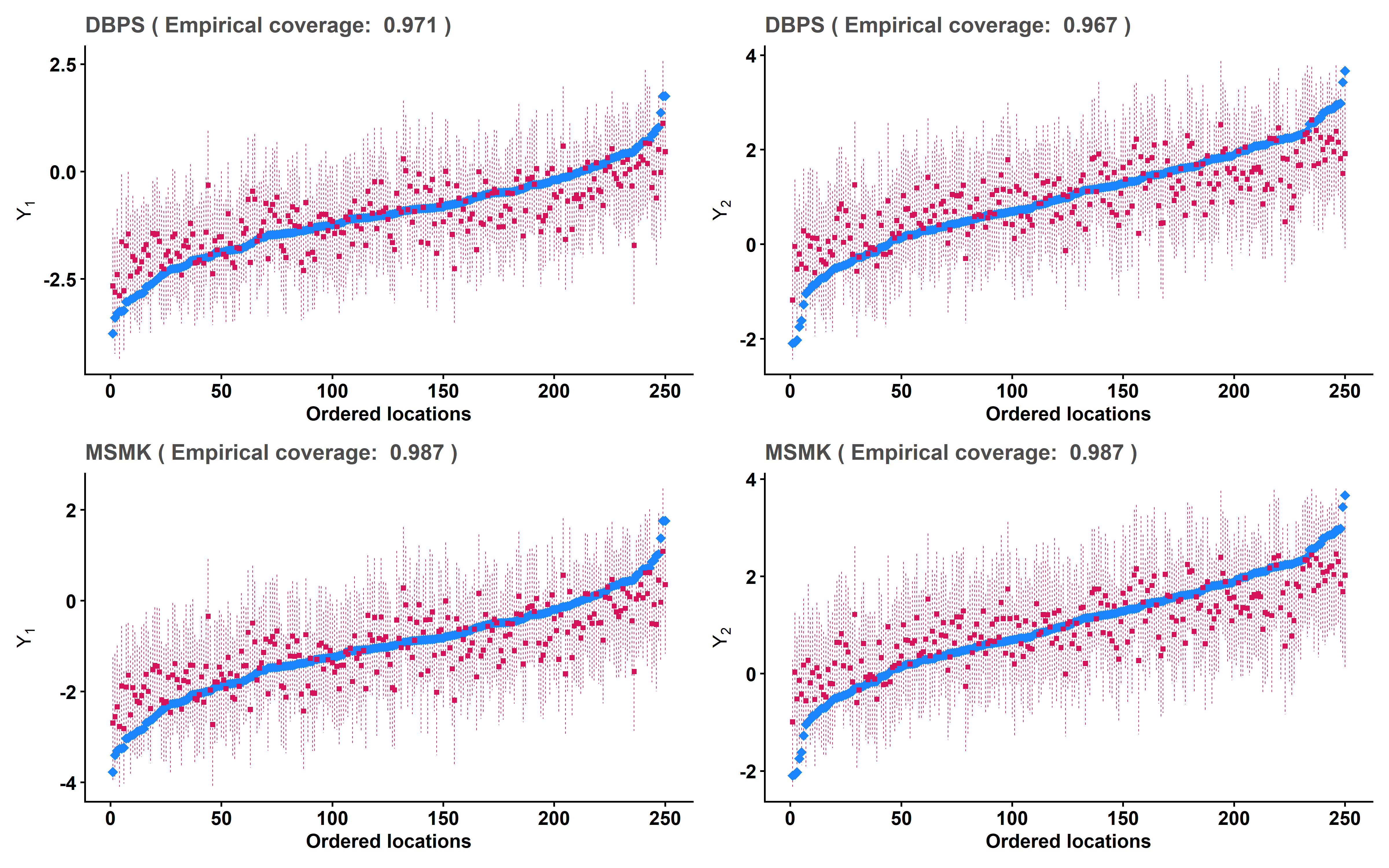}
    \caption{ From top to bottom: comparison between posterior predictive intervals for the predicted response from \textsc{double bps} and \textsc{msmk}, with empirical coverage. For $n=5000$, $K=10$.}
    \label{fig:UC_5_500m}
\end{figure}

\begin{figure}[t]
    \centering
    \includegraphics[scale=0.425]{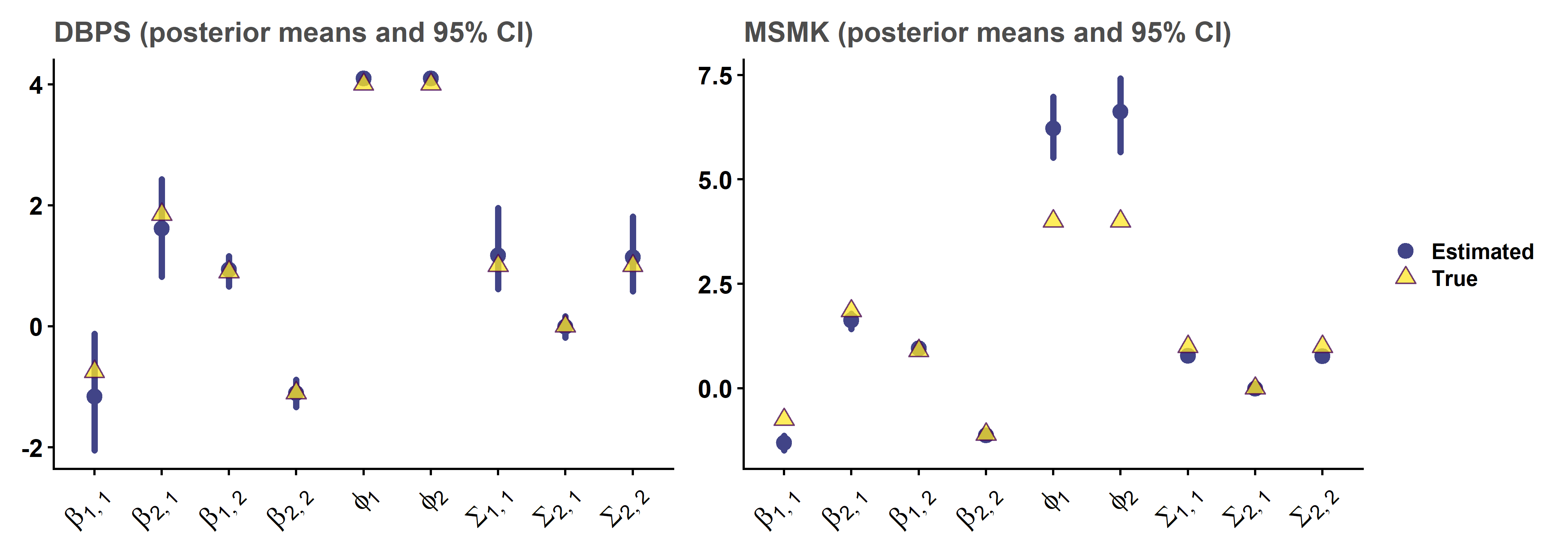}
    \caption{ From left to right: comparison between posterior credible intervals for the parameters recovered from \textsc{double bps} and \textsc{msmk}. For $n=5000$, $K=10$.}
    \label{fig:post_5_500m}
\end{figure}

Figure~\ref{fig:UC_5_500m} reports 95\% posterior predictive intervals for the response. The empirical coverage is impressive. We notice slightly narrower intervals from \textsc{double bps}, indicating a better calibration of uncertainty quantification with respect to \textsc{msmk}. Moreover, Figure~\ref{fig:UC_5_500m} reveals superior \textsc{map} estimates for the \textsc{double bps}. Finally, Figure~\ref{fig:post_5_500m} presents the recovery of parameter estimates. As seen in predictive inference, the posterior credible intervals for parameters also deliver practically indistinguishable inference for the two modeling frameworks. In particular, both methods recover the true values for $\beta$ and $\Sigma$, while \textsc{double bps} reconstructs a better point estimate for range parameters $\phi$ using $\sum_{k=1}^K \hat{w}_{k}\sum_{j=1}^J \hat{z}_{j} \phi_j$.

\begin{figure}[!t]
    \centering
    \includegraphics[scale=0.385]{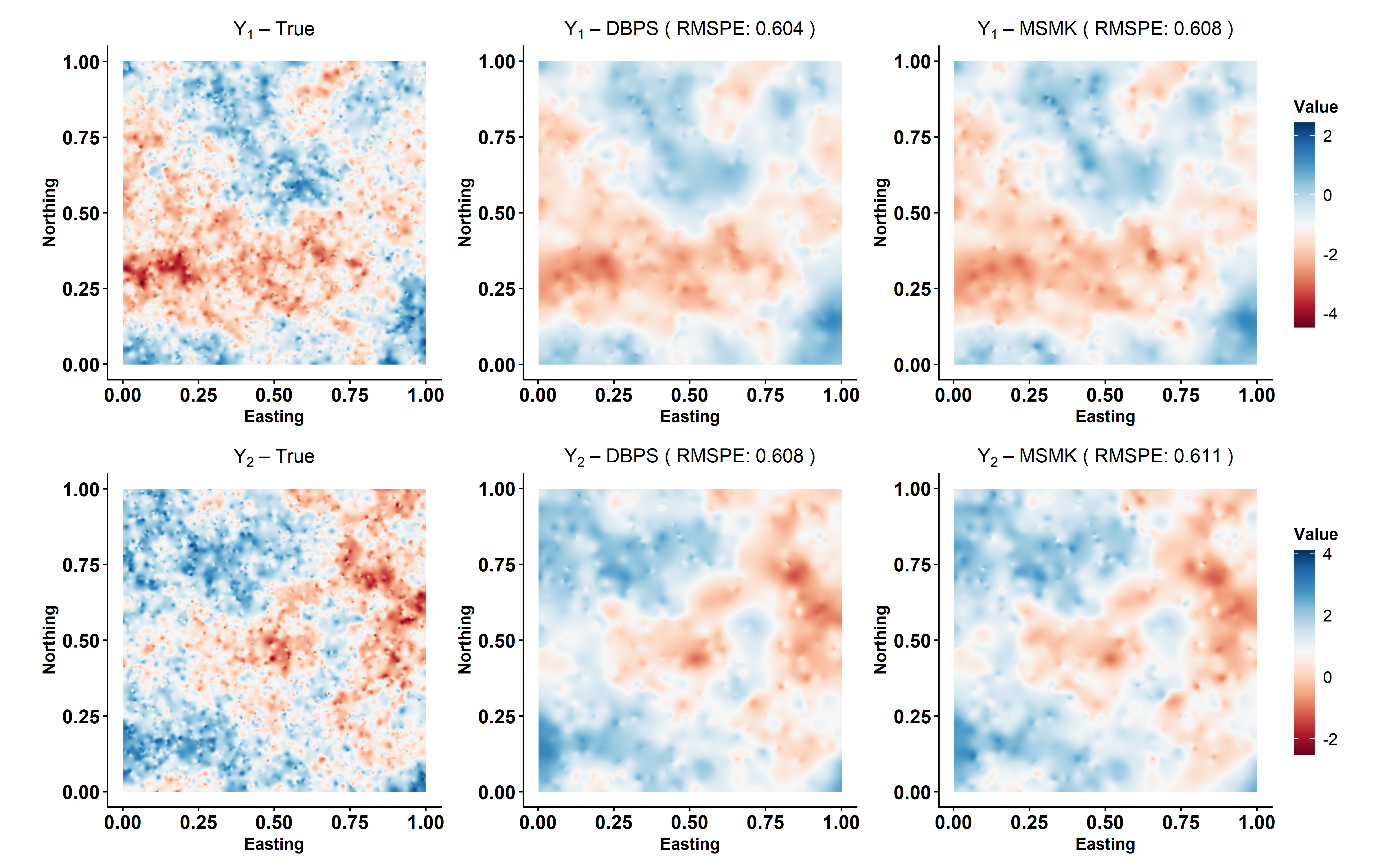}
    \caption{ From left to right: comparison between the true generated response surfaces, the surfaces predicted from \textsc{double bps} and \textsc{msmk} (posterior mean), with \textsc{rmspe}. For $n=5000$, $K=5$.}
    \label{fig:resp_5_1000m}
\end{figure}

\begin{figure}[!t]
    \centering
    \includegraphics[scale=0.385]{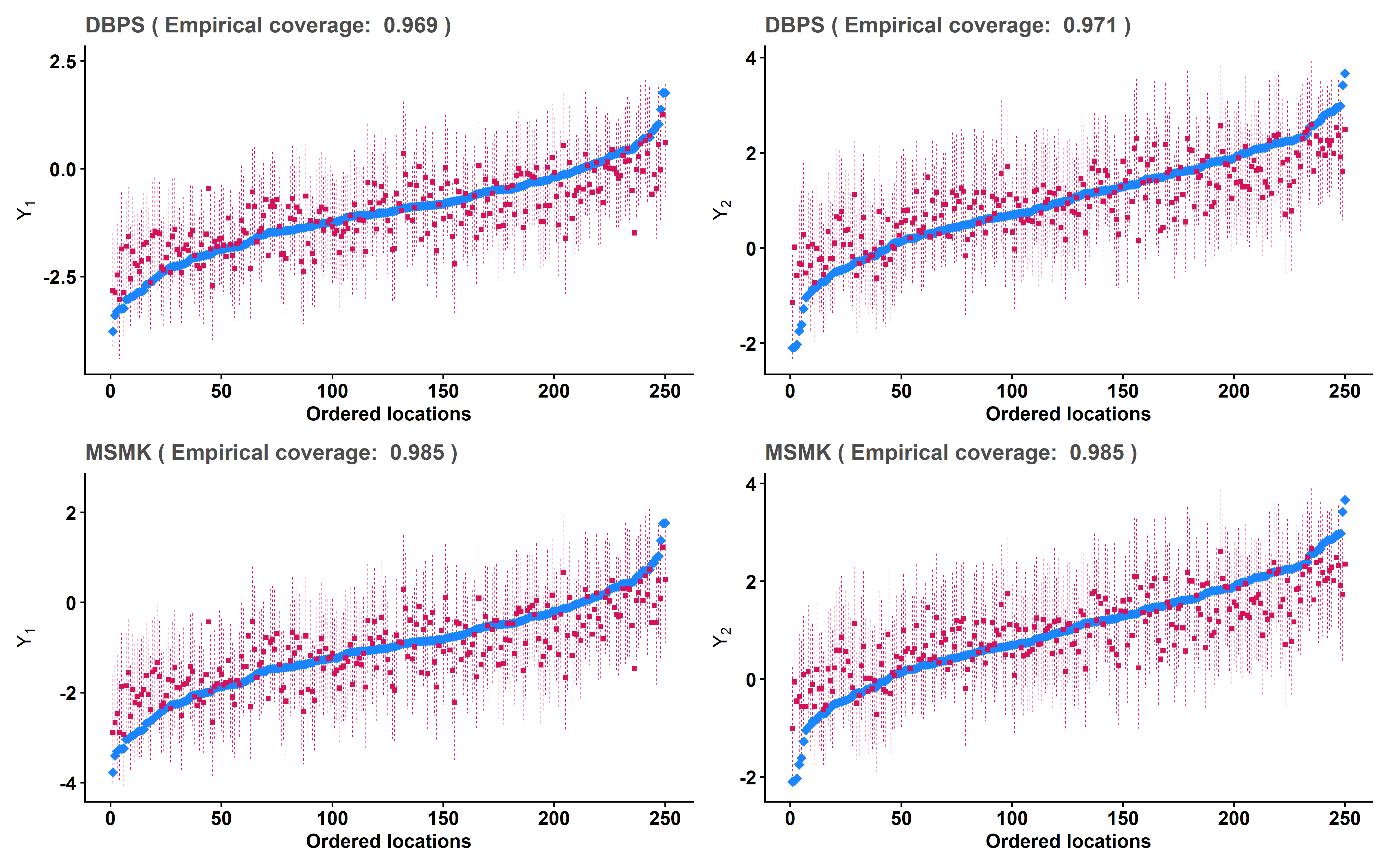}
    \caption{ From top to bottom: comparison between posterior predictive intervals for the predicted response from \textsc{double bps} and \textsc{msmk}, with empirical coverage. For $n=5000$, $K=5$.}
    \label{fig:UC_5_1000m}
\end{figure}

\begin{figure}[!t]
    \centering
    \includegraphics[scale=0.425]{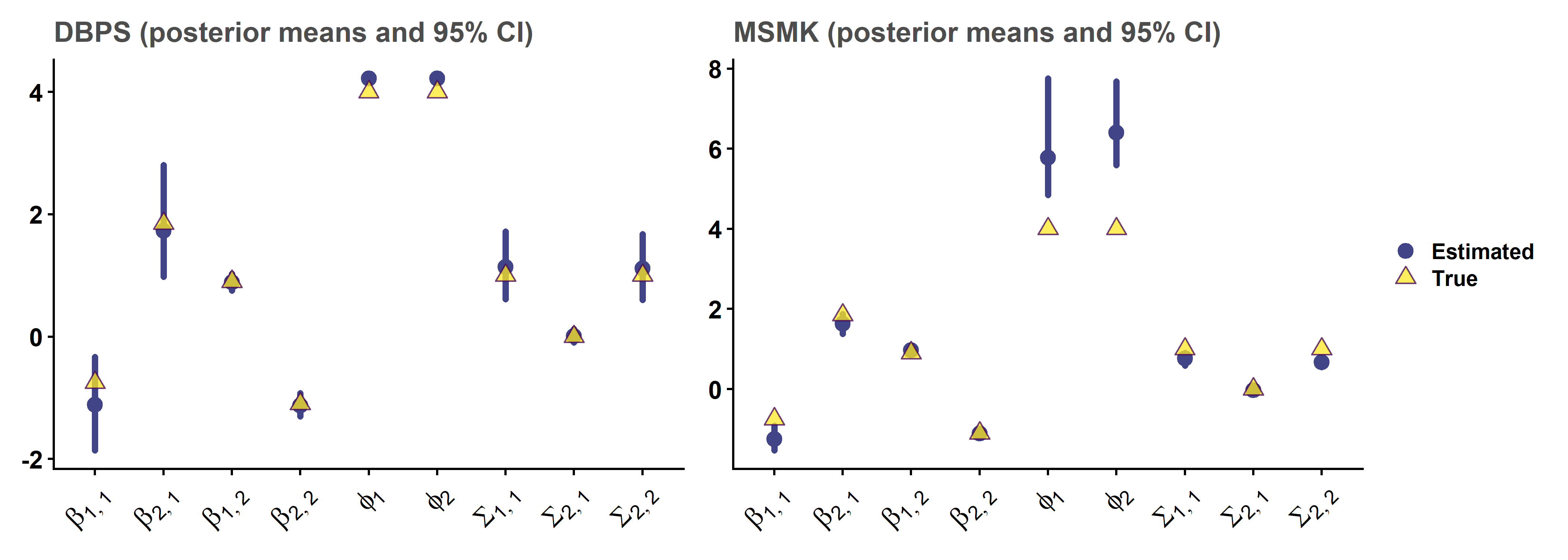}
    \caption{ From left to right: comparison between posterior credible intervals for the parameters recovered from \textsc{double bps} and \textsc{msmk}. For $n=5000$, $K=5$.}
    \label{fig:post_5_1000m}
\end{figure}

\begin{figure}[!t]
    \centering
    \includegraphics[scale=0.385]{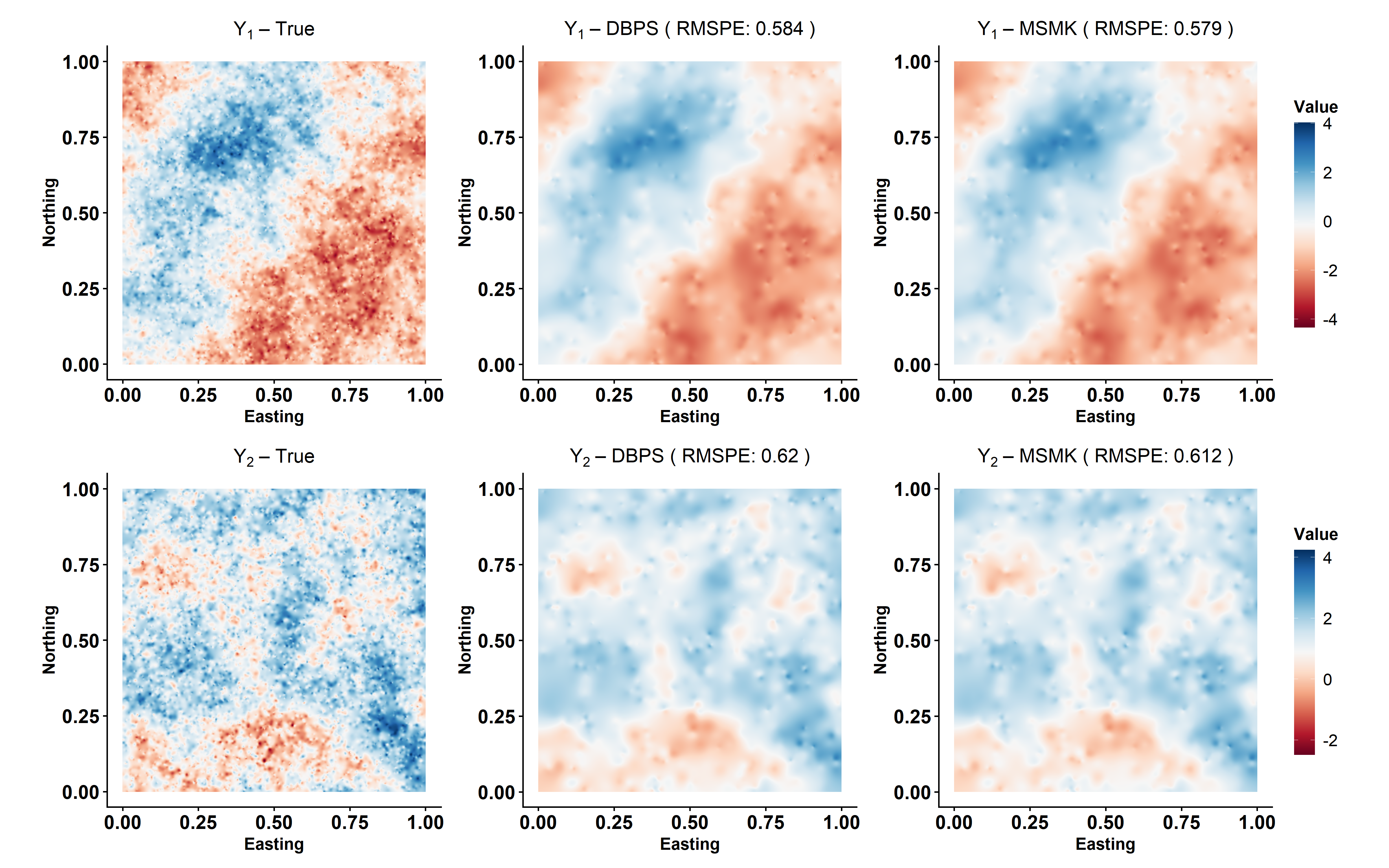}
    \caption{ From left to right: comparison between the true generated response surfaces, the surfaces predicted from \textsc{double bps} and \textsc{msmk} (posterior mean), with \textsc{rmspe}. For $n=10000$, $K=20$.}
    \label{fig:resp_10_500m}
\end{figure}

\begin{figure}[!t]
    \centering
    \includegraphics[scale=0.385]{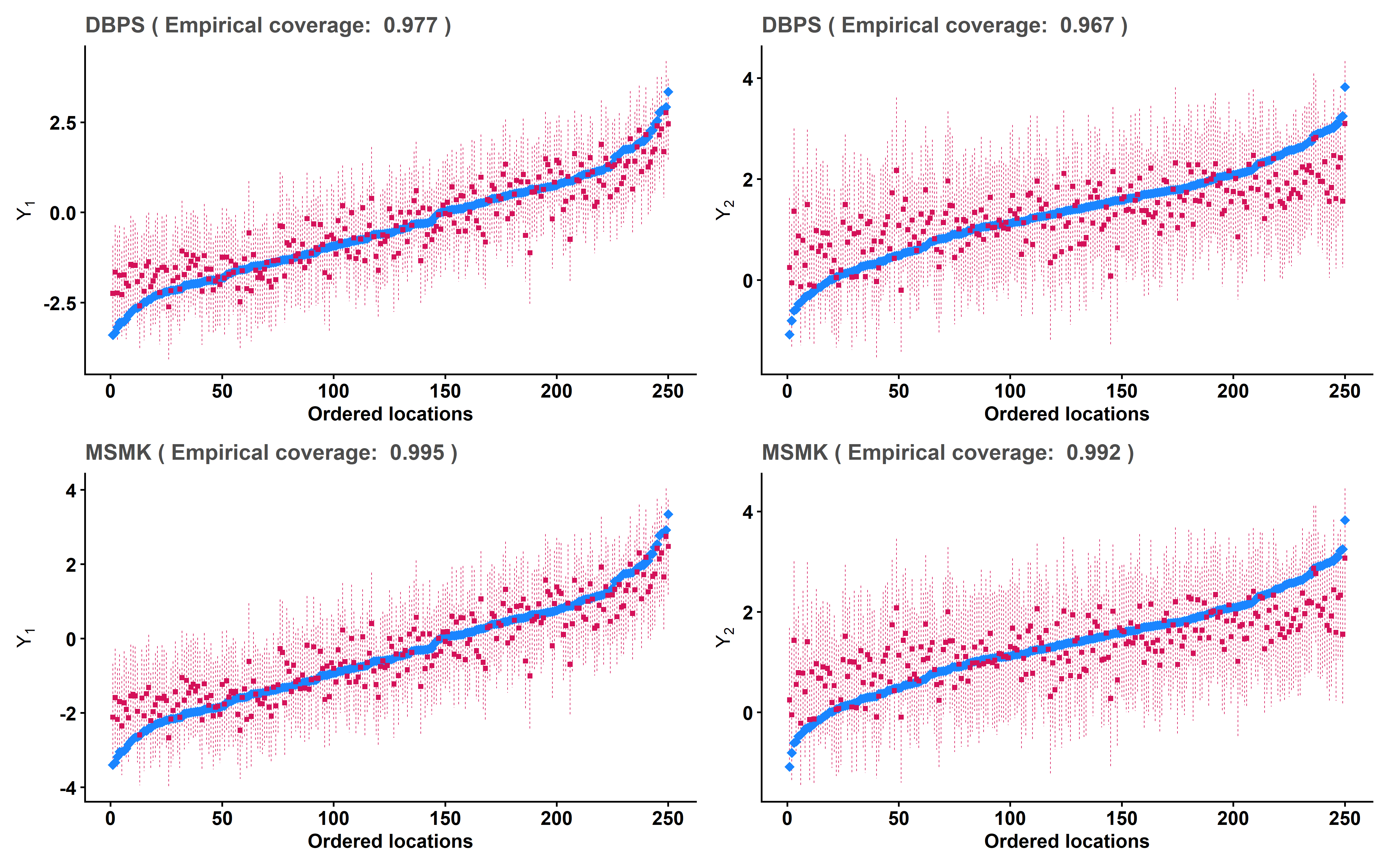}
    \caption{ From top to bottom: comparison between posterior predictive intervals for the predicted response from \textsc{double bps} and \textsc{msmk}, with empirical coverage. For $n=10000$, $K=20$.}
    \label{fig:UC_10_500m}
\end{figure}

\begin{figure}[!t]
    \centering
    \includegraphics[scale=0.425]{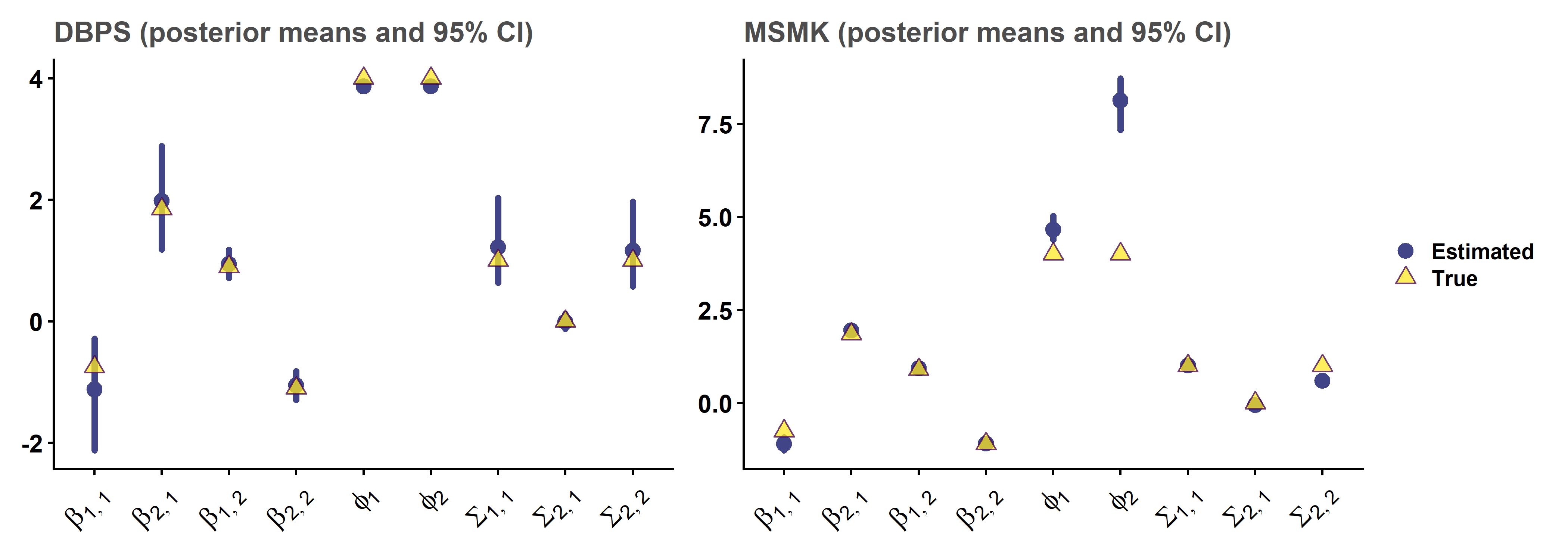}
    \caption{ From left to right: comparison between posterior credible intervals for the parameters recovered from \textsc{double bps} and \textsc{msmk}. For $n=10000$, $K=20$.}
    \label{fig:post_10_500m}
\end{figure}

Figure~\ref{fig:resp_5_1000m} shows the estimated response surface using \textsc{double bps} for $\{n=5000, K=5\}$. This plot reveals no noteworthy differences compared to the results obtained from \textsc{msmk}. Although the \textsc{rmspe} values for \textsc{dbps} are slightly better, the differences are negligible when compared with the response scale, leading us to conclude that both approaches exhibit almost identical predictive performance.
Regarding the first scenario from Section~\ref{sec:sim3}, it is unsurprising that a doubled number of locations per partition leads to improved performance. Figure~\ref{fig:post_5_1000m} provides empirical support to this. While \textsc{msmk} shows some degree of spatial over-smoothing, as indicated by higher estimated values of $\phi$ compared to the true ones, \textsc{double bps} does not exhibit such deviations. Both the \textsc{map} estimates and the posterior inferences for \textsc{double bps} show strong performance. 

Figure~\ref{fig:UC_5_1000m} aligns with the previous conclusions, highlighting slightly wider predictive credibility intervals for \textsc{msmk} compared to \textsc{double bps}. For both approaches, the empirical coverage remains impressive.

\begin{figure}[!t]
    \centering
    \includegraphics[scale=0.385]{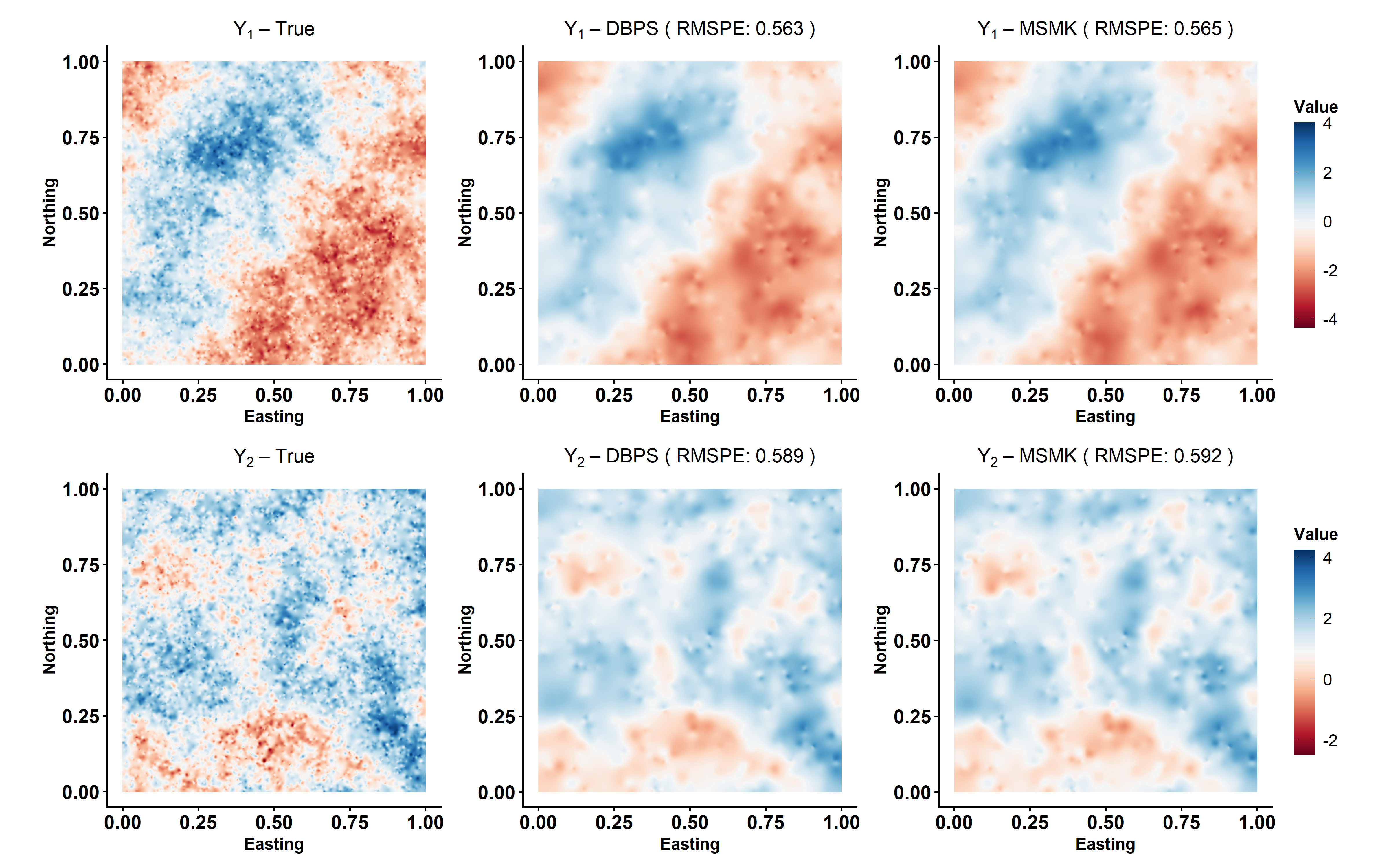}
    \caption{ From left to right: comparison between the true generated response surfaces, the surfaces predicted from \textsc{double bps} and \textsc{msmk} (posterior mean), with \textsc{rmspe}. For $n=10000$, $K=10$.}
    \label{fig:resp_10_1000m}
\end{figure}

\begin{figure}[!t]
    \centering
    \includegraphics[scale=0.385]{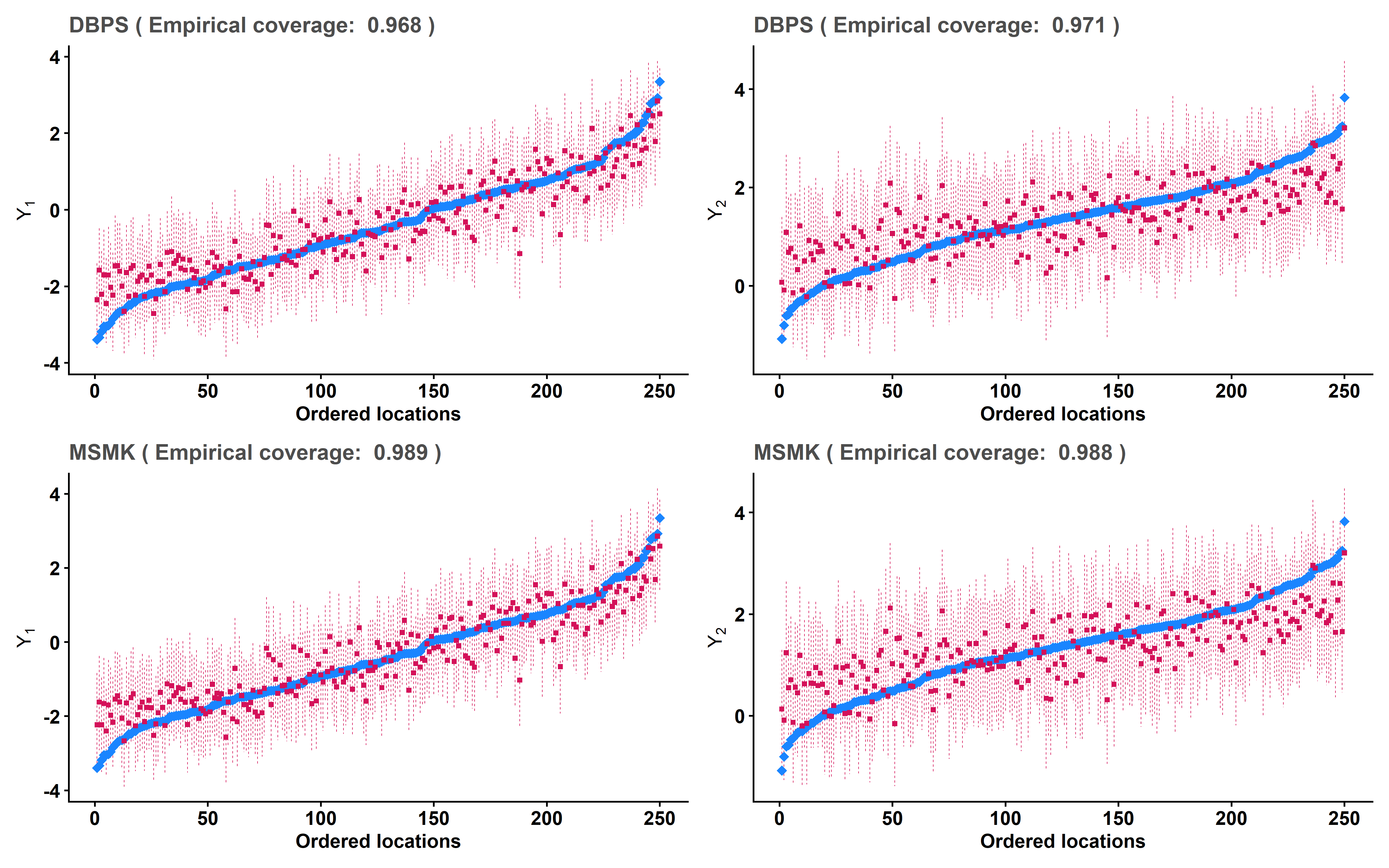}
    \caption{ From top to bottom: comparison between posterior predictive intervals for the predicted response from \textsc{double bps} and \textsc{msmk}, with empirical coverage. For $n=10000$, $K=10$.}
    \label{fig:UC_10_1000m}
\end{figure}

\begin{figure}[!t]
    \centering
    \includegraphics[scale=0.425]{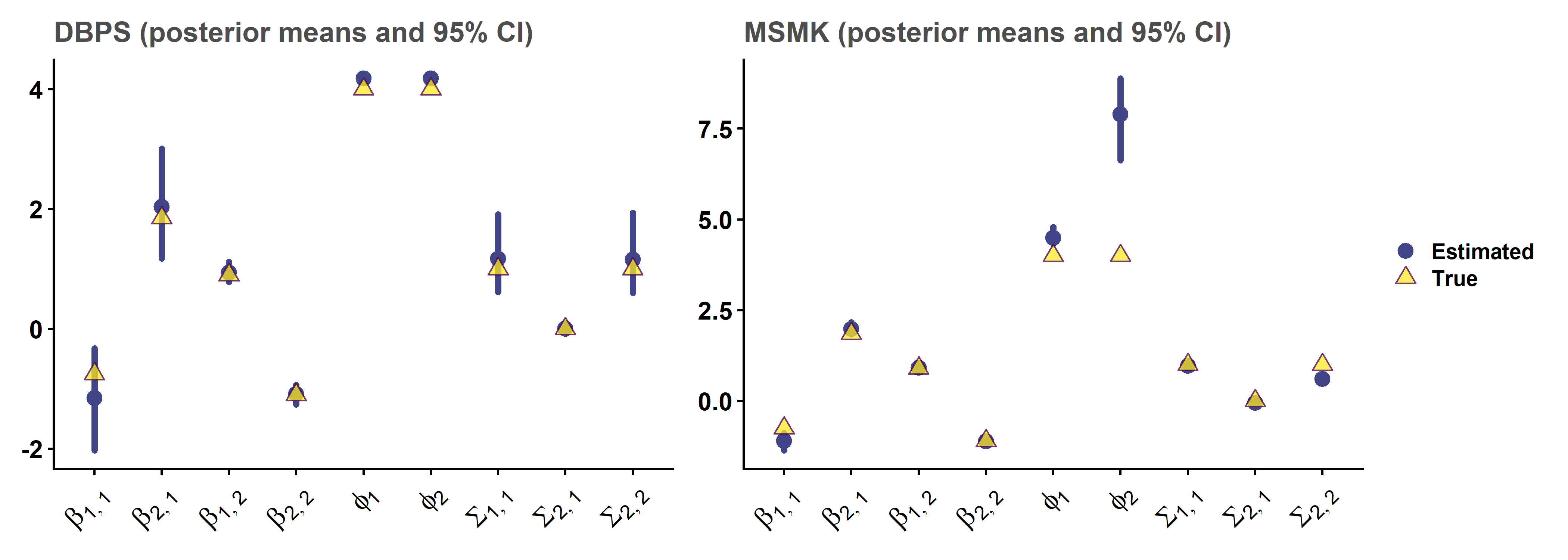}
    \caption{ From left to right: comparison between posterior credible intervals for the parameters recovered from \textsc{double bps} and \textsc{msmk}. For $n=10000$, $K=10$.}
    \label{fig:post_10_1000m}
\end{figure}

Despite focusing on the larger dataset later in this section, where $n=10,000$, the results increasingly appear to depend on the size of the subsets rather than the number of partitions. This highlights the influence of local inferences on global performance, with the number of partitions playing a lesser role.

For the case $\{n=10,000, K=20\}$, Figure~\ref{fig:resp_10_500m} compares the estimated posterior mean surfaces.
The performances of \textsc{double bps} and \textsc{msmk} are nearly equivalent. The most noticeable difference is in Figure~\ref{fig:UC_10_500m}, where \textsc{msmk} produces wider predictive credible intervals for the response variable. However, in Figure~\ref{fig:post_10_500m}, \textsc{double bps} demonstrates superior posterior \textsc{map} estimates, including for the range parameters, which are typically challenging to identify.
Together, Figures~\ref{fig:UC_10_500m}, \ref{fig:post_10_500m}, and \ref{fig:resp_10_500m} present comparable findings to the $\{n=5,000, K=10\}$ setting, reinforcing the importance of subset learning over the overall dataset size, as expected.

The simulation experiment concludes with the $\{n=10,000, K=10\}$ setting. Once again, Figures~\ref{fig:resp_10_1000m}, \ref{fig:UC_10_1000m}, and \ref{fig:post_10_1000m} show results consistent with previous settings, particularly with $\{n=5,000, K=10\}$, highlighting the critical role of partition size in achieving reliable posterior inferences.
In Figure~\ref{fig:post_10_1000m}, despite the stronger posterior performance, \textsc{double bps} exhibits wider posterior credible intervals compared to \textsc{msmk}. Conversely, in Figure~\ref{fig:UC_10_1000m}, \textsc{double bps} achieves predictive empirical coverage closer to the nominal level $95\%$.

In conclusion, notwithstanding indistinguishable posterior inferences between \textsc{dbps} and multivariate \textsc{smk}, Table~\ref{tab:sim3} clearly illustrates that the primary advantage of Bayesian predictive stacking lies in its enormous computational efficiency. This speedup is crucial for delivering feasible Bayesian inference for large datasets within GeoAI systems.

%-------------------------------------------
\subsection{Monte Carlo Approximation for Upper Bound Simulations}\label{sec:upperbound_sim}

We perform empirical investigations of the upper bound presented and detailed in Section~\ref{sec:upperbound}, for different values of $K$ and $J$ ceteris paribus.

We approach the problem of approximating the expectation in Equation~\eqref{eq:upperbound} with a Monte Carlo integration. The approximation takes the form
\begin{equation*}
\mathbb{E}_{p_{k,j}}\Bigg[\frac{\sum_{j=1}^{J}\hat{z}_{k,j}\; p(y\mid\mathscr{D}_{k},\mathscr{M}_{j})}{p_{t}(y\mid\mathscr{D})}\Bigg] \; \approx \; \frac{1}{L}\sum_{l=1}^{L}\Bigg[\frac{\sum_{j=1}^{J}\hat{z}_{k,j}\; p(y_{l}\mid\mathscr{D}_{k},\mathscr{M}_{j})}{p_{t}(y_{l}\mid\mathscr{D})}\Bigg],
\end{equation*}
where $y_{l}\sim p(y_{l}\mid\mathscr{D}_{k},\mathscr{M}_{j})$ for $l=1,\dots,L$. We then devise the Algorithm~\ref{alg:upperbound} to approximate the upper bound for the \textsc{kl} divergence between the \textsc{double bps} posterior predictive and the true one. 

\begin{figure}[!t]
    \centering
    \includegraphics[scale=0.5]{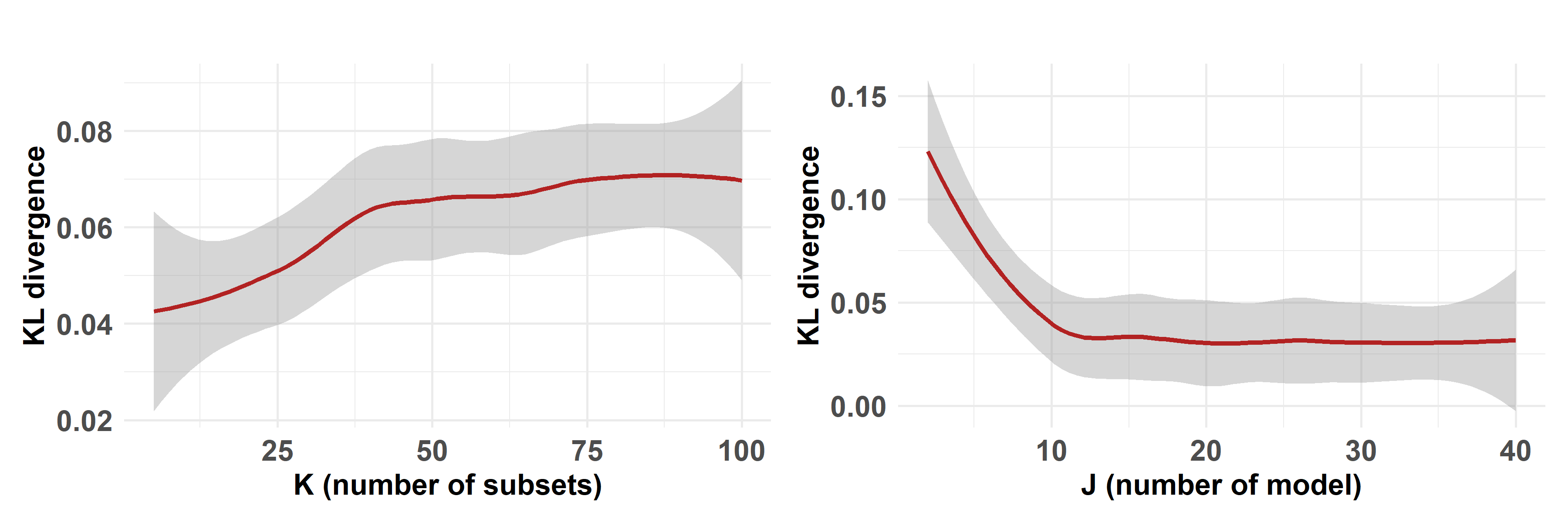}
    \caption{Upper bound behavior for growing values of $K$, and $J$}
    \label{fig:upperbound}
\end{figure}

To provide a meaningful interpolation, we consider $20$ points for each parameter regulating $\widehat{ub}(n, K, J)$. We let vary $K\in\{5,100\}$, $J\in\{2,40\}$, while $n=1000$ was fixed. Then, we remove data dependency by considering $M=10$ replications for each evaluation setting.
We perform different simulations for any of $\{K,J\}$ ceteris paribus, for the other.
The panels in Figure~\ref{fig:upperbound} show how $D_{KL}(\hat{P}\parallel P_{t})$ vary with $K$, and $J$, respectively.

\begin{algorithm}[!t]
\caption{Approximating upper bound for Kullback-Liebler divergence}\label{alg:upperbound}
\vspace{0.25em}
\textbf{Input:} $Y$ outcomes matrix; $X$ predictors matrix; $\hat{w} = \{\hat{w}_{k}: k =1,\dots,K\}$: Stacking weights between subsets; $\hat{z} = \{\hat{z}_{k}=\{\hat{z}_{k,j}\}: k \in \{1,\dots,K\}, j \in\{1,\dots,J\} \}$: Stacking weights within subsets; $\hat{p}_{k,j}(\,\cdot\,),p_{t}(\,\cdot\,)$ approximated and true predictive distributions $\forall j=1,\dots,J, k=1,\dots,K$; $K$: Number of subsets; $J$: number of competitive models in each subset; $n$: number of locations; $L$: number of Monte Carlo samples.\\
\textbf{Output:} $\widehat{ub}(n, K, J)$: approximated value for the upper bound, for a given set $\{n,K,J\}$.\\
\begin{algorithmic}[1]
\For{$k=1,\dots,K$}
\For{$j=1,\dots,J$}
\State Draw $L$ samples $\{y_{l}:l=1,\dots,L\}$ from $\hat{p}(\;\cdot\mid\mathscr{D}_{k},\mathscr{M}_{j})$
\For{$l=1,\dots,L$}
\For{$j=1,\dots,J$}
\State Evaluate $p_{k,j,l}=\hat{p}(\;y_{l}\mid\mathscr{D}_{k},\mathscr{M}_{j})$
\EndFor
\State Evaluate $p_{t,l}=p(y_{l}\mid\mathscr{D})$
\State Compute $r_l=\frac{\sum_{j=1}^{J}z_{k,j}\,p_{k,j,l}}{p_{t,l}}$
\EndFor
\State Compute $e_{k,j}=\frac{1}{L}\sum_{l=1}^{L} r_{l}$
\EndFor
\EndFor
\State Compute $c_k=\sum_{k=1}^{K}\hat{w}_{k}\;\sum_{j=1}^{J}\hat{z}_{k,j}\;e_{k,j}$
\State \textbf{return} $\widehat{ub}(n, K, J)= \log \prod_{k=1}^{K}\;c_{k}^{\hat{w}_{k}}$
\end{algorithmic}
\end{algorithm}

%-------------------------------------------
\subsection{Subset Size Sensitivity }\label{sec:sim5_sens}

The methodological novelty introduced in Section~\ref{sec:AccelerMultivarMod} can be summarized in three main steps, as illustrated in Figure~\ref{fig:double_stacking}. First, we partition the original, often massive, dataset into $K$ smaller subsets. The number of locations in each partition is a critical decision, seriously impacting inferential, predictive, and computational outcomes. Accordingly, a trade-off arises between computational resources and performance.

To address this, we conduct a simulation analysis to assess the sensitivity of the results to subset size. This section aims to investigate how predictive performance (in terms of \textsc{rmspe}), and runtime (in seconds) change as the number of locations within each partition grows. Intuitively and theoretically, as the dimension of the subsets grows, we expect predictive performance to improve, while runtime increases polynomially with $n$. To enhance the comparability of the results, we apply min-max normalization to each variable, defined as $\tilde{x} = \frac{x - \min(x)}{\max(x) - \min(x)}$. This normalization scales all variables to the interval $[0,1]$, facilitating a more direct graphical comparison.

We use a multivariate synthetic dataset comprising $n=5,000$ locations, $q=2$ simulated responses, and $p=2$ predictors to explore the sensitivity to subset size. This dataset is generated from the model in Equation~\eqref{eq:MatrixNormHier}, with parameters $\beta= \begin{bsmallmatrix} -0.75 & 1.85 \\ 0.90 & -1.10
\end{bsmallmatrix}$ and $\Sigma = \begin{bsmallmatrix}
    1.00 & -0.30 \\ -0.30 & 1.00
\end{bsmallmatrix}$. The predictor matrix $X$ includes an intercept and $p-1$ columns generated from a standard uniform distribution over $[0,1]$. The range parameter for the exponential spatial covariance function, and the proportion of spatial variability, are fixed at $\phi=4$ and $\alpha=0.8$, respectively. We set prior information as follows: $m_{0}=0_{p\times q}$, $M_{0}= 10\mathbb{I}_p$, $\Psi_{0}=\mathbb{I}_{q}$, and $\nu_{0}=3$. These specifications and prior information remain constant, allowing only the number of locations in each subset to vary. In performing \textsc{double bps} detailed in Section~\ref{sec:AccelerMultivarMod}, we consider $J=9$ competitive models characterized by $\alpha \in \{0.70, 0.80, 0.90\}$ and $\phi \in \{3, 4, 5\}$.

\begin{figure}[!t]
    \centering
    \includegraphics[scale=0.50]{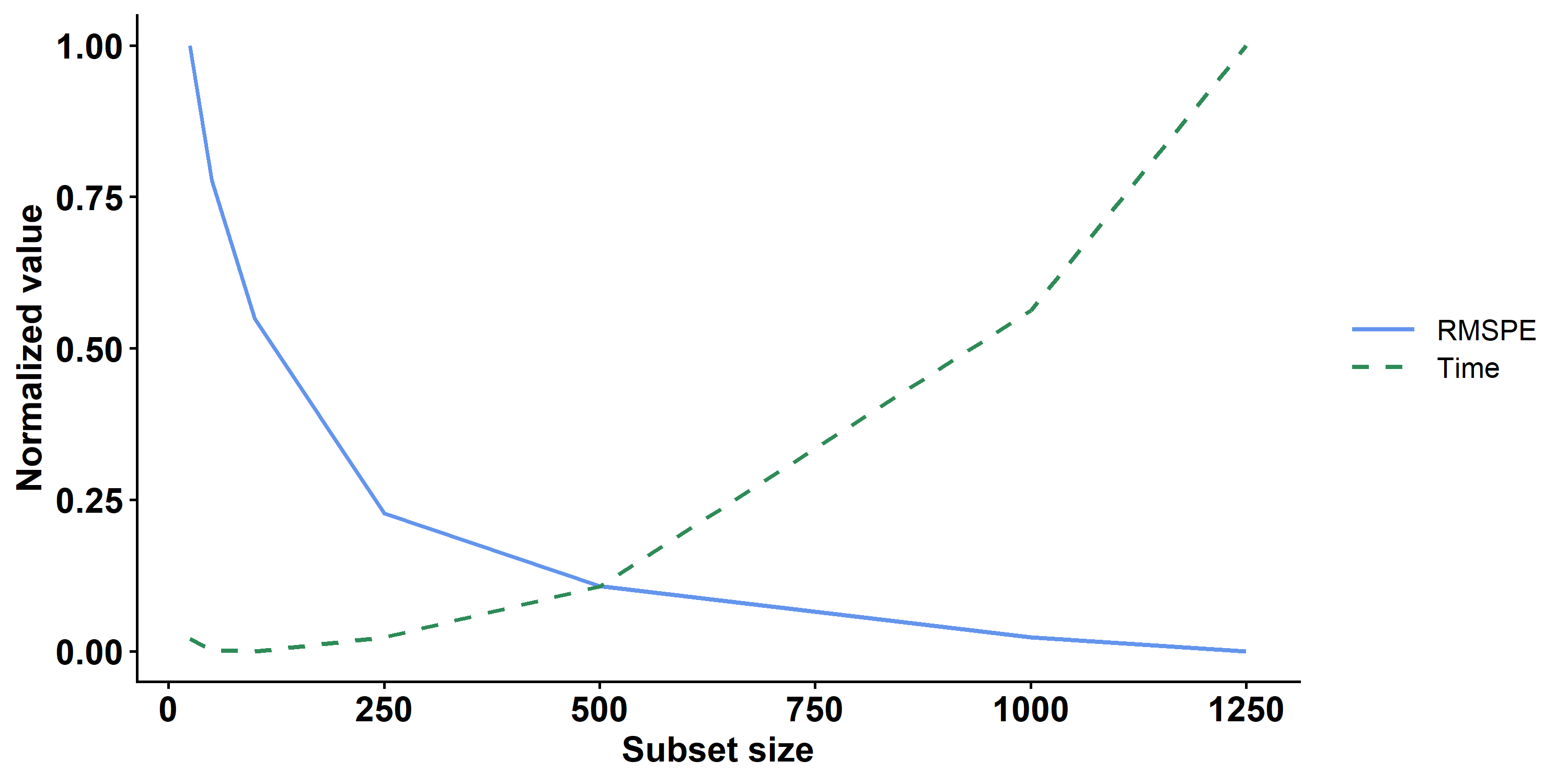}
    \caption{ Comparison between average \textsc{rmspe} (solid line) and model fitting time (dashed line) across various subset dimensions (both min-max normalized).
}
    \label{fig:subset_sens}
\end{figure}

We focus our sensitivity analysis by selecting the following set of partition sizes: \{25, 50, 100, 250, 500, 1000, 1250\}, which correspond to the number of partitions $K\in$\{200, 100, 50, 20, 10, 5, 4\}. Figure~\ref{fig:subset_sens} illustrates the two curves resulting from this sensitivity analysis.

As anticipated, the behavior of the two curves aligns with theoretical expectations across most scenarios. Specifically, the total time required to fit the model increases monotonically, exhibiting more than linear growth in the number of locations within each partition, as shown by the dashed line in Figure~\ref{fig:subset_sens}. Conversely, the root mean square prediction error (\textsc{rmspe}) decreases with partition size until it stabilizes at a ``plateau'' from approximately $500$ units onward. The intersection of the two normalized curves in Figure~\ref{fig:subset_sens} identifies a trade-off between predictive performance and computational effort close to a subset size of approximately $500$ units.

However, Figure~\ref{fig:subset_sens} also raises an important question: how much predictive error is acceptable? The two quantities, although normalized for comparison, differ significantly in their scales. More precisely, the trade-off is asymmetric: doubling the number of locations per partition yields a moderate reduction in \textsc{rmspe}, while the runtime can increase dramatically, rising at least quadratically with $n$. For all these reasons, we generally opt for a subset size of $500$ locations in both our simulation studies and data applications. Nonetheless, we should not overlook the opportunity to reduce this size, accepting a compromise in predictive performance to achieve even faster global Bayesian inference for exceptionally large GeoAI applications.

%%%%%%%%%%%%%%%%%%%%%%%%%%%%%%%%%%%%%%%%%%%%
\section{Exploratory Data Analysis}\label{sec:eda}

We illustrate more exploratory analysis on Vegetation Index data presented in Section~\ref{sec:dataset}. This section starts from the model-based non-spatial association among response variables, and concludes by presenting machine-generated supportive exploratory data insight used in Section~\ref{sec:dataappl}.

We investigate non-spatial association between \textsc{ndvi} and red reflectance fitting the Bayesian multivariate regression model, defined by Equation~\eqref{eq:MatrixNormLik}. The model comprises two predictors: an intercept and the solar zenith angle for the million locations in the training set. More details on modeling and prior distribution are provided in Section~\ref{sec:dataappl}, where a comparison of predictive performances is presented. Table~\ref{tab:eda_nonsp_assoc} reports quantile estimates for marginal variances, i.e., the diagonal elements of $\Sigma$, and the correlation between \textsc{ndvi} and \textsc{rr}. Strong negative values for correlation are estimated, showing an intense inverse relationship between the two spatially dependent outcomes.

\begin{table}[t!]
\centering
\small
\begin{tabularx}{\textwidth}{Y Y Y}
\toprule
$\Sigma_{\textsc{ndvi}}$ & $\Sigma_{\textsc{rr}}$ & $\varrho_{\textsc{ndvi},\textsc{rr}}$ \\
\midrule

\cellcolor{gray!10} 0.2208 (0.2202, 0.2215)
& \cellcolor{gray!10} 0.1549 (0.1545, 0.1553)
& \cellcolor{gray!10} -0.9049 (-0.9052, -0.9046)
\\

\bottomrule
\end{tabularx}
\caption{Non-spatial association between response variables. 50 (2.5, 97.5) quantile estimates using Bayesian multivariate linear regression.}
\label{tab:eda_nonsp_assoc}
\end{table}

Hereafter, we present results from the fully automated exploratory spatial data analyses that complement Section~\ref{sec:dataappl}. Variograms are employed in both analyses to extract ``guidelines'' on spatial parameters such as spatial variability proportion $\alpha$,
% signal-to-noise ratio $\delta^{2}$, 
and spatial range $\phi$, which are essential for setting up the \textsc{double bps} framework for GeoAI applications. Variogram fitting, used to gather parameter values required for \textsc{double bps}, is fully automated and requires no human intervention, except for specifying the grid width.

\begin{figure}[!t]
    \centering
    \includegraphics[scale=0.5]{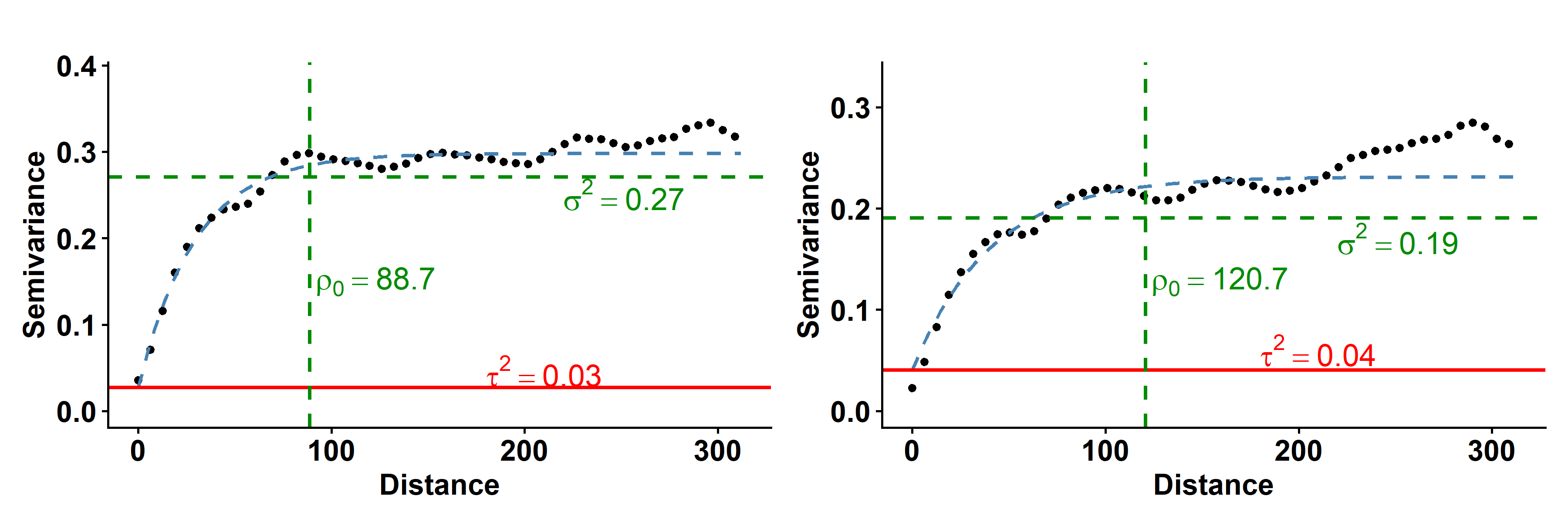}
    \caption{From left to right: sample variograms of \textsc{ndvi}, and Red Reflectance.}
    \label{fig:vario_ndvi_rr}
\end{figure}

First, we use independent sample variograms for \textsc{ndvi} and \textsc{rr}, based on $31,875$ randomly sampled locations. For \textsc{ndvi}, the empirical variogram estimates the nugget $0.03$, a sill of $0.27$, and a practical range of approximately $88$ based upon automated weighted least squares. This corresponds to significant spatial correlation up to about $10,000$ kilometers. The proportion of spatial variability is computed as $\alpha = \sigma^{2}/(\tau^{2} + \sigma^{2}) = 0.27/(0.03 + 0.27) \approx 0.9$, resulting in $0.909$ without rounding. Finally, the spatial range parameter is estimated as $\phi = 0.067$ based upon the distance beyond which the spatial correlation drops to less than $0.05$; see the left panel of Figure~\ref{fig:vario_ndvi_rr}.

For \textsc{rr}, the variogram parameters include a nugget effect of $0.04$, a sill of $0.19$, and a practical range of approximately $120$, which corresponds to around $13,000$ kilometers. The slightly higher nugget effect for \textsc{rr} suggests greater measurement error or micro-scale variability compared to \textsc{ndvi}. The proportion of spatial variability is estimated as $\alpha = 0.19/(0.04 + 0.19) \approx 0.825$. The practical range for \textsc{rr} is more extended than that of \textsc{ndvi}, indicating that \textsc{rr} values remain spatially correlated over a greater distance. Concluding the exploratory spatial data analysis, we select a spatial range of $\phi = 0.049$ for \textsc{rr}.

The combined analysis of the variograms for \textsc{ndvi} and \textsc{rr} provides essential information about the spatial variance proportion and range parameters, which are critical for informing artificially intelligent geospatial modeling systems. This analysis results in $\alpha \in \{0.825, 0.909\}$ and $\phi \in \{0.049, 0.067\}$. These findings help improve the accuracy of spatial predictions, enhancing ecological interpretations and increasing computational efficiency by avoiding excessively misspecified model specifications.

The entire exploratory analysis workflow, designed to gather critical insights for improving the \textsc{double bps} methodology, is fully automated. Human input is then minimized; the only required user input is the number of grid values for each spatial parameter.

\begin{figure}[!t]
    \centering
    \includegraphics[scale=0.55]{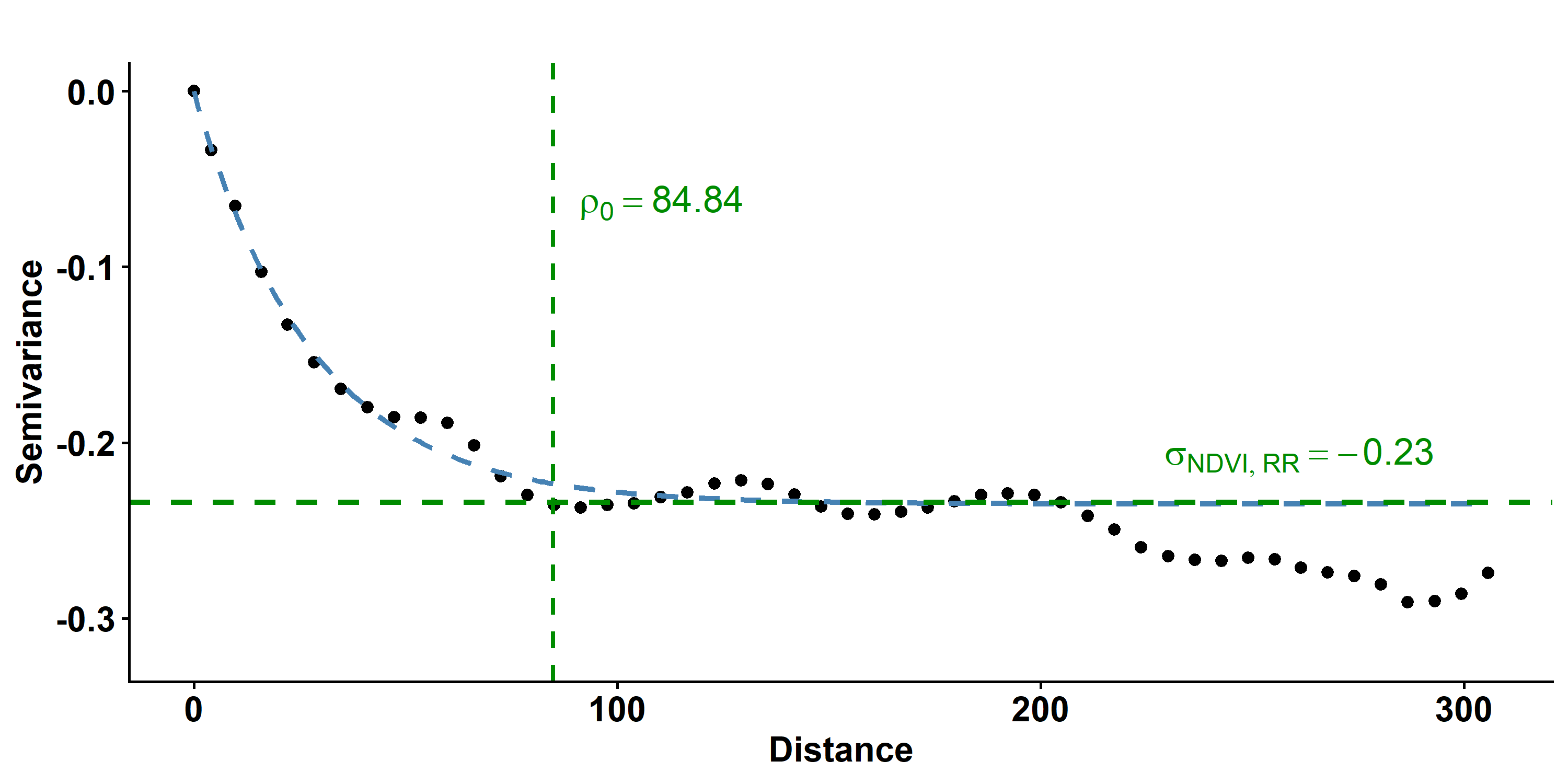}
    \caption{Sample cross-variogram between \textsc{ndvi}, and Red Reflectance.}
    \label{fig:vario_cross}
\end{figure}

Using the same subsample composed of $31,875$ locations, we also investigated the cross-variogram; see Figure~\ref{fig:vario_cross}. This will help us gain insights into spatial cross-dependencies.

The empirical cross-variogram depicts negative values, providing an estimate for the sill of $-0.23$, and a practical range of approximately $85$ based upon automated weighted least squares. Similarly to individual variogram analysis, this shows a significant (negative) spatial correlation that withstands up to several thousand kilometers, suggesting a clear and well-defined negative spatial correlation structure between \textsc{ndvi} and red reflectance. The negative cross-variogram mirrors the negative correlation found using the non-spatial model \eqref{eq:MatrixNormLik}. The presence of strong (negative) spatial correlation among these indices is not surprising, as their definitions are strictly related, and both are based on spectral reflectance measurements acquired in the visible and near-infrared regions. Intuitively, the negative correlation emerged considering that healthy vegetation, which reflects high levels of biomass (\textsc{ndvi}), has strong chlorophyll absorption abilities, then revealing low red reflectance. Conversely, an increase in red reflectance corresponds to stressed (or low) vegetation, which results in low levels of the normalized difference vegetation index.
The non-spatial and spatial negative associations are fully consistent with the nature of these indices and with the literature \citep{tucker_red_1979,sellers_canopy_1985}.

%%%%%%%%%%%%%%%%%%%%%%%%%%%%%%%%%%%%%%%%%%%%
\vskip 0.2in
\bibliography{JMLR/ref}

\end{document}